\documentclass[11pt]{article}

\usepackage{setspace}
\usepackage[margin=1.5in]{geometry}

\usepackage[english]{babel}
\usepackage{amsmath}
\usepackage{amsfonts}
\usepackage{amssymb}
\usepackage{graphicx}
\usepackage{tikz-cd}
\usepackage{mathrsfs}
\usepackage{xfrac}
\usepackage{url}
\usepackage{color}
\usepackage[colorlinks=true, citecolor=blue]{hyperref}
\usepackage{enumitem}
\usepackage[usestackEOL]{stackengine}
\usepackage{accents}

\numberwithin{equation}{section}
\date{\today}

\def\a{\alpha}   
   
\def\t{\theta}
\def\be{\begin{equation}}
\def\ee{\end{equation}}
\def\ba#1\ea{\begin{align}#1\end{align}}
\def\no{\nonumber\\ }
\def\ra{\rangle}
\def\la{\langle}
\def\lla{\left\langle\!\left\langle}
\def\rra{\right\rangle\!\right\rangle}
\def\diff{\text{\sl Diff}^+\!(S^1)}
\def\adiff{\mathfrak{diff}(S^1)}
\def\ddiff{\mathfrak{diff}^*(S^1)}
\def\whddiff{\accentset{\circ}{\mathfrak{diff}}^*(S^1)}
\def\psl{\text{\sl PSL}(2, \bb R)}
\def\apsl{\mathfrak{psl}(2, \bb R)}
\def\vira{\text{\sl Vira}}
\def\avira{\mathfrak{vira}}
\def\dvira{\mathfrak{vira}^*}
\def\Ad{\text{Ad}}
\def\ad{\text{ad}}
\def\coad{\text{coad}}

\def\ads{\text{\sl AdS}_3}
\def\mink{\text{\sl Mink}_3}

\def\lads{\ell_\text{\sl AdS}}
\def\bms{\mathfrak{bms}_3}
\def\BMS{\text{BMS}_3}
\def\timesext{\times_{\!\text{\it ext}}}

\def\mod{\text{ mod }}
\def\tr{\text{\emph{Tr}}}

\newcommand{\ca}[1]{\mathcal{#1}}
\newcommand{\bb}[1]{\mathbb{#1}}
\newcommand{\fr}[1]{\mathfrak{#1}}
\newcommand{\scr}[1]{\mathscr{#1}}
\newcommand{\ii}{\imath}

\newcommand{\wt}[1]{\widetilde{#1}}
\newcommand{\bs}[1]{\boldsymbol{#1}}
\newcommand{\wh}[1]{\widehat{#1}}
\newcommand{\ol}[1]{\overline{#1}}
\newcommand{\un}[1]{\underline{#1}}
\newcommand{\ac}[1]{\accentset{\circ}{#1}}

\newcommand{\sdplus}{\mathbin{\put(5.7,2.7){\oval(9.5,6.5)[l]}+\!}}

\graphicspath{{Figures/}}

\title{\bf Quantization of causal diamonds\\ in (2+1)-dimensional gravity\\Part II: Group-theoretic quantization
\vspace{5mm}
}
\date{\today}
\author{Rodrigo Andrade e Silva\thanks{rasilva@umd.edu}
\vspace{-5mm}
\\
\and
{\it \small Center for Fundamental Physics,
University of Maryland}\\
{\it\small College Park, MD, 20742, USA}\\
{\it \small and}\\
{\it \small Perimeter Institute for Theoretical Physics}\\
{\it\small Waterloo, ON, N2L 2Y5, Canada}}

\begin{document}
\begin{titlepage}
\maketitle
\thispagestyle{empty}

\begin{abstract}
We develop the non-perturbative reduced phase space quantization of {\sl causal diamonds} in (2+1)-dimensional gravity with a nonpositive cosmological constant. In Part I we described the classical reduction process and the reduced phase space, $\widetilde{\mathcal P} = T^*(\text{\sl Diff}^+\!(S^1)/\text{\sl PSL}(2, \mathbb R))$, while in Part II we discuss the quantization of the phase space and quantum aspects of the causal diamonds. Because the phase space does not have a natural linear structure, a generalization of the standard canonical (coordinate) quantization is required. In particular, as the configuration space is a homogeneous space for the $\text{\sl Diff}^+\!(S^1)$ group, we apply Isham's group-theoretic quantization scheme. We propose a quantization based on (projective) unitary irreducible representations of the $\text{BMS}_3$ group, which is obtained from a natural prescription for extending $\text{\sl Diff}^+\!(S^1)$ into a transitive group of symplectic symmetries of the phase space. We find a class of suitable quantum theories labelled by a choice of a coadjoint orbit of the Virasoro group and an irreducible unitary representation of the corresponding little group. The most natural choice, justified by a Casimir matching principle, corresponds to a Hilbert space realized by wavefunctions on $\text{\sl Diff}^+\!(S^1)/\text{\sl PSL}(2, \mathbb R)$ valued in some unitary irreducible representation of $\text{\sl SL}(2, \mathbb R)$.  A surprising result is that the twist of the diamond corner loop is quantized in terms of the ratio of the Planck length to the corner perimeter.
\end{abstract}

\end{titlepage}

\tableofcontents

\section{Introduction}
\label{sec:intro}

In this paper, which is the sequel of \cite{e2023quantization}, we continue the development of a non-perturbative canonical quantization of causal diamonds in (2+1)-dimensional gravity.
Causal diamonds refer to the class of finite-sized, globally-hyperbolic spacetimes whose Cauchy slices have the topology of a ball.
More precisely, the system of interest is (2+1)-dimensional Einstein-Hilbert gravity with a nonpositive cosmological constant in the domain of dependence of a topological spatial disc, where the (induced) metric on the boundary of the disc is fixed.
The general description of the problem and the fundamental motivations are explained in considerable detail in the introduction of Part I \cite{e2023quantization}. In addition, Part I is devoted to the classical aspects of the problem, including the study of the structure of constraints and gauge transformations, and the symplectic reduction of the phase space. The reduced phase space is shown to be $T^*(\diff/\psl)$. Here, in Part II, we study the quantization of this phase space, applying Isham's group-theoretic method of quantization \cite{isham1984topological,isham1989canonical}, and discuss some general aspects of the resulting quantum theory.
We remark that a succinct summary of the main results is also presented in \cite{e2023causal}.

For reference, we recollect that there is a rich literature on (2+1)-dimensional gravity systems, including work on spacetimes with closed spatial slices (where the reduced phase space is finite-dimensional) \cite{witten19882+,witten1989topology,moncrief1989reduction,moncrief1990solvable,fischer1997hamiltonian,ashtekar19892+,hosoya19902+,Carlip:1998uc, carlip2005quantum,mondal2020thurston}, on spacetimes with finite timelike boundary \cite{kraus20213d,adami2020sliding,ebert2022field}, and on asymptotically $\ads$ spacetimes\cite{brown1986central,freidel20042+,carlip2005conformal,witten2007three,maloney2010quantum,Scarinci:2011np,kim2015canonical, cotler2019theory}. 
As explained in Part I, our work aspires to add to this literature by considering two essential points together: the first is to better understand how to describe quantum gravity in finite regions of spacetime, which we do by considering the class of spacetimes consisting of causal diamonds (a topic that has received some recent interest, such as in \cite{chandrasekaran2019symmetries, de2016entanglement,banks2021path,jacobson2019gravitational,jacobson2022entropy}); and second we wish to carry out a fully nonperturbative quantization, and more generally continue the exploration of a program for quantizing gravity non-perturbatively by explicitly reducing the phase space via a CMC (constant-mean-curvature) gauge-fixing for time (pioneering papers include \cite{moncrief1990solvable,moncrief1989reduction,fischer1997hamiltonian}, and a contemporary one is \cite{witten2022note}).

\subsection{Summary}
\label{subsec:sum}

We start with a summary of the contents of the present paper. The goal is to provide a quick guide to the main ideas and results, in view of the extensive nature of the paper. 

\vskip 0.2cm
\noindent\emph{\underline{Note}:} Appendix \ref{app:GSandC} contains a compilation of the main symbols, definitions and conventions used in the text.
\vskip 0.2cm

\noindent By solving all the constraints of general relativity and eliminating the associated gauge ambiguities, we have shown in Part I that the the fully reduced phase space for the causal diamonds, with fixed boundary metric, is 
\be
\wt P = T^*(\diff/\psl)
\ee
with the natural symplectic form coming from the cotangent bundle structure.
The problem then becomes quantizing a non-gauge theory. Canonical quantization is a remarkable tool discovered by Dirac that is based on the principle that the classical theory and its corresponding quantum theory share (roughly) the same underlying algebraic structure of observables (Sec.~\ref{subsec:canons}). In its original and simplest incarnation, one takes a complete set of canonically conjugate coordinates $x^i$ and $p_j$, satisfying the Poisson algebra $\{x^i, p_j\} = \delta^i{}_j$, and then defines the quantum theory in such a way that self-adjoint operators $X^i$ and $P_j$ form an irreducible representation, on a Hilbert space, of the analogous algebra $\frac{1}{i\hbar}[X^i, P_j] = \delta^i{}_j$, where $\hbar$ is the Planck constant (which has dimensions of angular momentum, like $[X,P]$). This rule has worked magnificently in many different scenarios, from the quantization of Newtonian particles in Euclidean space to relativistic fields in Minkowski space. 
There is, however, an important limitation: it only produces a sensible quantum theory if the phase space has a natural linear structure, $\ca P  = \bb R^{2n}$, so that there is a natural class of global coordinates $x^i$ and $p_j$, ranging from $-\infty$ to $+\infty$, which can be used to produce a preferred quantum theory. In particular, notice that from the assumption that $P$ is self-adjoint it follows that $e^{aP/i\hbar}$ is a well-defined bounded operator for any $a \in \bb R$; and if $X$ has an eigenvector $|\psi\ra$ with eigenvalue $x \in \bb R$, then it follows from the algebra $[X,P] = i\hbar$ that $e^{aP/i\hbar}|\psi\ra$ is another eigenvector of $X$ with eigenvalue $x+a$; therefore, if the spectrum of $X$ (and similarly that of $P$) is not empty (i.e., the theory is non-trivial), then it must be the whole real line. For this reason, when the phase space has a non-trivial topology, or lacks a natural global chart of ``Cartesian coordinates'', a more sophisticated method of quantization is necessary. 

In our case, while $T^*(\diff/\psl)$ happens to be topologically contractible, it appears not to have a natural linear structure. Opportunely, it does admit a natural group action, as the configuration space $\ca Q := \diff/\psl$ is a homogeneous space for the group $\diff$, so it is reasonable to employ Isham's group-theoretic method of quantization \cite{isham1984topological,isham1989canonical}. In this method (Sec.~\ref{subsec:isham}), one takes a transitive\footnote{A group $G$ acts transitively on a manifold $\ca M$ if for any two points $x,\, x'\in \ca M$ there exists $g \in G$ such that $x' = gx$.} group $\wt G$ of symplectic symmetries of the phase space and uses it to define a set of classical observables and their corresponding quantum operators, where the Hilbert space is constructed to carry a (projective) irreducible unitary representation of the group. The justification comes from the core principles of canonical quantization: any element $\xi$ in the Lie algebra $\wt{\fr g}$ of $\wt G$ generates a Hamiltonian flow $X_\xi$ on the phase space, to which is associated a Hamiltonian charge $H_\xi$ (solution of $d H_\xi = - i_{X_\xi}\omega$); given any basis $\xi_i$ of $\wt{\fr g}$, $i = 1\ldots \text{dim}(\wt{\fr g})$, we have a set of classical charges $H_i := H_{\xi_i}$ whose Poisson algebra is homomorphic to $\wt{\fr g}$, i.e., $[\xi_i, \xi_j] = c^k_{ij} \xi_k \Rightarrow \{H_i, H_j\} = c^k_{ij} H_k$ (we assume that the group has been extended, if necessary, to incorporate a non-trivial central charge that could otherwise subvert this homomorphism); moreover, the transitivity of $\wt G$ implies that this set of charges is complete in the sense that the specification of their value determines the phase space point (up to, possibly, a discrete ambiguity); therefore, the underlying group of symmetries ensures that there is a well-grounded, complete and algebraically-closed set of classical charges on the phase space, which can then be quantized according to the traditional rule, i.e., $H_i \mapsto \wh H_i$ where $\wh H_i$ are self-adjoint operators on a Hilbert space forming an irreducible representation of the algebra $\frac{1}{i\hbar}[\wh H_i, \wh H_j] = c^k_{ij} \wh H_k$. In this context, $\wt G$ is called the {\sl canonical group}, and the $H$'s are {\sl canonical charges}. 

As an example, in the case of a particle on the Euclidean line (i.e., phase space $\ca P = T^*\bb R$), notice that $p$ generates translations in $x$ while $x$ generates translations in $-p$. So instead of thinking of the coordinates $x$ and $p$ as the elementary ingredients for quantization, reverse the logic and think of this $\bb R^2$ group of symmetries, $(x,p) \mapsto (x + a, p - b)$, as the starting point, i.e., where now $x$ and $p$ are {\sl derived} from the group as the associated canonical charges. In this case, the corresponding Poisson algebra does introduce a non-trivial central charge, $\{x, p\} = 1$, implying that we need to consider a central extension of $\bb R^2$, namely the Heisenberg group $H(3) = \wh{\bb R^2}$. Accordingly, the quantum theory is constructed in terms of a unitary irreducible representation of $\wt G := H(3)$, which is unique (up to unitary equivalence) and precisely the familiar one realized by square-integrable $\bb C$-valued wavefunctions on $\bb R$.

To construct a natural canonical group for the causal diamonds we start with the group $G = \diff$ acting on the configuration space $\ca Q = \diff/\psl$ from the left as $\delta_\phi[\psi] := [\phi\circ\psi]$, where $\phi,\,\psi \in \diff$ and $[\psi] \in \ca Q$. This action is transitive on $\ca Q$ and naturally lifts to a $G$-action $\wt\delta_\phi$ on the cotangent bundle $\wt{\ca P} = T^*\ca Q$, where 1-forms $p$ at $[\psi]$  are pulled-back by $\phi^{-1}$ to a 1-form at $[\phi\psi]$, $\wt \delta_\phi p := \delta_{\phi^{-1}}^* p$. While this action  indeed defines a symplectic symmetry of $\wt{\ca P}$, it is not transitive on $\wt{\ca P}$ because it does not move points ``vertically'' (i.e., along the fibers of $T^*\ca Q$). 
There is a general prescription (Sec.~\ref{subsec:cotbund}) to extend such a group $G$, acting transitively on $\ca Q$, to a group $\wt G$ acting transitively on $T^*\ca Q$ as symplectomorphisms. The core step is to find a representation of $G$ on some vector space $V$ with the property that at least one of the $G$-orbits is diffeomorphic to $\ca Q$. In this setting, every dual vector $\alpha \in V^*$ defines a ``momentum translation'' as follows: as $\ca Q$ is embedded in $V$, tangent vectors to $\ca Q$ can be identified with vectors $v$ in $V$, and thus $\alpha$ defines a 1-form field on $\ca Q$ mapping $v \in T\ca Q$ to $\alpha(v) \in \bb R$; this 1-form field can then be used to translate $p \in T^*\ca Q$ as $p \mapsto p - \alpha$, which thus defines a group action of $V^*$ on $\wt{\ca P}$. This vertical action is a symplectomorphism, and most importantly the combined $G$ and $V^*$ actions form a transitive group  $\wt G := V^* \rtimes G$ of symplectomorphisms of the phase space $T^*\ca Q$. 

With this general machinery at hand, we can return to our causal diamonds (Sec.~\ref{subsec:cangroup}). Despite the apparent promise of starting with the $\diff$ action, we could not find a linear representation of it with the required property of having an orbit diffeomorphic to $\diff/\psl$. Fortunately, it happens that the coadjoint representation of the Virasoro group does have an orbit diffeomorphic to $\diff/\psl$. The Virasoro group, being a central extension of $\diff$ by $\bb R$, $\vira = \wh{\diff}$, can just as well be used as the group of ``configuration translations'', $G = \vira$, where the central element acts trivially on the configuration space, i.e., $\delta_{(\phi, a)}[\psi] := [\phi \circ \psi]$. That is, the configuration space can be equivalently expressed as the homogeneous space $\ca Q = \vira/(\psl \times \bb R)$. For completeness and compatibility of notation, we review some basic facts about the Virasoro group (Sec.~\ref{subsec:Vira}), and properties of its adjoint (Sec.~\ref{subsubsec:ViraAd}) and coadjoint (Sec.~\ref{subsubsec:ViraCoad}) representations, with particular interest in describing how $\diff/\psl$ is embedded as a coadjoint orbit of Virasoro into $V = \dvira$ (the dual Lie algebra of $\vira$). We leave some details on the construction of the Virasoro group, and its adjoint and coadjoint representations, to App.~\ref{app:Vira}.
Thus, we propose (Sec.~\ref{subsubsec:CanonVira}) that the quantization of causal diamonds should be based on the canonical group that is the semi-direct product
\be
\wt G = V^* \rtimes G = (\avira^*)^* \rtimes \vira
\ee
where $\vira$ acts as ``configuration translations'' (i.e., lifted from the action on the configuration space, thus moving points ``laterally'' on $T^*\ca Q$) and the abelian normal subgroup $(\avira^*)^* \sim \avira$  acts as ``momentum translations'' (i.e., moving points vertically on $T^*\ca Q$). In more detail, this group acts on the phase space as follows: when $\ca Q$ is realized as a coadjoint orbit of Virasoro, a point  $[\psi] \in \ca Q$ is identified with a vector in $\dvira$; a point $p$ in phase space (which is, say, a 1-form at $[\psi]$) is identified with a pair $(\wh\rho,[\psi])$, where $\wh\rho$ is some dual vector in $\dvira$ (i.e., an element of $(\dvira)^* \sim \avira$); then a group element $(\wh \xi; \wh\phi) \in \avira \rtimes \vira$ acts on $p$ as $\Gamma_{(\wh \xi, \wh\phi)} (\wh\rho, [\psi]) = (\ad_{\wh\phi}\wh\rho - \wh\xi, [\phi\psi])$, where $\ad$ denotes the adjoint action of $\vira$ on $\avira$. 
To close this section, we describe (Sec.~\ref{subsubsec:Liftaction}) how to lift the group action to the partially-reduced phase space $\wh{\ca S} = \diff \times \whddiff$ (whose structure is simpler to manipulate) and also to the constrained ADM phase space $\ca P$ (characterized in terms of geometrical variables, which can therefore be useful in extracting the physical meaning of the symmetries and the associated charges). 

The next step (Sec.~\ref{subsec:cancharges}) is to evaluate the canonical charges  associated with $\wt G = \avira \rtimes \vira$, denoted by $H_{(\wh\eta; \wh\xi)}$, where $(\wh\eta; \wh\xi) \in \wt{\fr g} = \avira^c \sdplus \avira$ (in which $\avira^c$ denotes the algebra of the abelian group $\avira$, so $\avira^c$ is isomorphic to $\avira$ as a vector space but has a commutative algebraic structure).  This can be obtained from the general formulas derived in Sec.~\ref{subsec:cotbund} for phase spaces with a cotangent bundle structure. The $\vira$ part of $\wt G$ act as configuration (or lateral) translations and therefore the corresponding charges are interpreted as ``momentum variables'', denoted by $P_{\wh\xi} := H_{(0;\wh\xi)}$; and the (abelian) $\avira$ part of $\wt G$ act as momentum (or vertical) translations so the corresponding charges are interpreted as ``configuration variables'', denoted by $Q_{\wh\eta} := H_{(\wh\eta;0)}$. These charges can be decomposed in a convenient Fourier basis: note that elements $\wh\xi$ of $\avira$  are characterized by a vector field on $S^1$ plus a central component, $\wh\xi = \xi(\theta)\partial_\theta + x \wh c$, and the vector field can be expanded in a Fourier series, $\xi(\theta) = \sum_{n\in \bb Z} \xi_n e^{in\theta}$; the canonical charges can then be decomposed in terms of momentum modes $P_n := H_{(0; e^{in\theta}\partial_\theta)}$, a momentum central charge $P_R := H_{(0;\wh c)}$, configuration modes $Q_n := H_{(e^{in\theta}\partial_\theta;0)}$ and a configuration central charge $Q_T := H_{(\wh c;0)}$. Also, the additive constant ambiguities in the definition of the canonical charges can be chosen so that no additional central charges appear in the Poisson algebra, in which we take $P_R = 0$ and $Q_T = 1$. The
Poisson algebra of the canonical charges then becomes
\begin{align}
&\{P_n, P_m\} = i (n - m) P_{n+m} \nonumber\\
&\{Q_n, P_m\} = i (n - m) Q_{n+m} - 4\pi i n^3 \delta_{n+m, 0} \nonumber\\
&\{Q_n, Q_m\} = 0
\end{align}
Notably, this corresponds to a $\bms$ algebra, which is known to be the algebra of asymptotic symmetries at the null infinity of asymptotically-flat spacetimes in 2+1 dimensions \cite{barnich2007classical}. 

Since these charges will become the basic operators in the quantum theory, it is important to understand their physical meaning (Sec.~\ref{subsec:chargeinterpretation}). In fact, as the Hilbert space is always assumed to be separable (i.e., to have a countable topological basis) in quantum mechanics, there is nothing distinguishing about the Hilbert space itself (other than its dimension) and all the information characterizing a particular theory lies in the way that relevant physical observables are represented on the Hilbert space. 
The canonical charges derived from Isham's method have explicit formulas in term of variables describing the reduced phase space, but since these are quite abstract the underlying spacetime meaning of the charges must be uncovered. We explain (Sec.~\ref{subsubsec:chargeinterpretationmom})  that the $P$ charges are associated with diffeomorphisms acting non-trivially at the boundary and their value are related to Fourier modes of $K_{ab} t^a s^b$, the components of the extrinsic curvature $K_{ab}$ of the CMC slice along the tangent ($t^a$) and normal ($s^a$) unit vectors at the boundary of the disc. Of notable mention, the charge $P_0$ generates uniform rotation of the boundary (i.e., $SO(2)$ isometries of the boundary metric) and therefore can be interpreted as the spin (or angular momentum) of the diamond; moreover, its value is simply related to the twist (Sec.~\ref{subsubsec:chargeinterpretationtwist}) of the corner loop.\footnote{The {\sl twist} of a loop embedded in a three-dimensional space is the integrated torsion (with respect to proper length) along the loop. The torsion measures how a normal frame that is carried along the curve, as close as possible to being parallel transported (more precisely, Fermi-Walker transported) rotates around the axis aligned with the curve. The twist can also be interpreted as measuring the holonomy for (Fermi-Walker) transporting a frame along the loop, i.e., after the completion of the loop the transported frame will return boosted with respect to the original frame, and the boost angle is equal to the twist.} The physical interpretation of the $Q$ charges, on the other hand, has been far more elusive (Sec.~\ref{subsubsec:chargeinterpretationpos}). We know some properties that they must satisfy, and can speculate on what they could measure, but their precise meaning will be left for future examination.

Finally, we arrive at the quantum theory (Sec.~\ref{sec:quantumtheory}), constructed from a (projective) unitary irreducible representation of canonical group $\wt G = \avira \rtimes \vira$ or rather, as revealed from the Poisson algebra of canonical charges, the group $\text{BMS}_3$. 
Since this group is a semi-directed product of a group $G = \vira$ with an abelian group $\avira$, Mackey's theory of induced representations \cite{mackey1969induced} can be employed to obtain the unitary irreducible representations of $\wt G$ \cite{oblak2017bms} (modulo possible limitations associated with the infinite dimensionality of the group \cite{mackey1963infinite}). The result (Sec.~\ref{subsec:genaspectsquantum}) is that the quantum theory is characterized by a choice of coadjoint orbit of Virasoro together with a choice of (projective) unitary irreducible representation of the corresponding little group (i.e., the subgroup of $\vira$ that leaves any particular point of the orbit fixed).  The most natural choice is to take the orbit diffeomorphic to the configuration space, in which case the little group is $\psl \times \bb R$, where $\bb R$ comes from the central element of $\vira$ and is assumed to be represented trivially. 
We justify this choice from the Casimir matching principle (Sec.~\ref{subsec:monocasi}), which proposes that the value of Casimir operators should be matched between the classical and the quantum.\footnote{Classically, Casimir operators Poisson-commute with a complete algebra of observables and therefore are constant functions on the phase space. Quantum-mechanically, Casimir operators commute with all other operators and therefore are realized as a multiple of the identity in any (complex) irreducible representation. At either level, the Casimir takes the same value on any physical state, so it is natural to assume that the value of a quantum Casimir should match with with the value of its classical counterpart. (This  principle leads to Dirac's electric-magnetic charge quantization when quantizing a particle on a sphere from Isham's perspective~\cite{e2021particle}.)}
In particular, we consider two types of Casimir operators related to the monodromy-class and winding number of the coadjoint orbits of Virasoro.
With this refinement, the Hilbert space is realized (Sec.~\ref{subsec:naturalorbit}) by wavefunctions on $\ca Q = \diff/\psl$ valued in unitary irreducible representations of $\psl$.
The quantum theory can also be directly constructed from the algebra, which is the quantization of the Poisson algebra above,
\begin{align}
&[\wh P_n, \wh P_m] =  (m - n)\wh P_{n+m} \nonumber\\
&[\wh Q_n, \wh P_m] =  (m - n) \wh Q_{n+m} + 4\pi  n^3 \delta_{n+m, 0} \nonumber\\
&[\wh Q_n, \wh Q_m] = 0 
\end{align}
but note that $\wh P_n$ and $\wh Q_n$ are not expected to be self-adjoint but rather to satisfy $(\wh P_n)^\dag  = \wh P_{-n}$ and $(\wh Q_n)^\dag = \wh Q_{-n}$, 
mimicking the classical relations $(P_n)^* = P_{-n}$ and $(Q_n)^* = Q_{-n}$. We use this to derive the spectrum of $\wh P_0$.

We discuss two important observables in the quantum theory. First, we show (Sec.~\ref{subsec:spinquantization}) that the twist $\ca T$ of the corner loop, also related to the spin of the diamond, is quantized in the quantum theory according to
\be
\ca T = \frac{16\pi^2 \ell_P}{\ell} (n + s)\,,\quad n\in \bb Z
\ee
where $s = 0$ or $1/2$ and $\ell$ is the total length of the corner loop and $\ell_P$ is the Planck length. Lastly, the charge $\wh Q_0$, which happens to commute with the spin, can be shown (Sec.~\ref{subsec:Q0spec}) to have a continuum spectrum that is bounded from above and unbounded from below, attaining a maximum value of $2\pi$ at a state described by a wavefunction $\Psi([\psi])$ localized at the $\psl$-class of the identity $[\psi] = [I]$.

It is worthwhile to notice that this accomplishes only the kinematical part of the quantization, i.e., characterizing the Hilbert space and the manner that the canonical observables are represented on it. A complete quantization would also involve the description of the dynamics, such as successfully representing the time-evolution Hamiltonian (discussed in Part I) and possibly other relevant observables which are expressible in terms of the canonical ones. The limitations and future prospects of our quantization are considered in the discussion (Sec.~\ref{sec:discussion}).

In addition to App.~\ref{app:GSandC} --- {\sl glossary, symbols and conventions} --- we include four other appendices. 
App.~\ref{app:Vira} provides further details on the general construction of the Virasoro group.
App.~\ref{app:topologyQ} describes the topology of the configuration space, $\ca Q = \diff/\psl$.
App.~\ref{app:mackey} offers a general review of Mackey's theory of induced representation and the associated notion of systems of imprimitivity. 
App.~\ref{app:projrep} discusses projective representations and its relationship to central extension (by 2-cocycles) of the group.

\section{Group-theoretic quantization}
\label{sec:canquant}

Canonical quantization is a fantastical tool discovered by Dirac that allows one to infer the quantum theory underlying a given classical theory. It is based on the idea that the dynamical laws of classical and quantum theories, and the underlying mathematical structures describing those laws, are fundamentally analogous. 
The traditional approach to quantization usually involves a choice of conjugate coordinates, $x$'s and $p$'s, on the phase space which are promoted to quantum operators satisfying the canonical commutation relations. However, this approach does not apply directly to phase spaces with nontrivial topologies, or lacking a natural linear structure, since they do not admit a (natural) global coordinate chart. Since the reduced phase space of the diamond, $T^*(\diff/\psl)$, does not have a natural linear structure, we must resort to more general canonical quantization schemes. In particular,  as this phase space is the cotangent bundle of a homogeneous space, it is natural to consider Isham's group-theoretic approach to quantization \cite{isham1984topological,isham1989canonical} in which the quantum theory is based on a transitive group of symmetries of the phase space. This section contains a brief introduction to the basic principle of canonical quantization, followed by a general review of Isham's group-theoretic quantization scheme, which is finally specialized to the case of phase spaces with a cotangent bundle structure.

\subsection{The canons of quantization}
\label{subsec:canons}

Canonical quantization posits that the quantum theory should retain the dynamical laws of the classical theory, to the highest possible extent, while replacing the classical notion of kinematics by the appropriate quantum notion, where states go from points in a phase space to vectors (or rather, rays) in a Hilbert space and observables go from real functions on the phase space to self-adjoint operators on the Hilbert space. More precisely, classical observables are real functions $f$ on a phase space, endowed with a Poisson algebra coming from the symplectic structure,
\be
f *_C g := \{f, g\}
\ee
in a corresponding quantum theory, observables are self-adjoint operators $\wh f$ on a Hilbert space, endowed with a commutator algebra 
\be
\wh f *_Q \wh g := \frac{1}{i \hbar}[\wh f, \wh g]
\ee
where the complex unit is included so that the product returns a self-adjoint operator and $\hbar$ is a dimensional constant (interpreted as the Planck constant). The principle of canonical quantization is that given an appropriate subalgebra of observables in the classical theory, $\ca A_C$, there would be an homomorphism $\fr q$ to a corresponding subalgebra of operators in the quantum theory, $\ca A_Q$. That is, $\fr q: \ca A_C \rightarrow \ca A_Q$ is a linear association from classical observables to quantum observables, $f \mapsto \wh f := \fr q(f)$, such that $\fr q(f *_C g) = \fr q(f) *_Q \fr q(g)$, or, in the more familiar notation, $[\wh f, \wh g] = i\hbar\, \wh{\{f,g\}}$. The intention is to preserve, to a sensible extent, the dynamical structure of the theory. Namely, the time evolution of a classical observable $f$ under the Hamiltonian $H$ is given by
\be
\frac{df}{dt} = \{f, H\}
\ee
while, in the quantum theory, the time evolution (in the Heisenberg picture) of an observable $\wh f$ under the Hamiltonian $\wh H$ is given by
\be
\frac{d\wh f}{dt} = \frac{1}{i\hbar} [\wh f, \wh H]
\ee
indicating that mapping $\{f, H\} \mapsto \frac{1}{i\hbar} [\wh f, \wh H]$ is precisely what is needed to ensure that $df/dt$ is mapped to $d\wh f/dt$, so that the {\sl quantization map} is preserved in time. However, as explained below, there are often obstructions in constructing such a quantization map, consistently, for all observables. Accordingly, different quantization schemes propose different manners to construct a ``most consistent'' quantization map.

In general, the classical subalgebra of observables $\ca A_C$ is assumed to be complete in the sense that the specification of the value of all $f \in \ca A_C$ completely determines the state (i.e., point in the phase space), possibly up to a finite ambiguity (i.e., any point in the phase space has a neighborhood whose points are uniquely determined by the values of the $f$'s). Such a subalgebra of observables, when selected as the basis for quantization, will be called a {\sl canonical algebra}, and its elements {\sl canonical observables} or also {\sl canonical charges}. It would be desirable if this homomorphism could be taken to satisfy some basic properties, such as: 

$(i)$ Mapping the constant unit function, $u:\ca P \rightarrow \{1\} \subset \bb R$, to the identity operator, i.e., $\fr q(u) = 1$; 

$(ii)$ Commutativity with composition by real functions, i.e., for any real function $\phi : \bb R \rightarrow \bb R$, $\fr q(\phi(f)) = \phi(\fr q(f))$. (A.k.a., the von Neumann rule.)

\noindent One may consider these properties desirable for the following reasons. First, an observable that takes the same value $1$ at every classical state should correspond to an operator whose spectrum contains the single value $1$, which is the identity. Second, if one has designed a  classical experiment to measure a given observable $f$, then the same experiment can also be used to measure any other function $\phi$ of $f$ by simply running it and acting on the output data with $\phi$; at the quantum level one could expect that the same principle is true, in that if this experiment is run and the system is found to be in an eigenvector $|\alpha\rangle$ of $\fr q(f)$ with eigenvalue $\alpha$, then the same experiment could be used to measure $\fr q(\phi(f))$, and the same state $|\alpha\rangle$ would also be an eigenvector of the corresponding quantum operator with eigenvalue $\phi(\alpha)$ --- but, by definition, the operator that has the same eigenvectors $|\alpha\ra$ as $\fr q(f)$, with corresponding eigenvalues $\phi(\alpha)$, is precisely $\phi(\fr q(f))$.
In principle, this second property would allow one to extend the quantization map to the entire algebra of observables, since any observable can be (locally) written in terms of a complete set of observables --- such an extension is not unique due to operator-ordering ambiguities. 
Lastly, the corresponding notion of completeness at the quantum level is that the algebra must be represented irreducibly in the Hilbert space. This is based on the correspondence between symplectomorphisms, at the classical level, with unitary transformations, at the quantum level. Namely, an algebra of classical observables is complete, in a sense equivalent to the discussed above, if the symplectomorphisms it generates act transitively on the phase space, i.e., any two states can be connected by ``evolving along'' some observable in the algebra; at the quantum level, an algebra is represented irreducibly if there is no (non-trivial) invariant subspace, which implies that by acting on a state with all elements of the algebra (and their products) the whole Hilbert space is spanned, i.e., any\footnote{Rigorously, the orbit of a state may not be  the full Hilbert space, but rather a dense subset.} two quantum states (rays) can be connected by the unitary exponentiation of some observable in the algebra.

Naturally one would first attempt quantizing the entire algebra of observables, but fundamental obstructions are generally encountered, as revealed by Groenewold-van Hove no-go theorem.\footnote{Strictly speaking, Groenewold-van Hove theorem, in its standard formulation, applies only for trivial phase spaces, $\ca P = \bb R^{2n}$. See \cite{ali2005quantization} for details. The result has been extended to other cases, and it is expected that this kind of obstruction is generic \cite{gotay2000obstructions}. However, some examples have been found where a full, unobstructed quantization is possible \cite{gotay1995full, gotay2001quantizing}.}
In general, there is no quantization map that commutes with the composition by real functions (i.e., that satisfies the von Neumann rule). Moreover, one would expect that a quantization is sensible only if it preserves the natural relationship between the classical and the quantum notions of algebra completeness discussed above. That is, if $\ca A_C$ is a complete algebra of classical observables and there exists a subalgebra $\ca A'_C \subset \ca A_C$ that is also complete at the classical level, then the quantization $\ca A_C \mapsto \ca A_Q$, where $\ca A_Q$ is represented unitarily and irreducibly on a Hilbert space, naturally induces a unitary representation of $\ca A'_Q$ on the same Hilbert space, by restricting to the corresponding subalgebra of $\ca A_Q$. It would be physically reasonable if any such restricted representations were also irreducible. Nonetheless, it is not true in general that there exists an irreducible representation of $\ca A_Q$ such that all of its classically-complete subalgebras are represented irreducibly (and therefore are also quantum-complete).

The traditional proposal by Dirac applies well for linear phase spaces, i.e., those isomorphic to a vector space $\bb R^{2n}$, with some preferred set of global conjugate coordinates $x^i$ and $p_i$, satisfying $\{x^i, p_j\} = \delta^i{}_j$. These $2n$ coordinates, together with the constant function $1$, form a Heisenberg algebra of observables that is complete. Therefore, they could be used as basis for quantization, that is, the quantum theory would be based on a unitary irreducible representation of the Heisenberg algebra, $[X^i, P_j] = i\hbar \delta^i{}_j$. There are two important aspects of this quantization that must be stressed. First, the quantum theory is only sensible if the coordinates are global and range from $-\infty$ to $\infty$. The reason is that, according to Stone-von Neumann theorem \cite{hall2013quantum}, the Heisenberg algebra has a unique (up to unitary equivalence) irreducible unitary representation: the states are described by the familiar $\bb C$-valued, square-integrable wave-functions on $\bb R^n$, $\Psi(x)$, on which the canonical observables act as
\ba
&(X^i\Psi)(x) = x^i \Psi(x) \no
&(P_i\Psi)(x) = - i\hbar \frac{\partial\Psi}{\partial x^i}(x) \label{heisenbergrep}
\ea
The spectrum of $X^i$ is therefore the whole real line.\footnote{Another simple manner to see this is the following. Suppose there are self-adjoint operators $X$ and $P$ satisfying the algebra $[X, P] = i$. Since $P$ is self-adjoint, it can be exponentiated, so $e^{-iaP/\hbar}$ is a well-defined bounded unitary operator for any $a \in \bb R$. Now suppose that $X$ has a eigenvector $|x\ra$, with eigenvalue $x \in \bb R$. Then it derives from the algebra that $e^{-iaP/\hbar}|x\ra$ is an eigenvector of $X$ with eigenvalue $x + a$. Therefore, if the spectrum of $X$ is not empty (so the representation is non-trivial), then it must be the whole $\bb R$. Rigorously, $X$ does not have eigenvectors in the strict sense, since ``would-be'' eigenvectors associated with a continuum spectrum are non-normalizable, but it still has eigenvectors in a limiting sense, i.e., if $\lambda$ is in the spectrum then there exists a sequence of non-zero vectors $\Psi_n$ such that $\lim_{n\rightarrow \infty} |\!|(X - \lambda)\Psi_n|\!|/|\!|\Psi_n|\!| = 0$ \cite{hall2013quantum}.} Second, the quantization is only natural insofar as the coordinates are natural, for generally different choices of coordinates may lead to inequivalent quantum theories (e.g., because of operator ordering issues associated with non-linear transformations between coordinates). In the case of a particle on an Euclidean plane, one may argue that Cartesian spatial coordinates, selected by the flat metric, paired with their conjugate momenta, provide a natural basis for quantization. (In fact, different Cartesian systems are linearly related, so all such quantizations are equivalent.)

If the phase space is non-trivial, in the sense that there is no natural global coordinate system to employ, then one must be more careful to carry out the quantization. Two very simple examples are: a particle living on a half-line or on a sphere, where the phase spaces are respectively $T^* \bb R^+$ and $T^*S^2$. In the first case, it may appear that one could take coordinates $x > 0$ and $p$, still satisfying $\{x, p\} = 1$, but as we have seen the quantization would not be sensible since there is no unitary representation of this algebra in which the spectrum of $X$ is restricted to positive numbers. In the second case, the non-trivial topology of the sphere implies that there is no global system of coordinates. When there is no global chart, one could still try to quantize a local chart, which may be sensible in certain approximations, but global and topological aspects would be missed. In particular, the spin variable $S = J \cdot N$, where $J$ is the angular momentum and $N$ is the unit vector describing the (angular) position of the particle, turns out to be quantized in integer multiples of $\hbar/2$, dissimilar to what happens for a particle on a plane $\bb R^2$ (i.e., local chart) where the spin is continuous --- in fact, this quantization of spin is directly related to Dirac's quantization of the electric-magnetic monopoles, $eg = n\hbar/2$. It is worthwhile to note that the Stone-von Neumann theorem, establishing uniqueness of the representation, is very particular to the Heisenberg algebra, which in turn is very particular to trivial phase spaces --- in other cases, the kinematical part of the quantization (i.e., finding a self-adjoint representation of canonical variables) is itself a potentially complicated problem, and the structure of the Hilbert space is only determined after this stage is resolved (for example, in the case of a particle on the sphere, different spin values correspond to inequivalent quantum representations \cite{e2021particle}).

In view of these complications, different frameworks have been developed to employ the canonical principles of quantization to nontrivial phase spaces. Among them we can mention geometric quantization \cite{woodhouse1992geometric,dudley1970quantization,kostant1974definition,souriau1974modele,kirillov2001geometric}, deformation quantization \cite{sternheimer1998deformation,bayen1978deformation,bayen1978deformation} and group-theoretic quantization \cite{isham1984topological,isham1989canonical,isham1984groupi,isham1984groupii}. They all offer a prescription that attempts to find a reasonable balance minimizing the extent to which the classical and quantum algebras fail to be homomorphic, the von Neumann rule is violated, and the classical and quantum notions of completeness disagree. Due to the form of the reduced phase space of causal diamonds, possessing a natural group structure, we shall focus on Isham's group-theoretic formalism.

\subsection{Isham's quantization scheme}
\label{subsec:isham}

Isham proposed that one should identify
a transitive group of symplectic symmetries of the phase space and use it to generate both a special set of classical observables and their associated quantum self-adjoint operators \cite{isham1984topological,isham1989canonical}. 
Such a group will be called the {\sl quantizing group}, or also the {\sl canonical group}.
For intuition, see the introduction (Sec.~\ref{subsec:sum}) where we discuss 
the trivial case of a particle on the line, i.e., a phase space $T^*\bb R = \bb R^2$. 
%Instead of thinking about the coordinates $x$ and $p$ as the fundamental basis for quantization, consider that $p$ is a generator, through its symplectic flow, of spatial translations, $x \mapsto x + a$, while $x$ is similarly a generator of momentum translations, $p \mapsto p - b$. Therefore, $x$ and $p$ together generate a group $\wt G \sim \bb R^2$ of translation symmetries on the phase space, $(x,p) \mapsto (x + a, p - b)$. The new perspective is to take this group as the fundamental basis for quantization, and see $x$ and $p$ as being the charges generated by the group and construct the quantum theory in terms of a (projective) irreducible unitary representation of the group. Noticing that projective representations of a group are in correspondence with true representations of a centrally-extended group, and that in this case the appropriate extension is the Heisenberg group, the standard quantum theory, described by \eqref{heisenbergrep}, is recovered.
%In the general case, the choice of the canonical group may be ambiguous, but the system may have subsidiary structures that select a preferred canonical group (or class thereof), such as a metric on the configuration space that appears in the Hamiltonian.
Let us now review the general formalism underlying this approach. 

Consider a phase space $\wt{\ca P}$ with symplectic 2-form $\omega$. Assume that the phase space is a homogeneous space for some Lie group $\wt G$ of symplectic symmetries.\footnote{The tilde will typically indicate symbols referring to the phase space $\wt{\ca P}$, such as the group $\wt G$. Symbols referring to the configuration space $\ca Q$, such as the group of ``translations'' $G$ in the next subsection, will typically be denoted without accents. While the tildes cause an unfortunate clutter in this section, this choice of notation will prove convenient in the rest of the paper.} That is, there is a left action of $\wt G$ on $\wt{\ca P}$, $\Gamma_{\wt g} : \wt{\ca P} \rightarrow \wt{\ca P}$, that preserves the symplectic form, 
\be
\Gamma_{\wt g}^*\omega = \omega
\ee
for all $\wt g \in \wt G$, and is transitive, i.e., given any two points $p, p'\in \wt{\ca P}$ there exists $\wt g \in \wt G$ such that $p' = \Gamma_{\wt g}(p)$.
Each element $\wt\xi$ in the Lie algebra $\wt{\fr g}$ of $\wt G$ induces a vector field $X_{\wt\xi}$ on $\wt{\ca P}$, as follows. The algebra $\wt{\fr g}$ is naturally identified with the tangent space of $\wt G$ at the identity, $T_e\wt G$, so ${\wt\xi}$ can be seen as a vector tangent to a curve $\wt g_t$ on $\wt G$ starting at $e$; acting with this curve on any point $p \in \wt{\ca P}$ defines a curve $p_t := \Gamma_{\wt g_t}(p)$ on $\wt{\ca P}$ starting at $p$, and therefore a vector $X_{\wt\xi}$ tangent to $p$. More formally,
\be
X_{\wt\xi} |_p := \phi_{p *}({\wt\xi})
\ee
where $\phi_p : \wt G \rightarrow \wt{\ca P}$ is defined by $\phi_p(g) = \Gamma_{\wt g}(p)$.
This map is an anti-homomorphism from $\wt{\fr g}$ into the algebra of vector fields on $\wt{\ca P}$, i.e., 
\be
[X_{\wt\xi}, X_{\wt\eta}] = X_{[{\wt\eta}, {\wt\xi}]}
\ee
As $\Gamma_{\wt g}$ preserves $\omega$, we have $\pounds_{X_{\wt\xi}} \omega = 0$, so $X_{\wt\xi}$ is a (locally) Hamiltonian field. Thus $d(\ii_{X_{\wt\xi}} \omega) = \pounds_{X_{\wt\xi}} \omega - \ii_{X_{\wt\xi}} d\omega = 0$, where $\ii$ denotes the interior product. That is, $\ii_{X_{\wt\xi}} \omega$ is closed and therefore locally exact, so that 
\be
dH_{\wt\xi} = -\ii_{X_{\wt\xi}} \omega
\ee
admits local solutions $H_{\wt\xi}$, called {\sl canonical charges}, defined up to addition of a constant function on $\wt{\ca P}$. In order to properly regard these charges as canonical observables, they need to be globally defined on $\wt{\ca P}$, so we assumed that $\wt G$ strictly generates {\sl globally} Hamiltonian fields on $\wt{\ca P}$. This set of charges will be (classically) complete as a consequence of the transitiveness of the group action.

The symplectic form endows the space of functions on the phase space with an algebraic structure, $\ca A_C$, where the product is given by the Poisson bracket:
since $\omega$ is non-degenerate, any function $f$ on $\wt{\ca P}$ can be associated with a unique vector field $X_f$ on $\wt{\ca P}$ via the relation 
$df = - \ii_{X_f}\omega$; the Poisson bracket between two functions, $f$ and $f'$, is defined by 
\be
\{f, f'\} := - \omega(X_f, X_{f'})
\ee
It is straightforward to show that the vector field associated with $\{f,f'\}$ is $-[X_f, X_{f'}]$, so there is an anti-homomorphism from the Poisson algebra of charges to the algebra of their associated vector fields.
For the canonical charges, $\{H_{\wt\xi}, H_{\wt\eta}\}$ is then associated with the vector field $-[X_{\wt\xi}, X_{\wt\eta}] = X_{[{\wt\xi}, {\wt\eta}]}$, which thus imply that $d(\{H_{\wt\xi}, H_{\wt\eta}\} - H_{[{\wt\xi}, {\wt\eta}]}) = 0$. 
Therefore the map ${\wt\xi} \mapsto H_{\wt\xi}$ is a homomorphism from $\wt{\fr g}$ into $\ca A_C$ up to central charges, i.e.,
\be
\{H_{\wt\xi}, H_{\wt\eta} \} = - \omega(X_{\wt\xi}, X_{\wt\eta}) = H_{[{\wt\xi}, {\wt\eta}]} + z({\wt\xi}, {\wt\eta})
\ee
where $z({\wt\xi}, {\wt\eta})$ is constant on $\wt{\ca P}$. 
If the central charge $z({\wt\xi}, {\wt\eta})$ is not trivial (i.e., it cannot be removed by a redefinition of the charges $H_{\wt\xi} \mapsto H_{\wt\xi} + f({\wt\xi})$, for some $f: \wt{\fr g} \rightarrow \bb R$), we can always extend the group by a central element so that the extended algebra, $\wh{\wt{\fr g}}$, with topology $\wt{\fr g} \oplus \bb R$, has product law $[({\wt\xi}; a), ({\wt\eta}; b)] = ([{\wt\xi}, {\wt\eta}]; z({\wt\xi}, {\wt\eta}))$. The new group (obtained by exponentiating $\wh{\wt{\fr g}}$)  has a natural action on $\wt{\ca P}$, where the central element acts trivially, and the new charges are related to the old ones simply by $H_{({\wt\xi}; a)} = H_{\wt\xi} + a$. Consequently, the map $({\wt\xi}, a) \mapsto H_{({\wt\xi}, a)}$ is a true homomorphism. In this way, if we start with a candidate canonical group that generates non-trivial central charges when acting on the phase space, then we can restart the process with the appropriately extended group as the canonical group. Thus, we will assume  $\wt G$ is chosen such that
\be\label{gAChomo}
\{H_{\wt\xi}, H_{\wt\eta} \} =  H_{[{\wt\xi}, {\wt\eta}]}
\ee
i.e., so that $\wt{\fr g} \rightarrow \ca A_C$ is a true homomorphism.\footnote{For example, in the case of $\wt{\ca P} = T^*\bb R$, the natural group to consider is $\bb R^2$ acting as translations on $\wt{\ca P}$, i.e., $(a,b) \in \wt G$ acting on $(x,p) \in \wt{\ca P}$ as $\Gamma_{(a,b)}(x,p) = (x + a, p + b)$. Its algebra $\wt{\fr g} = \bb R^2$ is commutative, and the associated charges can be taken as $H_{(\alpha,\beta)} = \alpha p - \beta x$, where $(\alpha, \beta) \in \wt{\fr g}$. A non-trivial central charge appears in the Poisson algebra, $\{H_{(\alpha, \beta)},  H_{(\alpha', \beta')}\} = \alpha \beta' - \beta \alpha'$. The algebra can be extended by the central element to $\wh{\wt{\fr g}}$ with product law $[(\alpha, \beta; \gamma), (\alpha' ,\beta' ; \gamma')] := (0; \alpha \beta' - \beta \alpha')$, known as the (3-dimensional) Heisenberg algebra, $\fr h(3)$. The exponentiation of this algebra defines the Heisenberg group, $H(3) = \wh{\bb R^2}$, with topology $\bb R^3$, whose product rule can be expressed as $(a,b;c) (a',b'; c') = (a + a', b + b'; c + c' + \frac{1}{2}ab' -\frac{1}{2} ba')$. Thus the quantization should be based on $\wt G = H(3)$, instead of the original candidate $\bb R^2$. This group acts on $\wt{\ca P}$ as $\Gamma_{(a,b;c)}(x,p) = (x + a, p + b)$ and the associated charges are taken to be $H_{(\alpha,\beta;\gamma)} = \alpha p - \beta x + \gamma$.}

The quantum theory is constructed from $\wt G$ as follows. 
Let $U: \wt G \rightarrow \text{Aut}(\ca H)$ be  
an irreducible unitary representation of $\wt G$ on 
a Hilbert space $\ca H$. As any algebra element ${\wt\xi} \in \wt{\fr g}$ generates a one-parameter subgroup of $\wt G$ through exponentiation, $t \mapsto \exp(t {\wt\xi})$, they can be associated with self-adjoint operators $\widehat H_{\wt\xi}$ on $\ca H$ via
\be
U(\exp t{\wt\xi}) =: e^{t\widehat H_{\wt\xi}/i\hbar}
\ee
It follows from this definition that the map ${\wt\xi} \mapsto \widehat H_{\wt\xi}$ is a homomorphism from $\wt{\fr g}$ into $\ca A_Q$, the algebra of self-adjoint operators on $\ca H$,
\be\label{gAQhomo}
\frac{1}{i\hbar}[\wh H_{\wt\xi}, \wh H_{\wt\eta}] = \wh H_{[{\wt\xi}, {\wt\eta}]}
\ee
Therefore, considering \eqref{gAChomo} and \eqref{gAQhomo}, we see that the the association between 
each classical charge $H_{\wt\xi}$ and the corresponding generator $\widehat H_{\wt\xi}$ of the unitary representation,
\begin{equation}
H_{\wt\xi} \mapsto \widehat H_{\wt\xi} \,,
\end{equation}
is a homomorphism from $\ca A_C$ into $\ca A_Q$. This is called the {\sl quantization map}, for the quantization based on the group $\wt G$.

The quantization
process is summarized as follows:
\begin{center}
\begin{tikzpicture}[commutative diagrams/every diagram]
\node (xi) at (0,2) {$\wt\xi \in \wt{\fr g}$};
\node (X) at (-1.5,0) {$X_{\wt\xi}$};
\node (U) at (1.5,0) {$U\big(\exp\wt\xi\big)$};
\node (H) at (-1.5,-2) {$H_{\wt\xi}$};
\node (whH) at (1.5,-2) {$\wh H_{\wt\xi}$};
\path[commutative diagrams/.cd, every arrow, every label]
(xi) edge node {} (X)
(X) edge node {} (H)
(xi) edge node {} (U)
(U) edge node {} (whH)
(H) edge[dashed] node {} (whH);
\end{tikzpicture}
\end{center}
{\sl A transitive group $\wt G$ of symplectomorphisms defines classical charges and their corresponding quantum operators. 
On the classical side, 
each element ${\wt\xi}$ in the Lie algebra $\wt{\fr g}$ induces a Hamiltonian vector field $X_{\wt\xi}$ on $\wt{\ca P}$, which in turn defines a Hamiltonian charge $H_{\wt\xi}$. On the quantum side, the group element $\exp {\wt\xi}$ is represented by a unitary operator on $\ca H$, whose self-adjoint generator is $\widehat H_{\wt\xi}$. The association $H_{\wt\xi} \mapsto \wh H_{\wt\xi}$ is, by construction, a homomorphism between $\ca A_C$ and $\ca A_Q$.}

As global phases are unphysical, the true space of quantum states is the {\sl ray space}, $\ca R := \ca H/U(1)$, corresponding to the quotient of the Hilbert space $\ca H$ by phases $e^{i\theta} \in U(1)$. Accordingly, symmetries must be represented by unitary operators up to a phase.
These are known as {\sl projective representations} of the group $\wt G$, more precisely defined as homomorphisms from $\wt G$ to the group of projective unitary operators on $\ca R$, $P\scr U(\ca H) := \{U \sim e^{i\theta}U; \text{ where } U \in \scr U(\ca H) \text{ and } \theta \in \bb R\}$.
It is a general theorem (reviewed in App.~\ref{app:projrep}) that projective irreducible unitary representations of $\wt G$ are in one-to-one correspondence irreducible self-adjoint representations of a central extension of the Lie algebra $\wt{\fr g}$ by 2-cocycles. As explained later, we will append to the quantization prescription a principle of Casimir matching, which implies that if $\wt{\fr g}$ is properly represented on the phase space as in \eqref{gAChomo}, without additional central charges, then the quantum theory should be constructed from irreducible representations of $\wt{\fr g}$, without further extensions.  Equivalently, the quantum theory should be constructed  from unitary irreducible representations of the universal cover of $\wt G$. In this manner, we can understand Isham's method as a prescription to generate an algebraically-closed complete set of classical observables, $H_{\wt\xi}$, associated to some symmetry structure of the phase space, which is then quantized according to the standard rule of canonical quantization, where the same algebra is irreducibly represented by 
corresponding self-adjoint operators $\wh H_{\wt\xi}$ on a Hilbert space.

We mentioned that the reason for requiring a {\sl transitive} action on the phase space is so that the set of generated charges is classically complete, i.e.,  that any function on $\wt{\ca P}$ can be (locally) expressed in terms of the $H_{\wt\xi}$'s. In fact, note that transitivity implies that, at any $p \in \wt{\ca P}$, 
any tangent vector $X \in T_p \wt{\ca P}$ is equal to $X_{\wt\xi}$ for some ${\wt\xi} \in \wt{\fr g}$. Then, for any $V \in T_p\wt{\ca P}$, $dH_{\wt\xi} (V) = -\omega(X_{\wt\xi}, V)$ will not vanish for at least one ${\wt\xi}$, since $\omega$ is non-degenerate. In words, there is no direction $V$ along which all charges are (locally) constant, 
confirming that any function on $\wt{\ca P}$ can be locally written in terms of the charges. A more geometrical way to see this involves the momentum map, $\mu : \wt{\ca P} \rightarrow \wt{\fr g}^*$, defined by $\mu(p)({\wt\xi}) := H_{{\wt\xi}}(p)$,
where $p \in \wt{\ca P}$ and ${\wt\xi} \in \wt{\fr g}$. The observation above implies that $\mu_*$ is injective\footnote{Since $\wt{\fr g}^*$ is a vector space, tangent vectors can be identified with elements of $\wt{\fr g}^*$ itself. In this way, $\mu_*: T_p\wt{\ca P} \rightarrow \wt{\fr g}^*$ is given by $\mu_*(V)({\wt\xi}) = dH_{\wt\xi}(V) = -\omega(X_{\wt\xi}, V)$, so $\mu_*(V)$ is zero if and only if $V=0$.} and therefore $\mu$ is an immersion. Consequently, any $p\in \wt{\ca P}$ has a neighborhood $\ca U \subset \wt{\ca P}$ such that $\mu|_{\ca U}$ is an embedding of $\ca U$ into $\wt{\fr g}^*$. 
Now, given any basis ${\wt\xi}_i$ for $\wt{\fr g}$, there is an associated coordinate system $\{\wt\a_i\}$ on $\wt{\fr g}^*$ defined by $\wt\a_i|_{\wt\a} := \wt\alpha({\wt\xi}_i)$, where $\wt\alpha \in \wt{\fr g}^*$. It follows that $\mu^*\wt\a_i = H_{{\wt\xi}_i}$. Given any smooth real function  $f:\ca U \rightarrow \bb R$, there exists a neighborhood $\ol{\ca U}$ of $\mu(\ca U)$ and a smooth real function $\ol f: \ol{\ca U} \rightarrow \bb R$, such that $f = \mu^* \ol f$. As $\ol f$ can be written in terms of the coordinates $\wt\a_i$, $f$ can be written in terms of the charges $H_{{\wt\xi}_i}$.\footnote{If $\ol f(\wt\alpha) = F(\wt\a_1, \wt\a_2, \ldots)$, for some $F: \bb R^{\text{\it dim}(\wt{\fr g})} \rightarrow \bb R$, then $f(p) = F(H_{{\wt\xi}_1}(p), H_{{\wt\xi}_2}(p), \ldots)$.}

The prescription, as formulated by Isham, is intended to provide a general framework for quantization and, as such, refers only to the minimal structure necessary for a ``kinematical quantization'', i.e., producing sensible homomorphism between a classically-complete subalgebra of observables to a corresponding quantum-complete subalgebra of self-adjoint operators.
The job is not finished, however, until a particular representation of the canonical observables is determined and other physically relevant observables (which are classically expressed in terms of the canonical ones) are also promoted to self-adjoint operators on the Hilbert space. In particular, the time-evolution Hamiltonian, describing the dynamics of the system, may often not be part of the canonical algebra, and need to be included in the quantum theory eventually. These additional observables that may need to be quantized often help selecting a preferred class of irreducible representations of the canonical algebra, therefore partially reducing the ambiguities in the quantization.

Another rule that we employ for filtering the representations is the {\sl Casimir matching principle}. 
Casimir operators are elements in the center of a universal enveloping algebra.\footnote{A universal enveloping algebra is a formal extension of a Lie algebra consisting of polynomials of the algebra elements, where the product is defined by treating the elements as operators in an abstract representation. For example, say that $\alpha$, $\beta$, $\gamma$ and $\delta$ are elements of $\fr g$, then $\alpha \beta + \gamma$ belongs to the universal enveloping algebra of $\fr g$ and the product with $\delta$ is given by $[\alpha \beta + \gamma, \delta]:= \alpha [\beta, \delta] + [\alpha, \delta]\beta + [\gamma, \delta]$.}
At the classical level, the completeness of the algebra implies that Casimir observables are constant on the phase space. At the quantum level, Schur's lemma implies that in any irreducible (complex) representation Casimir operators are multiples of the identity. At both levels, Casimirs take the same value for all states. Thus, if the classical system is to arise from a classical limit of the quantum system,  the eigenvalue of the  quantum Casimir 
observable should match the value of the corresponding classical Casimir.
In the study of the quantization of a particle on a sphere, in the presence of a magnetic monopole, from the perspective of Isham's method, this principle plays a key role in the argument leading to Dirac's charge quantization condition, $eg = n\hbar/2$ \cite{e2021particle}.

\subsection{Phase spaces with a cotangent bundle structure}
\label{subsec:cotbund}

Now we discuss a general method, also described by Isham \cite{isham1984topological,isham1989canonical}, for constructing a transitive group of symplectomorphisms of phase spaces that are the cotangent bundle of homogeneous manifolds. Consider a phase space $\wt{\ca P} = T^*\ca Q$, with the canonical symplectic form $\omega = d\theta$ associated with the cotangent bundle structure, where the configuration space $\ca Q$ is homogeneous space for some Lie group $G$. Let us denote the left $G$-action on $\ca Q$ by $\delta : G \rightarrow \text{\sl Diff}(\ca Q)$, or simply by 
\be
gx := \delta_g(x)
\ee
where $x \in \ca Q$. There is a natural lift of this action to the cotangent bundle, $\wt\delta : G \rightarrow \text{\sl Diff}(\wt{\ca P})$, given by
\be
\wt\delta_g(p) := \delta_{g^{-1}}^*p
\ee
which maps the fiber over $x = \pi(p)$, where $\pi : \wt{\ca P} \rightarrow \ca Q$ is the bundle projection, to the fiber over $gx$. That is, the lifted action satisfies the identity $\pi \circ \wt\delta_g = \delta_g \circ \pi$. The symplectic potential $\theta$ is invariant under such a transformation, which can be seen as follows. If $V \in T_p\wt{\ca P}$, then $\wt\delta_g^*\theta(V) = \theta(\wt\delta_{g*}V)$, and from the definition of $\theta$ this is $(\wt\delta_g(p))(\pi_*\wt\delta_{g*}V)$; from the definition of the lifted action this is equal to $\delta_{g^{-1}}^*p(\pi_*\wt\delta_{g*}V) = p(\delta_{g^{-1}*}\pi_*\wt\delta_{g*}V)$, and using the identity above we get $p(\delta_{g^{-1}*}\delta_{g*}\pi_*V) = p(\pi_*V) = \theta(V)$, that is, $\wt\delta_g^*\theta = \theta$. Consequently, $\wt\delta_g^*d\theta = d\wt\delta_g^*\theta = d\theta$, i.e.,
\be
\wt\delta_g^*\omega = \omega
\ee
so the group $G$, associated with $\ca Q$, acts as symmetries of the phase space. 

The group $G$ alone cannot serve as the quantizing group though, since it does not act transitively on the phase space. Rather, it acts only ``laterally'' on the phase space, so we must enlarge the group by including also transformations along the fibers of the bundle. A natural choice for a ``vertical'' transformation is to consider {\sl momentum translations} by 1-forms, that is, given a 1-form field $\alpha$ on $\ca Q$, define
\be
\zeta_\alpha(p) := p - \alpha
\ee
where $\alpha$ on the right-hand side is evaluated at $\pi(p)$. We can compute how the symplectic potential changes under this transformation as follows. Given any $V \in T_p\wt{\ca P}$, we have 
\be\label{zetaaction}
\zeta_\alpha^*\theta(V) = \theta(\zeta_{\alpha *} V) = (\zeta_\a p)(\pi_*\zeta_{\alpha *} V) = (p - \alpha)(\pi_*V) = (\theta - \pi^*\alpha)(V)
\ee
where we have used that $\zeta_\alpha$ acts vertically and thus satisfies the identity $\pi \circ \zeta_\alpha = \pi$. The symplectic 2-form then transforms like
\be
\zeta_\alpha^*\omega = \omega  - \pi^*d\alpha
\ee
Therefore, in order to have a symmetry we must restrict to closed 1-form fields, $d\alpha = 0$. As we shall see, it is necessary to restrict further to exact 1-forms, $\alpha = df$, where $f \in C^\infty(\ca Q, \bb R)$, so that the associated charges are globally defined on the phase space. 

The space of all exact 1-form fields on $\ca Q$, which is infinite-dimensional, is unnecessarily large for our purposes, as we are interested in constructing a ``minimal'' transitive group of symmetries acting on the phase space. 
That is, given the principle discussed in Sec.~\ref{subsec:canons} that the classical and quantum notions of algebra completeness must agree, we should always focus on canonically quantizing an algebra that has no proper classically-complete subalgebras, so that this principle is vacuously satisfied. (Note the analogy with the case of a trivial phase space, where only the Heisenberg algebra is canonically quantized, as opposed to the whole algebra of classical observables.)
If $\ca Q$ is $n$-dimensional, we would hope that ``just about'' $n$ generators would already suffice to produce a transitive action (of course, we need at least $n$). There is a natural algorithm to select an appropriate subset of exact 1-forms that acts transitively along the fibers of $T^*\ca Q$ {\sl and} is compatible with the $G$-symmetries, in the sense that their $\zeta$-action combines with $G$ into an extended group. It begins by finding a linear representation of $G$ on a vector space $V$ such that at least one $G$-orbit $\ca O$ in $V$ is diffeomorphic to $\ca Q$. Then, any dual vector $\alpha \in V^*$ can be naturally seen as an 1-form field on $V$, and it can be restricted to $\ca O \sim \ca Q$ to define a 1-form field on $\ca Q$. This 1-form field is exact because the function $f_\alpha : V \rightarrow \bb R$ defined by
\be
f_\alpha(v) := \alpha(v)
\ee
satisfies $\alpha = df_\alpha$ (note the slight abuse of notation here, where the subscript $\alpha$ of $f$ is an element of $V^*$, while $\alpha$ in the left-hand side is the corresponding 1-form field on $V$). The group of momentum translations generated  by $\alpha \in V^*$ in this way acts transitively along the fibers of $T^*\ca Q$, i.e., given any $p$ and $p'$ on the same fiber, $\pi(p) = \pi(p')$, there is always $\alpha \in V^*$ such that $p' = p - \alpha = \zeta_\alpha(p)$. We shall refer to this realization of the configuration space as an orbit in a vector space as the {\sl embedding realization}. 

Combining the ``horizontal'' and ``vertical'' transformations discussed above, we get a transitive group of symmetries of the phase space with structure $\wt G = V^* \rtimes G$, acting on the phase space as
\be\label{wtGaction}
\Gamma_{(\alpha, g)} (p) := \delta_{g^{-1}}^*p - \alpha
\ee
The product rule on $\wt G$ is defined so that $\Gamma$ composes appropriately,
\be\label{wtGprod}
(\alpha, g)(\alpha', g') = (\alpha + \delta_{g^{-1}}^*\alpha', gg')
\ee
where $\alpha$ and $\alpha'$ are seen as 1-form fields on $\ca Q$. More abstractly, $\a \mapsto \delta_{g^{-1}}^*\alpha$ corresponds to the dual representation of $G$ on $V^*$, which can be denoted by $\alpha \mapsto g\alpha$, so
\be\label{wtGprod2}
(\alpha, g)(\alpha', g') = (\alpha + g\alpha', gg')
\ee
where $\alpha$ and $\alpha'$ are seen as elements of $V^*$. 
This group, or an appropriate central extension of it, can be taken as the quantizing group.

Let us now compute the classical charges associated with the group $\wt G = V^* \rtimes G$. First we consider the $G$ part of the group. If $\xi \in \fr g$ is an element of the algebra of $G$, let $\wt X_\xi$ denote the vector field induced by $\xi$ on $\wt{\ca P}$. 
The associated charge $P_\xi$, interpreted as a ``momentum variable'', is defined by
\be
dP_\xi = - \ii_{\wt X_\xi} \omega = - \ii_{\wt X_\xi} d\theta = d[\theta(\wt X_\xi)]
\ee
where we have used that $\theta$ is invariant under the $G$ action, so $\pounds_{\wt X_\xi}\theta = \ii_{\wt X_\xi}d\theta + d\ii_{\wt X_\xi}\theta = 0$. Up to an additive constant, we can therefore choose
\be\label{Pxigen}
P_\xi = p(X_\xi)
\ee
where $X_\xi := \pi_* \wt X_\xi$.\footnote{Note that $\wt X_\xi$ does project nicely under $\pi$, as can be seen from the fact that we can alternatively define $X_\xi$ as the vector field induced by $\xi$ on $\ca Q$ (via the action of $G$ on $\ca Q$). In fact, let $\wt\phi_p(g) = gp$ and $\phi_x(g) = gx$, where $p \in \wt{\ca P}$ and $x \in \ca Q$, then $\wt X_\xi := \wt\phi_{p*}\xi$ and let us define $X_\xi := \phi_{x*} \xi$; so because of the relation $\pi(gp) = g\pi(p)$, which translates into $\pi \circ \wt\phi_p = \phi_{\pi(p)}$, we have $X_\xi = \pi_* \wt X_\xi$.} 
Second, we consider the $V^*$ part of the group. Since $V^*$ is a vector space, we can naturally identify it with is Lie algebra. If $\alpha \in V^*$, seen as an element of the Lie algebra of $V^*$, let $Y_\a$ denote the vector field induced by $\a$ on $\wt{\ca P}$. The associated charges $Q_\a$, interpreted as a ``configuration variable'', is defined by
\be
dQ_\a = - \ii_{Y_\a}\omega = - \ii_{Y_\a}d\theta = - \pounds_{Y_\a}\theta = - \left. \frac{d}{dt}\zeta_{t\a}^*\theta \right|_{t=0} = \pi^*\a = \pi^*df_\a = d(f_\a \circ \pi)
\ee
where we have used $\ii_{Y_\a}\theta = 0$, the result in \eqref{zetaaction}, and the relation $\alpha = df_\alpha$. Thus, up to an additive constant, we can choose
\be\label{Qagen}
Q_\a = f_\a \circ \pi
\ee
Note that in the embedding realization $x \in \ca Q$ is seen a vector in $V$ specifying a point in the orbit $\ca O \sim \ca Q$, so this expression for the charge, evaluated at $p \in \wt{\ca P}$, reads simply $Q_\a = \a(x)$, where $x = \pi(p)$. 

Now let us discuss the general form of the algebra of those charges. Since $\wt G$ has a semi-direct product structure, its Lie algebra is a semi-direct sum $\wt{\fr g} = V^* \sdplus \fr g$. If the elements of $\wt{\fr g}$ are denoted by $(\alpha; \xi)$, the product rule derived from \eqref{wtGprod} is given by
\be\label{wtGalggen}
[(\a; \xi), (\a'; \xi')] = (\pounds_{X_{\xi'}}\a - \pounds_{X_{\xi}}\a'; [\xi, \xi'])
\ee
in which $[\xi, \xi']$ is just the product rule in $\fr g$.   As we know, the Poisson algebra of the charges $P_\xi$ and $Q_\a$ should form a representation up to central charges of the underlying symmetry algebra, so we can generically write
\begin{align}
\{P_\xi, P_{\xi'}\} &= P_{[\xi, \xi']} + z(\xi, \xi') \nonumber\\
\{Q_\a, P_{\xi}\} &= Q_{\pounds_{X_{\xi}}\a} + z(\a, \xi) \nonumber\\
\{Q_\a, Q_{\a'}\} &= z(\a, \a')
\end{align}
where $z(\xi, \xi')$, $z(\a, \xi)$ and $z(\a, \a')$ are the 2-cocycle elements (which are constants on the phase space).\footnote{Note that we are using a simplified notation here, in which $z(\xi, \xi') := z((0; \xi), (0;\xi'))$, $z(\a, \xi) := z((\a; 0), (0; \xi))$ and $z(\a, \a') := z((\a; 0), (\a'; 0))$. In fact, this is what we mean whenever we write an element of $\fr g$ or $V^*$ in a place supposed to feature an element of $\wt{\fr g}$.} To be explicit, let us write the charges computed above as 
\begin{align}
&P_\xi = p(X_\xi) + c_\xi \nonumber\\
&Q_\alpha = f_\alpha \circ \pi + c_\alpha
\end{align}
where $c_\xi$ and $c_\alpha$ are generic constants (on the phase space) depending linearly on $\xi$ and $\alpha$, respectively. We wish to derive a formula for the 2-cocycles $z$ in terms of these additive constants $c$. 

Let us start with the $Q$ charges. Their Poisson brackets are
\begin{equation}
\{Q_\alpha, Q_{\alpha'}\} = - \omega(Y_\alpha, Y_{\alpha'}) = 0
\end{equation}
At the same time, $[\alpha, \alpha'] = 0$, so $Q_{[\alpha, \alpha']} = 0$, which gives
\begin{equation}
z(\alpha, \alpha') = 0
\end{equation}
regardless of $c$.

Now we consider the $P$ charges. Their Poisson brackets are
\begin{equation}
\{P_\xi, P_{\xi'}\} = - \omega(\widetilde X_{\xi}, \widetilde X_{\xi'}) = - d\theta (\widetilde X_\xi, \widetilde X_{\xi'})
\end{equation}
which can be manipulated as 
\begin{equation}
d\theta (\widetilde X_{\xi}, \widetilde X_{\xi'}) = - \ii_{\widetilde X_{\xi'}} d\theta (\widetilde X_{\xi}) = - \left( \pounds_{\widetilde X_{\xi'}}\theta - d\,\ii_{\widetilde X_{\xi'}}\theta \right) \widetilde X_{\xi} =   d\left( \theta(\widetilde X_{\xi'})\right) \widetilde X_{\xi}= \widetilde X_{\xi} \left( \theta(\widetilde X_{\xi'})\right)
\end{equation}
where the last term is read as the vector $\widetilde X_{\xi}$ deriving the scalar $\theta(\widetilde X_{\xi'})$.
Thus
\begin{equation}
\widetilde X_{\xi} \left( \theta(\widetilde X_{\xi'})\right) = \pounds_{\widetilde X_{\xi}} \left( \theta (\widetilde X_{\xi'}) \right) = \theta \left( \pounds_{\widetilde X_{\xi}} \widetilde X_{\xi'} \right) = \theta \left( [\widetilde X_{\xi}, \widetilde X_{\xi'}] \right)
\end{equation}
so we get
\begin{equation}
d\theta (\widetilde X_{\xi}, \widetilde X_{\xi'}) = \theta \left( [\widetilde X_{\xi}, \widetilde X_{\xi'}] \right)
\end{equation}
In addition,
\begin{equation}
\theta \left( [\widetilde X_{\xi}, \widetilde X_{\xi'}] \right) = p \left( \pi_* [\widetilde X_{\xi}, \widetilde X_{\xi'}] \right) = p \left( [X_{\xi}, X_{\xi'}] \right) = - p\left( X_{[\xi,\xi']} \right) 
\end{equation}
Hence,
\begin{equation}
\{P_\xi, P_{\xi'}\} = p\left( X_{[\xi,\xi']} \right)
\end{equation}
which leads to the 2-cocycle
\begin{equation}
z(\xi, \xi') = - c_{[\xi, \xi']}
\end{equation}
We see that we can eliminate this 2-cocycle by choosing $c_\xi = 0$.

Next we consider the mixed $Q$ and $P$ brackets, which give
\begin{equation}\label{QPbrackets}
\{ Q_\alpha, P_\xi \} = - \omega(Y_\alpha, \widetilde X_\xi) = -d\theta (Y_\alpha, \widetilde X_\xi) = - \ii_{Y_\alpha} d\theta (\widetilde X_\xi) = \pi^* \alpha (\widetilde X_\xi) = \alpha(X_\xi)
\end{equation}
and, since this should be a function on the phase space, it should be more precisely written as $\alpha(X_\xi) \circ \pi$. Thus,
\begin{equation}
z(\alpha, \xi) = \alpha(X_\xi) \circ \pi -  f_{\pounds_{X_\xi} \alpha} \circ \pi - c_{\pounds_{X_\xi} \alpha}
\end{equation}
This cocycle may be non-trivial, i.e., it may not be removable by a choice of $c_{\alpha}$. 
In that case, we must centrally extend $\wt G$ by this 2-cocycle, so that the Lie algebra of the new group is isomorphic to the Poisson algebra of the new charges.

\section{The canonical group for the diamond}
\label{subsec:cangroup}

Now we apply Isham's scheme to the reduced phase space of the diamond, $\wt{\ca P} = T^*(\diff/\psl)$. 
Being a cotangent bundle over a homogeneous space, $\ca Q = \diff/\psl$, we may try to construct the quantizing group for this phase space according to the algorithm in the previous section. The natural choice for the group $G$ here is $\diff$, acting on the left of $\ca Q$ as
\be\label{leftactionpsi}
\psi'[\psi] := [\psi' \circ \psi]
\ee
where $[\psi] = q(\psi)$ is the quotient by the right $\psl$ action. 
The definition above makes sense because the quotient acts from the right, so if $\chi \in \psl$, $\psi'[\psi\chi] = [\psi'\psi\chi] = [\psi'\psi] = \psi'[\psi]$. The next step is to find a representation of $\diff$, on a vector space $V$, such that at least one $G$-orbit $\ca O$ in $V$ is isomorphic to $\diff/\psl$. However, we are not aware of a representation of $\diff$ which has this property. On the other hand, it is well known that the coadjoint representation of the Virasoro group does have an orbit isomorphic to $\diff/\psl$~\cite{witten1988coadjoint,lazutkin1975normal,segal1981unitary,alekseev1989path}. Moreover, it is just as natural to choose $G$ as the Virasoro group, $\vira$, given that it is a central extension of $\diff$ and we can simply define its action on $\ca Q$ such that the central element acts trivially.

\subsection{The Virasoro group}
\label{subsec:Vira}

Let us begin by recalling the definition of the Virasoro group and its Lie algebra.
In App.~\ref{app:Vira} we offer further details on the general constructions of the group and its adjoint and coadjoint representations.
For reference see, e.g., \cite{guieu2007algebre,khesin2009geometry,ovsienko2004projective,kosyak2018regular,oblak2017bms}.

The Virasoro group, $\vira$, is a one-dimensional central extension of $\diff$ which can be defined from its Lie algebra, the Virasoro algebra, $\avira = \wh{\adiff}$, a central extension of $\adiff$. An element $\wh\xi$ of the Virasoro algebra is characterized by an element $\xi$ of $\adiff$ (which is identified with a vector field $\xi(\theta)\partial_\t$ on $S^1$), together with a central element component, $x \wh c$, where $x \in \bb R$ and $\wh c$ is a vector in the center of $\avira$. Thus, we write
\be
\wh \xi = \xi(\theta)\partial_\t + x \wh c \,\in\, \avira
\ee
Sometimes we may also write $\wh\xi = \xi + x \wh c$, hoping that it will be clear from the context whether $\xi$ is referring to the vector field $\xi(\theta)\partial_\t$ or just the real function $\xi(\t)$. The product rule in $\avira$ is defined by
\be\label{viraprod}
[\xi\partial_\t + x \wh c, \eta \partial_\t + y \wh c] = \left( \eta \xi' - \xi \eta' \right) \partial_\t + \wh c \int\!d\t \left( \eta \xi''' - \xi \eta''' \right)
\ee
where the prime denotes derivative with respect to $\theta$. Note that the $\partial_\t$ component in the right-hand side is {\sl equal} to the product of $\xi\partial_\t$ and $\eta\partial_\t$ in the algebra $\adiff$, which is
{\sl minus} the vector field brackets of these fields on $S^1$ (see the ``$\adiff$'' entry in App.~\ref{app:GSandC}). Also, note that $\wh c$ spans the center of the algebra. 

We shall take the topology of $\vira$ to be $\diff \times \bb R$, so that it can be characterized by pairs 
\be
\wh\psi = (\psi, r) \,\in\, \vira
\ee
where $\psi \in \diff$ and $r \in \bb R$.
Accordingly, the group is not simply-connected because $\diff$ has fundamental group $\bb Z$ (see App.~\ref{app:topologyQ}).
Later, when considering quantizations based on projective representations, we will discuss its universal cover, denoted by $\un\vira$.

The general product rule reads
\be
(\psi, a)(\phi, b) = (\psi\phi, a + b + W(\psi, \phi))
\ee
where $W : \diff \times \diff \rightarrow \bb R$ is the {\sl Bott 2-cocycle}, given by $W(\psi, \phi) = \int_{S^1} \log ( \psi \circ \phi)' d \log \phi'$, in which $\psi$ (and $\phi$) are identified with real functions on $\bb R$ satisfying $\psi(x+2\pi) = \psi(x) + 2\pi$ and the prime denotes derivative with respect to $x$ \cite{kosyak2018regular}.

\subsubsection{Adjoint orbits}
\label{subsubsec:ViraAd}

As always, the adjoint representation of a Lie algebra on itself is given by the algebra product, 
\be\label{adxieta0}
\ad_{\xi\partial_\t + x \wh c} ( \eta \partial_\t + y \wh c ) = \left( \eta \xi' - \xi \eta' \right) \partial_\t + \wh c \int\!d\t \left( \eta \xi''' - \xi \eta''' \right)
\ee
By ``exponentiation'' we obtain the representation of the Lie group on its Lie algebra (see the ``$\Ad_g$, $\ad_g$ and $\ad_\xi$'' entries in App.~\ref{app:GSandC}),
\be\label{advira}
\ad_{\wh\psi} \wh\eta = \psi_*(\eta\partial_\t) + \wh c \left( y + 2 \int\!d\t\, S[\psi](\t) \, \eta(\theta) \right)
\ee
where $\wh\psi = (\psi, r)$, $\wh\eta = \eta\partial_\t + y \wh c$ and $S$, called the {\sl  Schwarzian derivative} \cite{ovsienko2004projective}, is a map from $\diff$ into $C^\infty(S^1, \bb R)$ defined as
\be
S[\psi](\t) := \frac{\psi'''(\t)}{\psi'(\t)} - \frac{3}{2} \left( \frac{\psi''(\t)}{\psi'(\t)} \right)^2
\ee
Note that the adjoint action does not depend on the central component of $\wh\psi = (\psi, r)$. Accordingly, we can simplify the notation and just write $\ad_\psi$ instead of $\ad_{\wh\psi}$, while keeping in mind that this refers to the adjoint representation of $\vira$, not $\diff$.

Let us show that there are no orbits of the adjoint action which are isomorphic to $\diff/\psl$. To do so, define the {\sl little algebra} of $\wh\eta \in \avira$, with respect to the adjoint action, as
\be\label{littlealgad}
\fr h_\ad(\wh\eta) := \{ \wh\xi \in \avira \,,\,\, \ad_{\wh\xi}\wh\eta = 0\}
\ee
The {\sl little group} of the orbit of $\wh\eta$ under the adjoint action is the exponentiation of the little algebra $\fr h_\ad(\wh\eta)$. Since $\wh c$ is adjointly represented as the zero operator, it always belongs to the little algebra. If there exits some orbit isomorphic to $\diff/\psl$ then there must exist some little group isomorphic to $\psl \times \bb R$, where $\bb R$ refers to the center of $\vira$, and so there must exist some little algebra isomorphic to $\apsl \oplus \wh c$. As we shall see, however, there is no such little algebra for the adjoint action. Note that a necessary condition for $\wh\xi \in \fr h_\ad(\wh\eta)$ is
\be
\eta \xi' - \xi \eta' = 0
\ee
so in every open interval of $S^1$ in which  $\eta \ne 0$ we have $\xi = k \eta$ for some constant $k$. This condition is therefore also sufficient, since it implies $\eta \xi''' - \xi \eta''' = 0$. There are three cases to consider,

\noindent $\quad (i)$ If $\eta$ vanishes only at isolated points, and at least one of its derivatives do not vanish at each of those points, then by requiring that $\xi$ is smooth we get that $\xi = k \eta$ on the whole $S^1$. In this case the little algebra is two-dimensional, $\fr h_\ad(\wh\eta) = \eta \oplus \wh c$;

\noindent $\quad (ii)$ If $\eta$ vanishes only at isolated points, and all of its derivatives also vanish at $n$ of those points (called ``flat points''), then the smoothness of $\xi$ is not sufficient to relate $k$'s at different sections $R_i$ of $S^1$, where each $R_i$ is the closed interval between two consecutive flat points. In this case, $\xi$ can be piecewise-defined as $\xi(\t) = k_i \eta(\t)$, for $\t \in R_i$, and so the little algebra is $(n+1)$-dimensional, $\fr h_\ad(\wh\eta) = (\eta \oplus \eta \cdots \oplus \eta) \oplus \wh c$;

\noindent $\quad (iii)$ If $\eta$ vanishes in an open set of $S^1$, then $\xi$ is not restricted on that interval and the little algebra is infinite-dimensional;

\noindent Note that cases $(i)$ and $(iii)$ do not have the right dimension to be isomorphic to $\apsl \oplus \wt c$, but case $(ii)$ can have 4 dimensions if $n=3$. However, it does not have the right algebraic structure, since it is commutative.

\subsubsection{Coadjoint orbits}
\label{subsubsec:ViraCoad}

Now we consider the coadjoint representation of $\vira$ on its dual Lie algebra, $\dvira$, and the corresponding coadjoint representation of $\avira$ on $\avira^*$. Similar to our manner of characterizing elements of $\ddiff$, we characterize elements of $\avira^*$ as $\wt\a = \a + \a_0 \wt c$ where $\a := \a(\t) d\t^2$ is a quadratic form on $S^1$ and $\wt c$ is dual to $\wh c$ in the sense that $\wt c(\eta + y \wh c) = y$. The pairing of $\wt\a \in \avira^*$ with $\wh\eta \in \avira$ is defined as
\be
\wt\a (\wh\eta) := \int_{S^1} \alpha(\eta) + \alpha_0 y = \int\!d\t\, \alpha(\t) \eta(\t) + \alpha_0 y
\ee	
From the definition of the coadjoint action (see ``$\coad_g$'' entry in App.~\ref{app:GSandC}) we have
\be
\coad_{(\psi, r)} \wt\alpha \left(\wh\eta\right) = \wt\alpha \left(\ad_{(\psi, r)^{-1}} \wh \eta \right) =  \wt\alpha \left(\ad_{(\psi^{-1}, -r - W(\psi, \psi^{-1}))} \wh \eta \right)
\ee
which implies
\be
\coad_{(\psi, r)} \wt\alpha =  \psi_*(\alpha d\t^2)  + 2\a_0 S[\psi^{-1}] d\t^2 + \alpha_0 \wt c 
\ee
Note that $r$ also acts trivially through the coadjoint map, so we can use the shortened notation $\coad_\psi$. For later reference, we can evaluate the push-forward explicitly to get
\be\label{coadpsiexpl}
\coad_\psi (\alpha d\t^2+ \a_0 \wt c) \Big|_\t = \frac{\a(\psi^{-1}(\t)) - 2\a_0 S[\psi](\psi^{-1}(\t))}{\psi'(\psi^{-1}(\t))^2} d\t^2 + \alpha_0 \wt c 
\ee
where we have used \eqref{Schprop} to relate $S[\psi^{-1}]$ with $S[\psi]$. 

The coadjoint representation of the algebra is given by
\be\label{coadxialpha}
\coad_{\wh\xi} \wt\alpha  = - \left( 2 \a \xi' + \xi \a' + 2 \a_0 \xi''' \right) d\t^2
\ee

Now we show that there is a coadjoint orbit of $\vira$ which is isomorphic to $\diff/\psl$. Define the little algebra of $\wt\alpha \in \avira^*$, with respect to the coadjoint action, by
\be
\fr h_\coad(\wt\alpha) := \{ \wh\xi \in \avira \,,\,\, \coad_{\wh\xi}\wt\a = 0\}
\ee
Note that the central element, $\wh c$, is always in $h_\coad$. 
Let us particularize to the little algebra of $\wt\a_0 := \a_0 d\t^2+ a \wt c$, where $\a_0 \in \bb R$ is a constant. The equation for $\wh\xi \in \fr h_\coad(\a_0 d\t^2+ a \wt c)$ reads
\be
\a_0 \xi' + a \xi''' = 0
\ee
This differential equation (of the third order) has a general solution of the form
\be
\xi = h_0 + h_1 \cos w\t + h_2 \sin w\t
\ee
where $h_0$, $h_1$ and $h_2$ are integration constants and $w = \sqrt{\a_0/a}$. Note that if $w \in \bb Z$ then all terms are allowed, but otherwise only the $h_0$ term is allowed (since $\xi$ must be $2\pi$-periodic so as to be smooth on $S^1$). In particular, if $\a_0 = a$, so that $w=1$, then the little algebra is spanned by $\partial_\t$, $\cos\t\, \partial_\t$, $\sin\t\, \partial_\t$ and $\wh c$, which is precisely $\apsl \oplus \wh c$. Hence, the coadjoint orbit of
\be\label{alphaorbit}
\wt\varepsilon = d\t^2+ \wt c
\ee
is isomorphic to $\diff/\psl$. Rigorously speaking, this statement is speculative since we have only considered the little algebra, and thus we would like to confirm that the {\sl little group} of $\wt\varepsilon$ is indeed $\psl \times \bb R$. The little group of $\wt\a \in \avira^*$, with respect to the coadjoint action, is defined as
\be
H_\coad(\wt\a) := \{ (\psi, r) \in \vira \,,\,\, \coad_{(\psi , r)}\wt\a = \wt\a \}
\ee
Naturally the center of $\vira$, $(I, r)$, is always in the little group. Given formula \eqref{coadpsiexpl} for the coadjoint action, particularized to $\wt\varepsilon = d\t^2+ \wt c$, we have 
\be\label{betacoad}
\coad_\psi \wt\varepsilon \Big|_\t = \frac{1 - 2 S[\psi](\psi^{-1}(\t))}{\psi'(\psi^{-1}(\t))^2} d\t^2 + \wt c 
\ee
so the condition that $\psi$ is in $H_\coad(\wt\varepsilon)$ translates into
\be\label{psilittleHcoad}
S[\psi](\t) = \frac{1- \psi'(\t)^2}{2}
\ee
We now wish to show that $\psl \subset H_\coad(\wt\varepsilon)$. Recall that $\psl \subset \diff$ is defined as the boundary action of the group of holomorphic automorphisms of the complex unit disc $\bb D \subset \bb C$. That is, if $\psi \in \psl \subset \diff$, then there exists a holomorphic function $f : \bb D \rightarrow \bb D$ such that $f(e^{i\t}) = e^{i\psi(\t)}$. In fact, $f$ is a Mobius transformation with the general form
\be
f(z) = e^{ib} \frac{z - a}{1 - \bar a z}
\ee
where $b \in \bb R$ and $a \in \bb D$ (see App. ``The uniformization map'' in Part I). From the chain rule we have
\be
\psi' = \frac{z}{f} \frac{df}{dz}
\ee
in which $z = e^{i\t}$ is at the boundary of the disc, $\partial \bb D \sim S^1$. From this, we can show that
\be\label{Schderdisc}
S[\psi] := \frac{\psi'''}{\psi'} - \frac{3}{2} \left( \frac{\psi''}{\psi'} \right)^2 = \frac{1- \psi'^2}{2} - z^2 \text{\LARGE $\mathfrak{s}$}[f]
\ee
where $\text{\LARGE $\mathfrak{s}$}[f]$ is the Schwarzian derivative for holomorphic functions, defined analogously to \eqref{LambdarelSch} by
\be
\text{\LARGE $\mathfrak{s}$}[f](z) := \frac{d^3\!f/dz^3}{df/dz} - \frac{3}{2} \left( \frac{d^2\!f/dz^2}{df/dz} \right)^2
\ee
This derivative has the property that $\text{\LARGE $\mathfrak{s}$}[f] = 0$ if and only if $f$ is a Mobius transformation. Therefore $\text{\LARGE $\mathfrak{s}$}[f]$ vanishes for the $\psl$ transformations, and relation \eqref{Schderdisc} implies that $\psi$ satisfies \eqref{psilittleHcoad}, establishing that $\psl$ is indeed in the little group of $\wt\varepsilon = d\t^2+ \wt c$.\footnote{Note that only inclusion has been proven, $\psl \times \bb R \subset H_\coad(\wt\varepsilon)$, not equality. The fact that the little algebra is $\apsl \oplus \wh c$ implies that $H_\coad(\wt\varepsilon)$ cannot be a higher-dimensional extension of $\psl \times \bb R$. While in principle it could be a discrete extension, $\psl$ cannot be discretely extended within $\diff$.}

\subsection{The canonical group for the reduced phase space}
\label{subsubsec:CanonVira}

In view of the result above, and Isham's algorithm discussed in Sec.~\ref{subsec:cotbund}, it is possible to construct the quantum theory based on the group
\be\label{quantizinggroup}
\wt G = (\avira^*)^* \rtimes \vira
\ee
in which the Virasoro group acts as the generator of ``configuration translations'' on $\diff/\psl$ and the abelian group $(\avira^*)^* \sim \avira$ (whose product rule is given by vector addition, $\wh\eta  \wh\eta' := \wh\eta + \wh\eta'$) acts as the generator of ``momentum translations'' along the fibers of $T^*(\diff/\psl)$. 

We now describe explicitly how this group acts on the phase space. A point $[\psi]$ in the configuration space can be conveniently characterized by giving {\sl a} Virasoro element that maps $d\t^2+ \wt c$ into $[\psi]$. Since the central elements $(I, r)$ act trivially, we need only to use elements of $\diff$ in this characterization. In other words, the coadjoint representation of the Virasoro group provides a explicit projection map $\psi \mapsto [\psi]$, from $\diff$ to $\diff/\psl$, given by
\be\label{qcoadrealization}
[\psi] := \coad_\psi (d\t^2+ \wt c) = \frac{1 - 2 S[\psi](\psi^{-1}(\t))}{\psi'(\psi^{-1}(\t))^2} d\t^2 + \wt c
\ee
The coadjoint action on $[\psi]$ is, as expected, just the right-action by $\diff$ proposed in \eqref{leftactionpsi}, 
\be
\phi[\psi] := \coad_{\phi} [\psi] = \coad_{\phi}\coad_\psi (d\t^2+ \wt c) = \coad_{\phi\psi} (d\t^2+ \wt c) = [\phi\psi]
\ee
so that we have $\delta_{(\phi, r)} = \coad_\phi$. 
Elements of the group will be denoted by
\be
(\wh\eta; \wh\phi) \in \wt G
\ee
where $\wh\eta \in \avira$ given the natural identification $(\avira^*)^* \sim \avira$.
The action of $\wt G$ on the phase space is, according to \eqref{wtGaction}, 
\be\label{Gammaactiondef0}
\Gamma_{(\wh\eta; \wh\phi)} (p) = \delta^*_{\wh\phi^{-1}}p - \wh\eta = \coad^*_{\wh\phi^{-1}}p - \wh\eta
\ee
in which,  $p$, a 1-form at $[\psi] \in \ca Q \subset \avira^*$, can be seen as the restriction of an element of $(\avira^*)^* \sim \avira$, naturally identified with a 1-form field on $\avira^*$, to $\ca Q$. In this view, $\coad^*_{\wh\phi^{-1}}$ is simply the dual representation of $\coad$ on $(\avira^*)^*$, which is naturally identified with the adjoint representation on $\avira$, so that
\be\label{Gammaactiondef}
\Gamma_{(\wh\eta, \wh\phi)} (p) = \ad_\phi p - \wh\eta
\ee
The group product rule then reads,
\be\label{cangroupprodrule}
(\wh\xi; \wh\psi)(\wh\xi', \wh\psi') = (\wh\xi + \ad_{\psi}\wh\xi', \wh\psi\wh\psi')
\ee
so, evidently, the semi-direct product is defined with respect to the adjoint action of $\vira$ on $\avira$.

The algebra of the group, $\wt{\fr g} = \avira^c \sdplus \avira$, is given, as in \eqref{wtGalggen}, by
\be
[(\wh \eta; \wh\xi), (\wh \eta'; \wh\xi')] = (\pounds_{X_{\wh\xi'}}\wh\eta - \pounds_{X_{\wh\xi}}\wh\eta'; [\wh\xi, \wh\xi'])
\ee
where $[\wh\xi, \wh\xi']$ is the product in $\avira$. Note that the abelian factor, associated with the momentum translations, is denoted as $\avira^c$ to emphasize that, while this factor is isomorphic to $\avira$ as a vector space, it is not the usual $\avira$ algebra but rather a commutative version of it, as $[(\wh \eta; 0), (\wh \eta'; 0)] = 0$.
It is straightforward to show that 
\be
\pounds_{X_{\wh\xi}}\wh\eta = - \ad_{\wh\xi} \wh\eta = [\wh\eta, \wh\xi ]
\ee
so the product rule on $\wt{\fr g}$ reads
\be\label{wtGprodrule}
[(\wh \eta; \wh\xi), (\wh \eta'; \wh\xi')] = ([\wh\eta, \wh\xi'] - [\wh\eta',\wh\xi]; [\wh\xi, \wh\xi'])
\ee
where the commutators appearing on the right-hand side refer to the product on $\avira$. Note that the elements $(\wh\eta; 0)$, generating momentum translations, form an abelian (normal) subalgebra, and the elements $(0; \wh\xi)$, generating configuration translations, form a Virasoro subalgebra. 

It is convenient to express this algebra in the usual {\sl harmonic basis}, where the functions $\xi(\t)$ and $\eta(\t)$ are expanded in their Fourier modes. That is, we define the basis as
\begin{align}
&L_n = (0; e^{in\t} \partial_\t) \,,\quad R = (0; \wh c) \nonumber\\
&K_n = (e^{in\t} \partial_\t; 0) \,,\quad T = (\wh c; 0) \label{harmonicbasis}
\end{align}
where $n \in \bb Z$. In this basis, the algebra product reads
\begin{align}
&[L_n, L_m] = i (n - m) L_{n+m} - 4\pi i n^3 \delta_{n+m, 0} R \nonumber\\
&[K_n, L_m] = i (n - m) K_{n+m} - 4\pi i n^3 \delta_{n+m, 0} T \nonumber\\
&[K_n, K_m] = 0 \nonumber\\
&[R, \#] = 0 \nonumber\\
&[T, \#] = 0
\end{align}
where $\delta$ is the Kronecker delta and ``$\#$'' denotes ``any element'', so $R$ and $T$ form the center of the algebra.

\subsection{Lifted group action on pre-phase spaces}
\label{subsubsec:Liftaction}

We have described how the canonical group $\wt G = (\avira^*)^* \rtimes \vira$ acts on the reduced phase space $\wt{\ca P} = T^*(\diff/\psl)$, but it is also useful to understand how this group action can be lifted to an action on the partially-reduced phase space $\wh{\ca S} = \diff \times \whddiff$ and, ultimately, on the original (constrained) phase space $\ca P$ coordinatized by ADM variables. (See Part I \cite{e2023quantization} for the necessary definitions.)

The action on $\wt{\ca P}$ by the group element $(\wh\eta; \wh\phi)$, denoted by $\Gamma_{(\wh\eta, \wh\phi)}$ and defined in \eqref{Gammaactiondef}, can be ``pulled-back'' or lifted to an action on $\wh{\ca S}$, denoted by $J^*\Gamma_{(\wh\eta; \wh\phi)}$ and defined in such a way that the following diagram is commutative
\begin{center}
\begin{tikzcd}
\wh{\ca S} \arrow{d}[swap]{J} \arrow{r}{J^*\Gamma} & \wh{\ca S} \arrow{d}{J} \\
\wt{\ca P}  \arrow{r}{\Gamma} & \wt{\ca P}
\end{tikzcd}
\end{center}
that is, $J \circ (J^*\Gamma) = \Gamma \circ J$, where the subscript label $(\wh\eta; \wh\phi)$ of $\Gamma$ and $J^*\Gamma$ has been omitted. Of course, being a group action, the map $(\wh\eta; \wh\phi) \mapsto J^*\Gamma_{(\wh\eta; \wh\phi)}$ is required to be a homomorphism from $\wt G$ to the group of diffeomorphisms on $\wh{\ca S}$. 

In the definition of $\Gamma$, \eqref{Gammaactiondef0}, let $p = J(\psi, \ac\sigma)$ so
\be
\Gamma_{(\wh\eta, \wh\phi)}  J(\psi, \ac\sigma) = \coad^*_{\phi^{-1}}J(\psi, \ac\sigma) - \wh\eta
\ee
Note that it maps the 1-form $J(\psi, \ac\sigma)$ at the point $[\psi] = q(\psi)$ to a 1-form at $\phi[\psi] = \coad_\phi q(\psi)$, which is clear from the pull-back coadjoint map (also recall that $\wh\eta$, seen as an element of $(\dvira)^*$, defines a 1-form at $\phi[\psi]$ by restricting it to the configuration space $\ca Q$, seen as the coadjoint orbit of $\wt\varepsilon = d\theta^2 + \wt c$). Now let $\wt X$ be a vector tangent to $\ca Q$ at $[\phi\psi]$ defined as
\be
\wt X := q_* l_{\phi\psi*} \xi
\ee
where $\xi \in \adiff$. The whole tangent space at $[\phi\psi]$ is spanned by these vectors. Applying the 1-form $\Gamma_{(\wh\eta, \wh\phi)}  J(\psi, \ac\sigma)$ to $\wt X$ gives
\be
\Gamma_{(\wh\eta, \wh\phi)}  J(\psi, \ac\sigma) \wt X = J(\psi, \ac\sigma) (\coad_{\phi^{-1}*}q_* l_{\phi\psi*} \xi) - \wh\eta (\wt X)
\ee
Let us focus on the first term on the right-hand side. 
An identity that will be useful is $q \circ l_\phi = \coad_\phi \circ q$ which can be shown as follows
\be
q \circ l_\phi (\psi) = q(\phi\psi) = \coad_{\phi\psi} \wt\varepsilon = \coad_\phi \coad_\psi \wt\varepsilon = \coad_\phi \circ q (\psi)
\ee
where we have used formula \eqref{qcoadrealization} for $q: \diff \rightarrow \diff/\psl \subset \dvira$.
The derivative of this identity allows us to get
\be
J(\psi, \ac\sigma) (\coad_{\phi^{-1}*}q_* l_{\phi\psi*} \xi) = J(\psi, \ac\sigma) (q_* l_{\phi^{-1}*}l_{\phi\psi*} \xi) = J(\psi, \ac\sigma) (q_* l_{\psi*} \xi) = \ac\sigma(\xi)
\ee
where the definition of $J$ has been used in the last step. Now let us consider the second term, $\wh\eta(\wt X)$. Define the notation $\coad\,\wt\varepsilon (\psi) := \coad_\psi \wt\varepsilon$,
so that $q = \coad\,\wt\varepsilon$. Note that, at the identity $\psi = I$, $q_* = (\coad\,\wt\varepsilon)_*$ is precisely equal to the coadjoint action of $\avira$ on $\dvira$, i.e., $(\coad\,\wt\varepsilon)_* \xi = \coad_\xi \wt\varepsilon$. Then
\be
\wh\eta(\wt X) = \wh\eta (q_* l_{\phi\psi*} \xi) = \wh\eta( \coad_{\phi\psi*} q_* \xi) = \coad^*_{\phi\psi} \wh\eta (\coad_\xi\wt\varepsilon)
\ee
As $\coad^*_{\phi\psi}$ is a linear map from $\avira$ into itself, the same argument used to go from \eqref{Gammaactiondef0} to \eqref{Gammaactiondef} applied here to replace $\coad^*_{\phi\psi}$ by $\ad_{(\phi\psi)^{-1}}$, yielding
\be
\wh\eta(\wt X) = \ad_{(\phi\psi)^{-1}} \wh\eta (\coad_\xi\wt\varepsilon)
\ee
We then define the map $\Upsilon : \diff \times \avira \rightarrow \ddiff$ by
\be\label{UpsilonDefabs}
\Upsilon_\psi \wh\eta (\xi) := \ad_{\psi^{-1}} \wh\eta (\coad_\xi\wt\varepsilon)
\ee
As the notation suggests, it is convenient to think of $\Upsilon$ as $\diff$-labeled linear map from $\avira$ to $\ddiff$. 
We then have,
\be\label{GammaJX}
\Gamma_{(\wh\eta; \wh\phi)}  J(\psi, \ac\sigma) \wt X = (\ac\sigma - \Upsilon_{\phi\psi}\wh\eta)(\xi)
\ee
We need to infer what $J^*\Gamma$ has to be in order for $J \circ J^*\Gamma_{(\wh\eta; \wh\phi)}(\psi, \ac\sigma) \wt X$ to reproduce the same result. We know that $J^*\Gamma_{(\wh\eta; \wh\phi)}(\psi, \ac\sigma)$ must project to a 1-form at $[\phi\psi]$ under $J$, which suggests the obvious guess
\be
J^*\Gamma_{(\wh\eta; \wh\phi)}(\psi, \ac\sigma) = (\phi\psi, \ac\sigma - \Upsilon_{\phi\psi}\wh\eta)
\ee
This can be easily confirmed by projecting it under $J$ and acting on $\wt X$, $J(\phi\psi, \ac\sigma - \Upsilon_{\phi\psi}\wh\eta)(\wt X) = (\ac\sigma - \Upsilon_{\phi\psi}\wh\eta)(\Xi(l_{\phi\psi*}\xi)) = (\ac\sigma - \Upsilon_{\phi\psi}\wh\eta)(\xi)$, matching with \eqref{GammaJX}.
Note that the image of $\Upsilon$ is actually contained in the $\whddiff$ subspace of $\ddiff$, since every $\xi \in \apsl$ is in the little group of $\wt\varepsilon$, so $\coad_\xi\wt\varepsilon = 0$. Thus, for every $\psi \in \diff$, $\Upsilon_\psi : \avira \rightarrow \whddiff$, and it makes sense to have $\Upsilon_{\phi\psi}\wh\eta$ subtracted from $\ac\sigma$ as it gives another element of $\whddiff$.

It is convenient to have a more explicit expression for $\Upsilon_\psi \wh\eta$, since its definition \eqref{UpsilonDefabs} is quite abstract. Using expression \eqref{advira} for $\ad_{\psi^{-1}} \wh\eta$ and \eqref{coadxialpha} for $\coad_\xi \wt\varepsilon$ we get
\be
\Upsilon_\psi \wh\eta (\xi) = - \int\!d\theta \frac{\eta(\psi(\theta))}{\psi'(\theta)} \left( 2 \xi'(\theta) + 2 \xi'''(\theta) \right) = 2\int\!d\theta \left[ {\left(\frac{\eta(\psi(\theta))}{\psi'(\theta)}\right)\!}' + {\left(\frac{\eta(\psi(\theta))}{\psi'(\theta)}\right)\!}''' \right] \xi(\theta)
\ee
where integration by parts was used in the last step. We have also used that $\psi^{-1}_*(\eta \partial_\theta)|_\theta = \frac{\eta(\psi(\theta))}{\psi'(\theta)} \partial_\theta$. Note that the central component of $\ad_{\psi^{-1}} \wh\eta$ does not contribute since $\coad_\xi \wt\varepsilon$ has no central component. Therefore we can read off
\be
\Upsilon_\psi \wh\eta = 2 \left[ {\left(\frac{\eta(\psi(\theta))}{\psi'(\theta)}\right)\!}' + {\left(\frac{\eta(\psi(\theta))}{\psi'(\theta)}\right)\!}''' \right] d\theta^2
\ee
which is manifestly belongs to $\whddiff$.

Let us briefly comment on how the group action lifts to the original (constrained) phase space, described in terms of metrics and extrinsic curvatures. The state $(\psi, \ac\sigma)$ corresponds to the metric $h_{ab} = \Psi_* e^\lambda \bar h_{ab}$ and the (traceless part of the) extrinsic curvature $\sigma^{ab} = \Psi_* e^{-2\lambda} \bar\sigma^{ab}$, where $\Psi$ is any extension of $\psi$ to the disc and $\lambda$ satisfies the associated Lichnerowicz equation. Under a transformation $(\psi, \ac\sigma) \mapsto (\psi', \ac\sigma')$ there will be corresponding transformations of $(i)$ $\Psi \mapsto \Psi'$, $(ii)$ $\bar\sigma^{ab} \mapsto {\bar\sigma'}^{ab}$ and also of the Weyl scalar $(iii)$ $\lambda \mapsto \lambda'$. Let us discuss each one of these changes under $(\psi, \ac\sigma) \mapsto J^*\Gamma_{(\wh\eta, \wh\phi)}(\psi, \ac\sigma) = (\phi\psi, \ac\sigma - \Upsilon_{\phi\psi}\wh\eta) =: (\psi', \ac\sigma')$: 

\vskip 0.7em
\noindent $(i)\,\,\,$ If $\Phi$ is any extension of $\phi$ to the disc, then $\Phi\Psi$ is an extension of $\phi\psi$ to the disc. We can therefore take $\Psi' = \Phi\Psi$.

\vskip 0.7em
\noindent $(ii)\,\,\,$ Recall that $\ac\sigma$ and $\bar\sigma^{ab}$ are related via
\be\label{sigmahatsigmabarrelation}
\ac\sigma(\wh\xi) = - \int  \vartheta_{\bar h} \bar\sigma^{ab} \pounds_\xi \bar h_{ab}
\ee
Since $\Upsilon_{\phi\psi}\wh\eta \in \whddiff$, we can use this same expression to define a symmetric traceless and divergenceless tensor $\bar\Upsilon^{ab}_{\phi\psi}\wh\eta$, that is, $\Upsilon_{\phi\psi}\wh\eta(\wh\xi) = -\int\! \vartheta_{\bar h} \bar\Upsilon^{ab}_{\phi\psi}\wh\eta \pounds_\xi \bar h_{ab}$. Then $\bar\sigma'^{ab} = \bar\sigma^{ab} - \bar\Upsilon^{ab}_{\phi\psi}\wh\eta$.

\vskip 0.7em
\noindent $(iii)\,\,\,$ The Weyl scalar $\lambda$ changes for two reasons. First the ``source term'' $\bar\sigma^{ab}\bar\sigma_{ab}$ in the Lichnerowicz equation changes, and second the boundary values $\lambda_\partial$ for that equation also changes. Note that $\lambda_\partial$ satisfies $e^{\lambda|_\partial} d\theta^2 = \psi^{-1}_* \gamma$, so as $\psi \mapsto \phi\psi$ we have $e^{\lambda'_\partial} d\theta^2 = (\phi\psi)^{-1}_* \gamma$. While there is no explicit formula for $\lambda'$, as it is a solution of this modified Lichnerowicz equation with modified boundary conditions, we know that the map $\lambda \mapsto \lambda'$ is well-defined due to the existence and uniqueness properties of solutions of that equation.
\vskip 0.7em

A very simple example to discuss is the $SO(2)$ rotation, i.e., the action of the group element $(\wh\eta, \wh\phi) = (0, r_\varphi)$ where $r_\varphi(\theta) = \theta + \varphi$ (modulo $2\pi$). In that case the pre-phase space point changes as $(\psi, \ac\sigma) \mapsto (r_\vartheta \psi, \ac\sigma)$. Note that $\bar\sigma^{ab}$ does not change. Also, $\lambda_\partial$ does not change (using the convention that $\gamma \propto d\theta^2$) since $e^{\lambda'_\partial} d\theta^2 = \psi^{-1}_* (R_\varphi)^{-1}_* \gamma = \psi^{-1}_* \gamma = e^{\lambda_\partial} d\theta^2$. Consequently the Weyl scalar $\lambda$ does not change. Therefore, if $R_\varphi$ is any extension of $r_\varphi$ to the disc, $h'_{ab} = (R_\varphi\Psi)_*e^\lambda \bar h_{ab} = (R_\varphi)_*h_{ab}$ and $\sigma'^{ab} = (R_\varphi\Psi)_*e^{-2\lambda} \bar\sigma h_{ab} = (R_\varphi)_*\sigma^{ab}$, showing that $(\wh\eta, \wh\phi) = (0, r_\varphi)$ really corresponds to a rotation of the initial data on the Cauchy slice.

\section{The canonical charges for the diamond}
\label{subsec:cancharges}

In this subsection we shall compute the charges associated with the canonical group, $\wt G = \avira \rtimes \vira$, and the corresponding Poisson algebra of them. As in \eqref{Pxigen}, the momentum charges are associated with elements $(0; \wh\xi)$ and given by
\be
P_{\wh\xi}(p) = p(X_{\wh\xi})
\ee
where $p \in T^*(\diff/\psl)$ and $X_{\wh\xi}$ is the vector field on $\diff/\psl$, evaluated at $[\psi] = \pi(p)$, induced by $\wh\xi \in \avira$. And, as in \eqref{Qagen}, the configuration charges are associated with elements $(\wt\eta; 0)$ and given by
\be\label{QchargewtG}
Q_{\wh\eta}(p) = \wh\eta(\pi(p))
\ee
where $\pi(p)$ is seen as a vector in $\avira^*$, as in \eqref{qcoadrealization}. 

Now we show that this choice of charges, with no constants added, form a legitimate representation of $\wt{\fr g}$, i.e., without additional central charges.\footnote{The reason for saying ``additional'' central charges is because $\wt{\fr g}$ already has a non-trivial center given by $(\wh c; 0) \oplus (0; \wh c)$.} In particular, this implies that there is no need to centrally extend the group $\wt G$ by a 2-cocycle. As explained in subsection \ref{subsec:cotbund}, the only 2-cocycle that can be non-trivial is the one mixing the $P$ and $Q$ parts of the algebra, i.e., $z((\wh\eta, 0), (0, \wh\xi))$. According to \eqref{QPbrackets},
\be
\{ Q_{\wh\eta}, P_{\wh\xi} \} =\wh\eta(X_{\wh\xi})
\ee
Since $z$ is constant on the phase space, it suffices to evaluated it at a single point. Most conveniently, we choose to evaluate it at $p = p_0 = 0$ on the fiber over $\wt\varepsilon = d\t^2 + \wt c$. At this point, $X_{\wh\xi} = \coad_{\wh\xi}\wt\varepsilon$, so we have
\be
\{ Q_{\wh\eta}, P_{\wh\xi} \}(p_0) =\wh\eta(\coad_{\wh\xi}\wt\varepsilon) = \coad_{\wh\xi}\wt\varepsilon (\wh\eta) = - \wt\varepsilon (\ad_{\wh\xi}\wh\eta) = \wt\varepsilon([\wh\eta, \wh\xi])
\ee
But from expression \eqref{QchargewtG} for the charge, and from the product rule \eqref{wtGprodrule}, we have
\be
H_{[(\wh\eta; 0), (0; \wh\xi)]} = H_{([\wh\eta, \wh\xi]; 0)} = Q_{[\wh\eta, \wh\xi]}
\ee
which evaluated at $p_0$ gives
\be
H_{[(\wh\eta; 0), (0; \wh\xi)]}(p_0) = [\wh\eta, \wh\xi](\wt\varepsilon) = \wt\varepsilon([\wh\eta, \wh\xi])
\ee
matching exactly with $\{ Q_{\wh\eta}, P_{\wh\xi} \}$. Therefore $z((\wh\eta; 0), (0; \wh\xi)) = 0$.

The harmonic basis in \eqref{harmonicbasis} defines a basis of charges given by
\begin{align}
P_n &:= H_{L_n} \nonumber\\
Q_n &:= H_{K_n}
\end{align}
together with the central charges $H_R$ and $H_T$. From the general expressions for the charges, we can compute explicitly the values of the central charges (which are constants on the phase space). We get, at a point $p \in T^*(\diff/\psl)$, 
\be
H_R = p(X_{\wh c}) = 0 
\ee
where $X_{\wh c} = 0$ because the coadjoint action of $\wh c$ is trivial. If $\psi$ is such that $\pi(p) = \coad_\psi \wt\varepsilon$, then
\be
H_T = \wh c (\pi(p)) = \wh c ( \coad_\psi \wt\varepsilon) = 1
\ee
where we have used that the $\wt c$ component of $\coad_\psi \wt\varepsilon$ is always $1\wt c$, for any $\psi$, as in \eqref{qcoadrealization}. In this way, the Poisson algebra of the canonical charges becomes
\begin{align}
&\{P_n, P_m\} = i (n - m) P_{n+m} \nonumber\\
&\{Q_n, P_m\} = i (n - m) Q_{n+m} - 4\pi i n^3 \delta_{n+m, 0} \nonumber\\
&\{Q_n, Q_m\} = 0 \label{canPoissonalgebra}
\end{align}
revealing that the algebra of momentum charges is just isomorphic to $\adiff$, after all. That is, the whole effect of having taken the Virasoro group as the group of ``configuration translations'', was not really to extend the algebra of configuration translations, but rather to introduce a central charge in the commutators mixing the configuration and momentum charges. It turns out that this algebra is equal to a  $\fr{bms}_3$ algebra of asymptotic symmetries of asymptotically flat three-dimensional gravity, as described in \cite{barnich2007classical,oblak2017bms}.

There are particularly explicit expressions for the charges on the partially-reduced phase space $\wh{\ca S} = \diff \times \whddiff$. (See Part I \cite{e2023quantization} for the necessary definitions.) We start by pulling back the $P$ charges under the $J$ map. If $p = J(\psi, \ac\sigma)$, we have
\be
P_{\wh\xi}(p) = J(\psi, \ac\sigma)(X_{\wh\xi}) = \ac\sigma(l^*_{\psi^{-1}} \overline X_{\wh\xi})
\ee
where $\overline X_{\wh\xi}$ is any vector at $\psi$ that projects to $X_{\wh\xi}$ under $q$. Given $\wt\a \in \avira^*$, define the coadjoint map from $\vira$ to $\avira^*$ as
\be
\coad\,\wt\a(\wh\psi) := \coad_{\wh\psi}\wt\a
\ee
Naturally, since the central elements of $\vira$ have a trivial coadjoint action, we can also think of the same map as being from $\diff$ to $\avira^*$ by simply replacing $\wh\psi$ by $\psi$. Note that derivative of this map gives the coadjoint action of $\avira$ on $\avira^*$,
\be
(\coad\,\wt\a)_*\wh\xi = \coad_{\wh\xi}\wt\a
\ee
In this notation, the quotient map $q : \diff \rightarrow \diff/\psl$, as realized in \eqref{qcoadrealization}, is simply written as
\be
q = \coad\,\wt\varepsilon
\ee
where $\wt\varepsilon = d\t^2 + \wt c$, and the vector field $X_\xi$ induced by $\xi \in \adiff \subset \avira$ at $\wt\a \in \avira^*$ is given by
\be
X_{\xi} = (\coad\,\wt\a)_*\xi 
\ee
If $\psi \in \diff$ is such that $\wt\a = \coad_\psi \wt\varepsilon$, we get
\be
X_{\xi} = \left(\coad (\coad_\psi \wt\varepsilon) \right)_*\xi 
\ee
Observe that, given any $\phi \in \diff$, we have
\be
\coad (\coad_\psi \wt\varepsilon) (\phi) = \coad_\phi \coad_\psi \wt\varepsilon = \coad_{\phi\psi} \wt\varepsilon = \coad\,\wt\varepsilon(\phi\psi) = \coad\,\wt\varepsilon \circ r_\psi (\phi)
\ee
where $r_\psi$ is the right-multiplication in $\diff$. Thus,
\be
X_{\xi} = (\coad\,\wt\varepsilon)_* r_{\psi*} \xi = q_*r_{\psi*} \xi = q_*l_{\psi*} l_{\psi^{-1}*} r_{\psi*} \xi = q_*l_{\psi*} (\ad^{\fr{diff}}_{\psi^{-1}} \xi)
\ee
where $l_\psi$ is the left-multiplication in $\diff$ and $\ad^{\fr{diff}}$ is the adjoint map in $\adiff$. This reveals that $\overline X_{\wh\xi}$ can be chosen as $l_{\psi*} (\ad^{\fr{diff}}_{\psi^{-1}} \xi)$, and therefore 
\be
P_{\wh\xi} = J(\psi, \ac\sigma)(X_{\wh\xi}) = \ac\sigma(\ad^{\fr{diff}}_{\psi^{-1}} \xi) = \coad^\fr{diff}_{\psi} \ac\sigma (\xi)
\ee
If $\ac\sigma \in \whddiff$ is seen as an element of $\avira^*$, $\ac\sigma = \ac\sigma + 0\wt c$, we have that $\coad^\fr{diff}_{\psi} \ac\sigma = \coad_\psi\ac\sigma$ (the two coadjoint maps always match when acting on elements with no $\wt c$ components). So we can write
\be\label{Pabs}
P_{\wh\xi} = \coad_{\psi} \ac\sigma (\xi) = (\psi_*\ac\sigma)(\xi)
\ee
or, a little bit more explicitly,
\be
P_{\wh\xi}(\psi, \ac\sigma) = \int\!d\t\, \frac{\ac\sigma(\psi^{-1}(\t))}{\psi'(\psi^{-1}(\t))^2} \xi(\t)
\ee
Now we pull-back the $Q$ charges under $J$, which is trivial since it is enough to write $\pi(p) = \coad_\psi \wt\varepsilon$, yielding
\be\label{Qabs}
Q_{\wh\eta}(p) = \wh\eta(\coad_\psi \wt\varepsilon)
\ee
or, more explicitly,
\be\label{Qexplicitexpr}
Q_{\wh\eta}(\psi, \ac\sigma) = \int\!d\t\, \frac{1 - 2 S[\psi](\psi^{-1}(\t))}{\psi'(\psi^{-1}(\t))^2} \eta(\t) + y
\ee
where $\wh\eta = \eta(\t)\partial_\t  + y \wh c$.

\subsection{Geometrical interpretation of the charges}
\label{subsec:chargeinterpretation}

Statements made about a quantum theory are only useful to the extent that one can interpret their classical consequences. That is, in order to extract physical information from a quantum theory one needs not only to know the properties of some self-adjoint operators, but it is important that one can match those operators to classical observables, which in principle could be tested and understood through classical experiments. So far we have already encountered one observable to which we could give a simple geometric interpretation: the CMC time-evolution Hamiltonian which corresponds, at a time $\tau$, to the area of the CMC slice with $K = -\tau$.
In the previous section the canonical charges were defined quite abstractly on the reduced phase space $\wt{\ca P}$, and were also given in terms of $(\psi, \ac\sigma) \in \wh{\ca S}$, which are still abstract parameters characterizing the diamond. In order to understand their physical meaning we should try to express them in terms of something more concrete like ADM variables, $(h_{ab}, K^{ab})$, for these have a simple geometrical meaning (the metric and extrinsic curvature of the CMC slices).

\vskip 0.2cm
\noindent\emph{\underline{Note}:} This section relies heavily on the content and notation of Part I \cite{e2023quantization}. Importantly, one should notice that in this section, exceptionally, $\Psi \in \text{\sl Diff}^+(\Sigma)$ denotes an extension of $\psi \in \diff$ to the Cauchy slice, and should not be confused with quantum wavefunctions that will appear later.
\vskip 0.2cm

\subsubsection{Momentum charges}
\label{subsubsec:chargeinterpretationmom}

Let us begin with the $P$ charges. Formula \eqref{Pabs}, in conjunction with \eqref{sigmahatsigmabarrelation}, gives
\be\label{Pxiintegral}
P_{\xi} = \ac\sigma(\psi^{-1}_*\xi) = - \int  \vartheta_{\bar h} \bar\sigma^{ab} \pounds_{\Psi^{-1}_*\xi} \bar h_{ab}
\ee
where in the last expression $\Psi$ is any extension of $\psi$ to a diffeomorphism of the disk and $\xi$ is, with a slight abuse of notation, any smooth extension of $\xi$ (at the boundary) to a vector field  on the disk. Now we insert factors of $e^\lambda$, where $\lambda$ is the solution of the associated Lichnerowicz equation, as follows
\be
P_{\xi} = \ac\sigma(\psi^{-1}_*\xi) = - \int  e^\lambda \vartheta_{\bar h} e^{-2\lambda}\bar\sigma^{ab} e^\lambda \pounds_{\Psi^{-1}_*\xi} \bar h_{ab}
\ee
Using relation $\vartheta_{e^\lambda \bar h} = e^\lambda  \vartheta_{\bar h}$ and $h_{ab} = \Psi_* e^\lambda \bar h_{ab}$ we have $e^\lambda \vartheta_{\bar h} = \vartheta_{\Psi^{-1}_* h}$. Also, notice that $e^\lambda \pounds_{\Psi^{-1}_*\xi} \bar h_{ab} =  \pounds_{\Psi^{-1}_*\xi} \left(e^\lambda\bar h_{ab}\right) -(\pounds_{\Psi^{-1}_*\xi} e^\lambda ) \bar h_{ab}$, but the second term would vanish when contracted with $\bar\sigma^{ab}$, so we can replace $e^\lambda \pounds_{\Psi^{-1}_*\xi} \bar h_{ab} =  \pounds_{\Psi^{-1}_*\xi} \left(e^\lambda\bar h_{ab}\right)$ in the integrand. Finally, from  $e^\lambda\bar h_{ab} = \Psi^{-1}_*h_{ab}$ and  $\sigma^{ab} =  \Psi_* e^{-2\lambda} \bar\sigma^{ab}$ we have $e^{-2\lambda}\bar\sigma^{ab} = \Psi^{-1}_* \sigma^{ab}$. Thus, we get
\be
P_{\xi} = - \int  \vartheta_{\Psi^{-1}_* h} \Psi^{-1}_* \sigma^{ab}  \pounds_{\Psi^{-1}_*\xi} \left(\Psi^{-1}_*h_{ab}\right) = - \int \Psi^{-1}_* \left[ \vartheta_h  \sigma^{ab}  \pounds_{\xi} h_{ab} \right]
\ee
and since integrals are invariant under diffeomorphisms,
\be
P_{\xi} =  - \int  \vartheta_h  \sigma^{ab}  \pounds_{\xi} h_{ab} 
\ee
As $\sigma^{ab}$ is transverse with respect to $h_{ab}$, this can be converted to a boundary integral
\be\label{Pxiintegralmod}
P_{\xi} =  - 2\int  \vartheta_h  \sigma^{ab}  \nabla_a \xi_b = - 2\int_\partial\!d\theta\,  \sigma^{ab}  n_a \xi_b
\ee
where $n^a$ is the unit (outward-pointing) vector field normal to the boundary. Finally, using $\sigma^{ab} := K^{ab} - \frac{1}{2} K h^{ab}$ and the orthogonality between $n$ and $\xi$ (at the boundary),
\be
P_{\xi} =  - 2\int_\partial\!d\theta\,  K_{ab}  n^a \xi^b
\ee
which is fully expressed in terms of geometrical quantities.

Since we are interested in the physical interpretation, it is worthwhile to reintroduce the physical scales, as explained in Sec.~ ``{\sl Reintroducing the physical scales}'' of Part I. We note that \eqref{Pxiintegral} remains unchanged, but there are factors of $\ell$ and $\ell_P$ in the transformation to $h_{ab}$ and $\sigma^{ab}$, so \eqref{Pxiintegralmod} becomes
\be
P_{\xi} =  - \frac{1}{8\pi \ell_P} \int_\partial\!ds\,  K_{ab}  n^a \xi^b
\ee
where $ds$ is the element of length with respect to $h_{ab}$. 
Now we wish to re-express $\xi$, which is an abstract vector on $S^1$, in terms of a vector associated with the physical boundary of the Cauchy slice.
%This subtly can be illustrated by considering $P_0 := P_{\partial_\theta}$. Note that $\xi = \partial_\theta$ is a unit vector with respect to $\bar h_{ab}$, but it is not unit with respect to $h_{ab}$. In particular, 
If $t^a$ is the unit vector tangent to the boundary of the diamond, $h_{ab} t^a t^b = 1$, and if we assume that the physical boundary metric is\footnote{There seems to be no reason not to make this assumption, since any $\gamma$ can be parametrized by proper length so that $\gamma = ds^2$, and then the angle coordinate $\theta$, with respect to which $\xi = \xi(\t)\partial_\t$, can be defined simply by $\t := 2\pi s/\ell$. Later, in Sec.~\ref{subsec:spinquantization}, we will argue that different choices lead to physically equivalent theories, even at the quantum level.}
\be\label{bdymetricconv}
\gamma = \left(\frac{\ell}{2\pi}\right)^2 d\t^2
\ee
then
\be\label{unittangentt}
t^a = \frac{2\pi}{\ell} (\partial_\t)^a
\ee
Therefore, with this choice $\xi^a = \xi(\t) (\partial_\t)^a = \xi(\t) (\ell/2\pi) t^a$, we have
\be\label{Pxigeometric}
P_\xi =  - \frac{\ell}{16\pi^2 \ell_P} \int_\partial\!ds\,  K_{ab}  n^a \xi(\theta) t^b
\ee
where $\theta = 2\pi (s/\ell)$, with $s$ being the length along the boundary from some reference point (a change of this reference point amounts to redefining the charges by phases, $P_n \mapsto e^{in\varphi} P_n$ and $Q_n \mapsto e^{in\varphi} Q_n$, which clearly leaves the canonical algebra invariant).

Let us discuss the $P_0$ charge in more depth, as it has some interesting features. 
From the expression above, with $\xi = \partial_\t$, we see that
\be\label{P0geometric}
P_0 =  - \frac{\ell}{16\pi^2 \ell_P} \int_\partial\!ds\,  K_{ab}  n^a t^b
\ee
Notably, $P_0$ generates a $SO(2)$ dynamical symmetry of the diamond, corresponding to rigid rotations of the boundary. It is therefore natural to call $P_0$ the {\sl spin} of the diamond. This can be seen in two ways. First, as $P_0 = P_{\partial_\theta}$, its Poisson flow is generated by $\partial_\theta$ acting on the configuration space $\diff/\psl$. More precisely, a finite rotation by the angle $\varphi$, acting on $S^1$ as $R_\varphi(\theta) = \theta + \varphi$, will act on the configuration space as $R_\varphi[\psi] = [R_\varphi \circ \psi]$. We can show explicitly that the Hamiltonian is invariant under such a transformation, so that $\{P_0, \wt H\} = 0$, implying that $P_0$ is conserved under time evolution. 
%\RS{Insert direct proof that $\wt H$ is invariant under rigid rotations.} 

Another way to see that $P_0$ corresponds to a rotational symmetry is by considering directly the ADM formalism. In this context the Hamiltonians takes the form
\be\label{GRhamiltonian}
H_\text{\sl bulk}[N, \vec N] = \frac{1}{16 \pi G} \int_{\Sigma} d^2\!x  \left[ N \sqrt{h}\left(K^{ab}K_{ab} - K^2 -  R + 2\Lambda\right) - 2 N_b \nabla_a \pi^{ab} \right]
\ee
up to boundary terms. The Hamiltonian, labelled by a lapse $N$ and a shift  $\vec N$ parameters, can be interpreted as generating spacetime diffeomorphisms along the vector $Nu^a + N^a$, where $u^a$ is the unit future-directed vector normal to the spatial slice. 
When the lapse and shift are non-trivial at the boundary, the expression above may not define a regular Hamiltonian function on the phase space, i.e., a function that can be associated with a regular Hamiltonian flow. More precisely, note that this association is defined via $\delta H = -\ii_X \Omega$, so if the symplectic form $\Omega$ is given by the integral of a symplectic current on the Cauchy slice without boundary terms, $\Omega = \int_\Sigma \delta \pi^{ab} \wedge \delta h_{ab}$, then the right-hand side of this relation gives $ \int_\Sigma (X^h)_{ab} \delta \pi^{ab} - (X^{\pi})^{ab} \delta  h_{ab}$, revealing that the left-hand side must be of the form $\delta H = \int_\Sigma A^{ab} \delta h_{ab} + B_{ab} \delta \pi^{ab}$ for a solution to exist. In other words, $H$ generates a regular symplectic flow if and only if it has a well-defined functional derivative. In this case we say that $H$ is {\sl symplectically differentiable}. 
Since the Poisson bracket is defined in terms of the associated Hamiltonian vectors, functions only have well-defined Poisson brackets if they are symplectically differentiable.
Accordingly, we may need to add appropriate boundary terms to expression \eqref{GRhamiltonian} in order to define regular Hamiltonians. 
A trick to find these boundary terms is to vary the bulk Hamiltonian $H_\text{\sl bulk}$ given in \eqref{GRhamiltonian} and, if this produces boundary terms, we add suitable boundary terms  $H_\partial$ to cancel it, so that $\delta H = \delta H_\text{\sl bulk} + \delta H_\partial$ is purely a bulk integral (not containing spatial derivatives of field variations). If the Hamiltonian does acquire a non-trivial boundary term in this manner, then its on-shell value will be $H \approx H_\partial$, implying that $\delta H = - \ii_X \Omega \ne 0$ and consequently the phase space transformation $X$ induced by $H$ is not gauge. In other words, the diffeomorphism generated by $N$ and $N^a$ would not be a gauge transformation but rather a non-trivial symmetry. 
%Therefore, one way to search for all diffeomorphisms that are symmetries is to consider all lapses and shifts which respect the boundary conditions, and then look for boundary terms appearing in the variation of the corresponding Hamiltonians. In the case of our diamond, since we should not allow for diffeomorphisms that flow points from the inside to the outside (i.e., crossing the surface of the diamond), we must impose that $N$ vanishes at the boundary and that $N^a$ is tangent to the boundary. Moreover, in order to preserve the induced metric $\gamma$ on the boundary, $N^a|_\partial$ must be an isometry of $\gamma$.
Let us consider the case where $N$ vanishes at the boundary and $N^a|_\partial$ is tangent to the boundary, i.e., $N^a \propto t^a$.\footnote{A diffeomorphism generated by $N^a$ would not in general preserve the boundary metric, unless $N^a \propto t^a$ by a constant factor. Nevertheless, in Sec.~\eqref{subsubsec:chargeinterpretationpos} we will argue that there is a way in which these Hamiltonians produce well-defined flows by virtue of the CMC gauge-fixing constraint.}
The variation of the term involving $N$ in \eqref{GRhamiltonian} gives
\be
\delta H_\text{\sl bulk} = \text{\sl ``bulk term involving $N$''} - \frac{1}{16\pi \ell_P} \int_\partial\!ds\, N n^a \left(\nabla^b \delta h_{ab} - h^{cd} \nabla_a \delta h_{cd} \right) + \cdots
\ee
where $n^a$ is the unit spatial outward-pointing normal vector field on the boundary, and the ``$\cdots$'' refers to terms involving $N^a$. Since we chose $N|_\partial = 0$, no boundary terms come from $N$. Now the term involving $N^a$ gives
\be
\delta H_\text{\sl bulk} = \text{\sl ``bulk terms''} -  \frac{1}{8\pi \ell_P} \int_\partial\!ds\, n_a N_b \left( h^{-1/2} \delta \pi^{ab} + h^{bc} \delta h_{cd} h^{-1/2} \delta \pi^{ad}\right) 
\ee
This boundary term is in fact exact (as a phase space form) and can be written as
\be
\delta H_\text{\sl bulk} = \text{\sl ``bulk terms''} - \delta \left[ \frac{1}{8\pi \ell_P} \int_\partial\!ds\, n_a N_b K^{ab} \right] 
\ee
which implies that the Hamiltonian must be ``corrected'' with a boundary term as follows
\be
H =  H_\text{\sl bulk} + \frac{1}{8\pi \ell_P} \int_\partial\!ds\, n_a N_b K^{ab}
\ee
Thus, if $N^a$ is non-zero at the boundary, the on-shell value of the corresponding Hamiltonian is non-trivial
\be
H[N^a|_\partial = \xi^a]_\text{\sl on-shell} = \frac{1}{8\pi \ell_P} \int_\partial\!ds\, n_a \xi_b K^{ab}
\ee
Using convention \eqref{bdymetricconv} for the boundary metric, so that $\xi^a = \xi(\t) (\partial_\t)^a = \xi(\t) (\ell/2\pi) t^a$, we get
\be\label{Hcornerdiff}
H[N^a|_\partial = \xi^a]_\text{\sl on-shell} = \frac{\ell}{16\pi^2 \ell_P} \int_\partial\!ds\, n_a \xi(\t) t_b K^{ab} = - P_\xi
\ee
In particular, when $N^a$ is an isometry of $\gamma$, $\xi = \partial_\t$, 
\be\label{Hspin}
H[N^a|_\partial = (\partial_\theta)^a]_\text{\sl on-shell} = \frac{\ell}{16\pi^2 \ell_P} \int_\partial\!ds\, n_a t_b K^{ab} = - P_0
\ee
Thus we see that this Hamiltonian, generator of rigid rotations of the boundary, corresponds to (minus) the charge $P_0 = P_{\partial_\theta}$,  justifying its denomination as the {\sl spin} of the diamond. 
(The reason for the minus sign is due to our conventions of how $\diff$, or $\vira$, acts on the phase space.)

We can show that $P_0$ is preserved under CMC time evolution (or, equivalently, the Hamiltonian generating CMC time evolution is rotation-invariant). In fact, $P_0$ is independent of the spatial slice altogether. Let $H_{\vec N}$ be the Hamiltonian corresponding to a shift $N^a$ that matches $(\partial_\theta)^a$ at the boundary and no lapse, which as we have just seen reduces to $-P_0$ on-shell; and let $H_N$ be the Hamiltonian corresponding to a lapse $N$ which deforms one Cauchy slice into another (thus vanishing at the boundary) and no shift. The change of $P_0$ with respect to slice change is then given by
\begin{align}
\dot P_0 = - \{ H_{\vec N}, H_N \} &= - \{ H_{\vec N}, \int\!d^2x\, N(x) \ca H_0(x) \} \nonumber\\
&= - \int\!d^2x\, N(x) \{ H_{\vec N}, \ca H_0(x) \} \nonumber\\
&=  \int\!d^2x\, N \pounds_{\vec N} \ca H_0 \nonumber\\
&= \int\!d^2x\, N \left( \nabla_a N^a \ca H_0 + N^a \nabla_a \ca H_0 \right) \nonumber\\
&= \int\!d^2x\, \nabla_a \left( N  N^a \ca H_0 \right) - \int\!d^2x\, N^a \nabla_a N \ca H_0  \nonumber\\
&= \int_\partial\!ds\, N n_a N^a (h^{1/2}\ca H_0) + H_{\pounds_{\vec N} N} \nonumber\\
&= H_{\pounds_{\vec N} N} \approx 0
\end{align}
In the first line we wrote $H_N$ explicitly in terms of the Hamiltonian constraint $\ca H_0$; in the third line we used that $H_{\vec N}$ generates spatial diffeomorphisms; in the fourth line we wrote the Lie derivative in terms of covariant derivatives (note that $\ca H_0$ is a density); in the fifth line we integrated by parts; in the sixth line we used Gauss' theorem and identified the bulk term as the Hamiltonian corresponding to the lapse $\pounds_{\vec N} N$; and in the last line we used that $N^a$ is tangent to the boundary (so $n_a N^a = 0$) leading to the conclusion that $\dot P_0$ vanishes on-shell (i.e., on the constraint surface).

\subsubsection{The spin/twist relationship}
\label{subsubsec:chargeinterpretationtwist}

After having geometrically expressed $P_0$ as an integral of the extrinsic curvature along the boundary and recognized it as the spin of the diamond, we now provide another nice interpretation for it as being the {\sl twist of the corner of the diamond}. First, let us define the twist of a curve embedded in a three-dimensional space. Consider a closed spacelike curve $q : S^1 \rightarrow \ca M$ embedded in an oriented Lorentzian space $\ca M$.\footnote{The concepts discussed here also apply naturally for Riemmanian spaces.} Let $t$ be the unit vector tangent to $q$ and $a^a := t^b \boldsymbol\nabla_b t^a$ be the acceleration vector. Since $t$ is normalized, $a$ is orthogonal to $t$, so let us define $n$ as the unit vector aligned with $a$ (assumed not be lightlike or zero). The plane $tn$ defines a frame along the curve, whose third normal vector can be defined as $u^a :=\boldsymbol\epsilon^{abc} t_b n_c$, where $\boldsymbol\epsilon$ is the volume element on $\ca M$. The {\sl torsion} quantifies the rate (with respect to proper distance) that this frame rotates along the curve, and can be defined as $\chi := \text{\sl sign}(n) u_a (t^b \boldsymbol\nabla_b n^a) = \text{\sl sign}(n) \boldsymbol\epsilon_{abc}t^b n^c (t^d \boldsymbol\nabla_d n^a)$, where $\text{\sl sign}(n) := n_a n^a$. The {\sl twist} $\ca T$ of the (closed) curve is then defined as the integrated torsion along the curve,
\be\label{twistdef}
\ca T := \text{\sl sign}(n) \int\!ds\, \boldsymbol\epsilon_{abc}t^b n^c (t^d \boldsymbol\nabla_d n^a)
\ee
where $ds$ is the element of length along the curve.\footnote{Some references include numerical factors like $1/2\pi$ in the definition of the twist, which we choose not to do.} What is interesting about the twist is that it can be evaluated with the expression above using {\sl any} unit normal vector $n$, not necessarily the one aligned with the acceleration. This can be seen by evaluating this expression with two unit normal vectors $n$ and $\wt n$. If both vectors are spacelike, we can write $\wt n = \cosh(\varphi) n + \sinh(\varphi) u$, where $\varphi : S^1 \rightarrow \bb R$ denotes the (hyperbolic) angle of rotation between $n$ and $\wt n$. We get
\be\label{twistnchange}
\boldsymbol\epsilon_{abc}t^b {\wt n}^c (t^d \boldsymbol\nabla_d {\wt n}^a) = \boldsymbol\epsilon_{abc}t^b n^c (t^d \boldsymbol\nabla_d n^a) - \frac{d\varphi}{ds}
\ee
If both $n$ and $\wt n$ are smooth along the curve, so that $\varphi$ is periodic, the integral of $d\varphi/ds$ will vanish, and therefore the twist can be defined as in \eqref{twistdef} for any (smooth) spacelike unit normal vector field $n$. One can verify that the definition also applies for timelike normal vectors, thanks to the factor of $\text{\sl sign}(n)$. Note that the twist is always well-defined, even when the torsion is not (the torsion is only well-defined if the acceleration is not vanishing and not lightlike). 

We can also provide a formula for the twist which refers to lightlike normal vectors. This may be pertinent since the two inward-pointing null vectors are the generators of the future and past horizons of the diamond. Let $k_\pm$ be the future ($+$) and past ($-$) inward-pointing null vector fields orthogonal to the curve, normalized like $(k_+)_a (k_-)^a = 1$. If $n$ is a smooth (outward-pointing) spacelike normal field along the curve, and $u$ is the (future-pointing) vector field normal to both $n$ and $t$, then we can write $k_\pm = (- n \pm u)/\sqrt{2}$. Now writing $\ca T = (\ca T + \ca T)/2$, where the first $\ca T$ is expressed as in \eqref{twistdef} with $n$ as the normal vector and the second $\ca T$ with $u$ as the normal vector, we see that
\be\label{twistdefnull}
\ca T = \int\!ds\, \boldsymbol\epsilon_{abc}t^b (k_\pm)^c t^d \boldsymbol\nabla_d (k_\mp)^a = \mp \int\!ds\, (k_\pm)_a t^b \boldsymbol\nabla_b (k_\mp)^a
\ee
Note that this involves both $k_+$ and $k_-$, essentially evaluating how much one changes with respect to the other.

Interestingly, relation \eqref{twistnchange} provides another interpretation for the twist, as follows. Let $\wt n$ be a unit (spacelike) normal Fermi-Walker transported\footnote{The Fermi-Walker transport gives a natural notion of ``parallel transporting'' a frame along a curve, while keeping it aligned with the curve. More precisely, this is defined by parallel-transporting a frame for an infinitesimal distance and then projecting the normal vectors orthogonally to the curve and the tangent vector along the curve. The Fermi-Walker transport coincides with the parallel transport if (and only if) the curve is a geodesic. The Fermi-Walker derivative is defined by $D^\text{FW}_q v^a := t^b \boldsymbol\nabla_b v^a + (a_b t^a - t_b a^a) v^b$, where $t$ and $a$ are respectively the tangent and acceleration vectors of the curve $q$, and it has the property that $D^\text{FW}_q v^a = 0$ if (and only if) $v$ is Fermi-Walker transported along the curve.} along the curve, thus satisfying
\be
D^\text{FW}_q {\wt n}^a = t^b \boldsymbol\nabla_b {\wt n}^a + a_b \wt n^b t^a = 0
\ee
When $\wt n$ comes back to the initial point, after a loop, it may not coincide with its original value. If $n$ is any smooth (spacelike) unit normal field along the curve, we can define the (hyperbolic) angle $\varphi$ of $\wt n$ with respect to $n$ as before. But now $\varphi$ is not necessarily periodic, so let $\Delta\varphi := \varphi_\text{final} - \varphi_\text{initial}$ be the total angle accumulation after a loop. Integrating both sides of \eqref{twistnchange} in $ds$ yields zero for the left-hand side (since $\boldsymbol\epsilon$ would be contracting with two $t$'s), $\ca T$ for the first term on the right-hand side and $-\Delta\varphi$ for the last term. Therefore, we have
\be
\ca T = \Delta\varphi
\ee
revealing that the twist $\ca T$ can be interpreted as a holonomy induced on the normal bundle of the curve, expressing how the parallel-transported normal frame comes back rotated after a loop. A similar result also holds for lightlike vectors, using \eqref{twistdefnull}. In this case, let $\wt k_\pm$ be a pair of Fermi-Walker transported inward-pointing null vectors orthogonal to the curve, normalized like $(\wt k_+)_a (\wt k_-)^a = 1$, and let $k_\pm$ be any smooth (normalized) pair of inward-pointing null vectors orthogonal to the curve. We can write $\wt k_\pm = e^{\chi_\pm} k_\pm$ for some scalars $\chi_\pm$. The normalization conditions imply that $\chi_- = - \chi_+$. From a relation analogue to \eqref{twistnchange}, for the integrand of \eqref{twistdefnull}, we conclude that
\be
\ca T = \mp \Delta\chi_\pm
\ee
where $\Delta\chi_\pm := (\chi_\pm)_\text{final} - (\chi_\pm)_\text{initial}$ is the phase difference accumulated along a loop. That is, the future and past null generators of the Cauchy horizon of the diamond come back boosted (rescaled with respect to each other) after being transported along the corner loop of the diamond.

Let us now see that $P_0$ is directly related to the twist of the boundary. Let $u$ be the unit timelike (future-pointing) normal vector to the CMC, so that $K_{ab} = \boldsymbol\nabla_b u_a$, where $\boldsymbol\nabla$ denotes the spacetime covariant derivative. Let $n$ be the unit outward-pointing normal vector to the boundary, parallel to the CMC, and $t^a$ be the unit vector tangent to the boundary, assumed to be oriented such that $\boldsymbol\epsilon(u, n, t) = 1$, where $\boldsymbol\epsilon$ is the spacetime orientation. Then \eqref{P0geometric} can be integrated by parts as
\be\label{P0twist}
P_0 =  - \frac{\ell}{16\pi^2 \ell_P} \int_\partial\!ds\,  \boldsymbol\nabla_b u_a  n^a t^b = \frac{\ell}{16\pi^2 \ell_P} \int_\partial\!ds\, u_a t^b  \boldsymbol\nabla_b n^a = \frac{\ell}{16\pi^2 \ell_P} \ca T
\ee
showing that the spin $P_0$ is proportional to the twist $\ca T$, with a proportionality factor involving the ratio between the boundary length and the Planck length (in units where $\hbar =1$). 

This relationship between twist and spin can be seen as an analogue, in lower dimensions, of a result obtained in
\cite{donnelly2021gravitational}, in which the edge mode structure of gravity in 3+1 dimensions is analyzed from the covariant phase space perspective. It is found 
that the charges generating volume-preserving diffeomorphisms of the corner ($S^2$) are related to the curvature of the connection on the normal bundle of the corner, as naturally defined from its embedding into the ambient spacetime. 
For our causal diamonds, volume-preserving diffeomorphisms of the corner ($S^1$) are precisely 
the isometries of the boundary metric.
In this case, the normal bundle of the corner also inherits a natural connection from the ambient spacetime, but one cannot define a non-trivial curvature tensor since the base space is 1-dimensional. Nevertheless, due to the non-trivial fundamental group of $S^1$ (unlike $S^2$), 
there is another kind of  invariant associated to the connection: the holonomy defined by 
parallel transporting the normal frame along the loop. As we have seen, this holonomy is described by a boost angle which is interpreted as the twist of the corner.

On a separate note, the twist of a curve has also appeared in an analysis of the holographic entanglement entropy in the context of two-dimensional conformal field theories with a gravitational anomaly \cite{castro2014holographic}, but it is unclear whether this has any relation with our work.

\subsubsection{Configuration charges}
\label{subsubsec:chargeinterpretationpos}

The interpretation of the $Q$ charges appears to be much less simple. Unlike the $P$ charges which are related to a local integration of $K_{ab}$ along the boundary, the $Q$ charges seem to be related in a non-local way with the curvature of the boundary as embedded in the CMC itself.

We can still establish some general properties of the $Q$ charges.
First, note from \eqref{Qexplicitexpr} that the $Q$'s do not depend on $\ac\sigma$, but only on $\psi$. In fact, it is a function on the configuration space so it depends only on the $\psl$-class of $\psi$, $[\psi] \in \diff/\psl$. 
A given spatial metric $h_{ab}$ uniquely determines one such equivalence class $[\psi]$, as can be seen from App.~``{\sl The uniformization map}'' of Part I. There we explain an algorithm for constructing a $\psi$ given a $h_{ab}$, but it contains an ambiguity associated with the conformal automorphisms of the unit complex disc which translates into a $\psl$ indeterminacy, $\psi \sim \psi\chi$, where $\chi \in \psl$.
On the other direction, a class $[\psi]$ determines a spatial metric up to boundary-trivial conformal transformations, $[h] = [\Phi_*\Theta h]$, where $\Phi \in \text{\sl Diff}^+(\Sigma)$ acts as the identity on the boundary and the function $\Theta \in C^\infty(\Sigma, \bb R^+)$ is $1$ at the boundary. This can be seen from  Sec.~``{\sl Reduction via conformal coordinates}'' of Part I, where we explain
how a given $\psi$ determines the boundary value of the Weyl factor $\Lambda$ from the condition on the boundary metric,
\be
h_{ab}|_\partial = \psi_* \Lambda_\partial\bar h_{ab} |_\partial
\ee
so the only ambiguities are on extending $\psi \mapsto \Psi$ and $\Lambda_\partial \mapsto \Lambda$ to the interior of the disc, producing the spatial metric
\be
h_{ab} = \Psi_* \Lambda \bar h_{ab} 
\ee
that is, one could instead have chosen to extend these variables as $\psi \mapsto \Phi \circ \Psi$ and $\Lambda \mapsto \Theta \Lambda$, where $(\Phi, \Theta)$ is boundary-trivial. Also, it is clear that could have started with $\psi\chi$ instead, and such a choice affects $\Lambda_\partial$; however, it does it in such a way that the conformal transformation $(\Psi, \Omega)$ should be replaced by $(\Psi, \Omega) \circ (\Xi, \Gamma)$, where $(\Xi, \Gamma)$ is a conformal isometry of $\bar h_{ab}$. Consequently, $(\Psi, \Omega) \circ (\Xi, \Gamma)$ defines the same metric $h_{ab}$. 
We thus conclude that the $Q$ charges 
depend only on the conformal class
of the spatial metric,
\be
Q_n([\psi]) = Q_n([h_{ab}])
\ee
with a slight abuse of notation.

It should be remarked that the fact that $Q_n$ depends only on the (conformal class of the) spatial metric does not mean that $Q_n$ cannot be expressed in a format that involves the extrinsic curvature. That is, $K^{ab}$ could appear as long as it is combined with other quantities to produce a conformal invariant of $h_{ab}$. This is plausible only because the canonical charges are defined intrinsically  on the reduced phase space, so any statements regarding their relationship with the ADM variables should be understood as being valid on-shell (i.e., assuming that the constraints hold), and consequently the extrinsic curvature is not independent from the spatial metric. 
In fact, it may well be that the extrinsic curvature appears in the most ``natural form'' of $Q$, i.e., the form with the most readily physical interpretation. 

A second property, of $Q_0$ in particular, is that it must be bounded from above, attaining a maximum value of $2\pi$ when $[\psi] = [I]$, and unbounded from below \cite{oblak2017bms}. In this configuration, $Q_n([I]) = \int\!d\theta\, e^{in\theta} = 2\pi \delta_{n0}$. 
This will be discussed in Sec.~\ref{subsec:Q0spec}, where we shall see that the same property also holds in the quantum theory.

Lastly, we comment on an alternative point of view, based on the ADM analysis of corner symmetries, that may be helpful in clarifying the meaning of the $Q$ charges. We shall refrain from going into much detail here since, to the extent that it applies to the present question, it is still an speculation.\footnote{Based on collaboration with Laurent Freidel, Luca Ciambelli and Ted Jacobson.}
The main point is to notice that, as proven in Sec.~``{\sl The CMC gauge is attainable}'' of Part I, the condition of fixing the induced boundary metric implies that all refoliations of the causal diamond can be achieved via {\sl gauge} transformations. Moreover, we have also proved that, within the considered class of causal diamond spacetimes, a regular foliation always exists defined by the CMC condition, $K = -\tau$, which provides a universal gauge-fixing of time. We can think of this gauge-fixing as introducing a constraint to the ADM phase space
\be
C_\tau[\phi] := \int_\Sigma\! d^2x\, \phi (K + \tau)
\ee
where $\phi$ is a scalar labeling this family of functions. 
This can be seen as a constraint conjugate to the generator of refoliations, i.e., $C_\tau[\phi]$ and $H[N, \vec 0]$, with $N|_\partial = 0$, form a family of second-class constraints.
By themselves, generic corner-deforming ADM Hamiltonians $H[N, \vec N]$, where $N$ and $\vec N$ are arbitrary at the boundary, would typically violate the boundary condition on the metric. But since $C_\tau[\phi]$ generates Weyl transformations on the spatial metric, we can always find linear combinations of $H[N, \vec N]$ and $C_\tau[\phi]$ that preserve the boundary condition.
We conclude that the following Hamiltonian is symplectically differentiable and, for the appropriate choice of $\phi$, generates a flow that preserves the boundary condition
\be\label{ADMchargesgen}
H[N, \vec N] = H_\text{bulk}[N, \vec N] + C_\tau[\phi(N, \vec N)] + 2 \int_\partial ds\, \left[ N^c t_c t^a n^b K_{ab} - N^c n_c t^a t^b K_{ab} - N A \right]
\ee
where $A$ is the (scalar) acceleration of the corner as embedded in the spatial slice, i.e., $t^a \bs \nabla_a t^b = - A n^b$.\footnote{In a rigorous treatment of the second-class constraints in the ADM phase space, this Hamiltonian, with the appropriate constraint terms added, is precisely what should be used to compute Dirac brackets as regular Poisson brackets, i.e., $\{H_\partial, \cdot\,\}_D \approx \{H, \cdot\,\}$.}
There is another manner to write this expression which is interesting. From the perspective of the diamond spacetime, there are three directions that are special at the corner: one is along the corner itself, $(N, \vec N) = (0, \xi t^a)$, and the others are along the null rays of the diamond future horizon, $(N, \vec N) = (\zeta^+, - \zeta^- n^a)$, and past horizon, $(N, \vec N) = (-\zeta^-, - \zeta^- n^a)$.
The charges associated with the horizon flows are related to the expansion parameters, $\Theta^\pm$, of the respective null generators; $\Theta^\pm$ is defined as the rate of change of length of a piece of corner, relative to its length, as it is transported along the null generators, $\pm u^a - n^a$.
Therefore, we have three species of charges, 
\ba
&\text{\sl Corner diffeomorphisms:}\,\, (N, \vec N) = (0, \xi t^a) \,\,\longrightarrow\,\,  H[\xi] = 2 \int_\partial ds\, \xi t^a n^b K_{ab} \no
&\text{\sl Future horizon flow:}\,\, (N, \vec N) = (\zeta^+, -\zeta^+ n^a) \,\,\longrightarrow\,\,  H[\zeta^+] = 2 \int_\partial ds\, \zeta^+ \Theta_+ \no
&\text{\sl Past horizon flow:}\,\, (N, \vec N) = (-\zeta^-, -\zeta^- n^a) \,\,\longrightarrow\,\,  H[\zeta^-] = - 2 \int_\partial ds\, \zeta^- \Theta_-
\ea
The charges associated with the corner diffeomorphisms are, as shown in \ref{Hcornerdiff}, identified with (minus) $P_\xi$.
We speculate that the $Q$ charges are related to some (possibly non-linear) function of the expansion parameters, i.e., the charges associated with horizon flows.
The investigation on this front is left to later work.

\section{The quantum theory}
\label{sec:quantumtheory}

The quantum theory is constructed based on {\sl a} (projective) unitary irreducible representation of the transitive symmetry group $\wt G = \avira \rtimes \vira$.
Note that, in this case, quantization amounts to finding a class of suitable quantum theories, in which the (complete) subalgebra of canonical observers can be properly represented on the Hilbert space --- as explained in Sec.~\ref{subsec:canons} this is unlike the case of a phase space with a vector space structure, where the quantization is  based on the Heisenberg group which has a unique unitary irreducible representation (given a fixed value of $\hbar$). 

Equivalently, as explained in App.~\ref{app:projrep}, projective unitary irreducible representations of a group are in one-to-one correspondence with (true) unitary irreducible representations of (the universal cover of) a central extension (by 2-cocycles) of the group, which in turn are in one-to-one correspondence with self-adjoint irreducible representations of the Lie algebra (of the extended group). Note however that, because of the Casimir matching principle discussed in Sec.~\ref{subsec:isham}, central elements of the algebra are to be represented with the same value as their classical counterparts, so if the Poisson algebra associated with the canonical group is truly homomorphic to the Lie algebra of the group, then we should not consider further central extensions of the algebra, even if in principle one is admissible. 
We showed in Sec.~\ref{subsec:cancharges} that the Poisson algebra generated by the action of $\avira \rtimes \vira$ reduces to
\begin{align}
&\{P_n, P_m\} = i (n - m) P_{n+m} \nonumber\\
&\{Q_n, P_m\} = i (n - m) Q_{n+m} - 4\pi i n^3 \delta_{n+m, 0} \nonumber\\
&\{Q_n, Q_m\} = 0 \label{canalgcla2}
\end{align}
so not only there is no need to extend $\avira \rtimes \vira$ further, but the central charge associated with the $\vira$ factor should be represented trivially. In this manner, even though we had to extend $\diff$ into $\vira$ for the purpose of constructing an appropriate transitive group of symplectomorphisms on the phase space, the algebra of configuration translations is reduced to the original $\adiff$ algebra when realized as a Poisson algebra of momentum charges on the phase space, so the ultimate effect of that extension was only to introduce a central charge in the mixed bracket between momentum and configuration charges. 

It is also worth noticing that the algebra above happens to be the $\bms$ algebra of asymptotic symmetries at null infinity of asymptotically-flat spacetimes in three dimensions, as obtained in \cite{barnich2007classical}. 
The group underlying this algebra is known as $\BMS$, or the {\sl Bondi–Metzner–Sachs group} in three dimensions. See that $\BMS$ is not a subgroup of $\avira \rtimes \vira$, but rather the latter is a central extension of the former. 

In this way, the quantum theory is therefore based on some unitary irreducible representation of the 
``quantized'' version of the classical Poisson algebra \eqref{canalgcla2}, in which the Poisson bracket $\{\,,\}$ is replaced by $-i [\,,]$,\footnote{Recall that we are using units where $\hbar = 1$. 
One would have to be careful in recovering $\hbar$ explicitly in these formulas. 
In particular, in Sec.~``{\sl Reintroducing the physical scales''} of Part I, we have essentially  borrowed the $\hbar$ from quantum mechanics in writing classical formulas for the metric $h_{ab}$ and extrinsic curvature $\sigma^{ab}$ in terms of the (dimensionless) quantities $\bar h_{ab}$ and $\bar\sigma^{ab}$. Had we not made the choice $\hbar = 1$, the physical symplectic form on the reduced phase space would actually be $\hbar \omega$, where $\omega$ is the (dimensionless) symplectic form associated with the cotangent bundle structure of $T^*(\diff/\psl)$. This $\hbar$ would appear in the definition of the canonical charges, and eventually cancel with the $1/i\hbar$ in the homomorphism to the quantum algebra.}
\begin{align}
&[\wh P_n, \wh P_m] =  (m - n)\,\wh P_{n+m} \nonumber\\
&[\wh Q_n, \wh P_m] =  (m - n)\, \wh Q_{n+m} + 4\pi  n^3 \delta_{n+m, 0} \nonumber\\
&[\wh Q_n, \wh Q_m] = 0 \label{quantumalgebra}
\end{align}
where $\wh P_n$ and $\wh Q_n$ are operators on the Hilbert space $\ca H$ corresponding, respectively, to the observables $P_n$ and $Q_n$. Note that they are not supposed to be self-adjoint since their classical counterparts are not real (as they are associated with complex Fourier modes of diffeomorphisms of $S^1$). Instead, these operators should satisfy 
the conjugation relations
\begin{align}
(\wh P_n)^\dag &= \wh P_{-n} \nonumber\\
(\wh Q_n)^\dag &= \wh Q_{-n}
\end{align}
which mimic the classical relations $(P_n)^* = P_{-n}$ and $(Q_n)^* = Q_{-n}$.
We shall use the terminology that such a representation of the algebra is {\sl unitary}, so that momentum and configuration variables labelled by real diffeomorphisms are self-adjoint operators and consequently the group of (real) symmetries is represented unitarily.

The unitary operators corresponding to the canonical transformations are obtained exponentiating the respective algebra elements. That is, we have
\be\label{QrepfromU}
U(\exp(t(\wh\eta; 0))) = e^{-i t\wh Q_{\wh\eta}}
\ee
\be\label{PrepfromU}
U(\exp(t(0; \wh\xi))) = e^{- i t\wh P_{\wh\xi}}
\ee
where $t$ is a real parameter, $\exp$ denotes the Lie exponential in $\wt G = \avira \rtimes \vira$ and elements of its Lie algebra are denoted by $(\wh\eta; \wh\xi) \in \avira^c \sdplus \avira$. 
The Casimir matching principle implies, from $T = 1$, that $\wh Q_{\wh c} = 1$, so
\be\label{CasimircondQ}
U(\exp(t(\wh c; 0))) = U(((I,t); (I,0)))  = e^{-it}
\ee
and also, from $R = 0$,  that
$\wh P_{\wh c} = 0$, so
\be\label{CasimircondP}
U(\exp(t(0; \wh c))) = U(((I,0);(I,t))) = 1
\ee
where $I \in \diff$. We have used here expression \eqref{viraexpdef} parametrizing elements of $\vira$ in terms of exponential of algebra elements.

If the representation is constructed directly at the group level, as in Sec.~\eqref{subsec:genaspectsquantum} below, then the configuration and momentum operators are obtained by differentiating the unitary operators, 
\be\label{QdeffromU}
\wh Q_{\wh\eta} = i \left. \frac{\partial}{\partial t} U(\exp(t(\wh\eta; 0))) \right|_{t=0} 
\ee
\be\label{PdeffromU}
\wh P_{\wh\xi} = i \left. \frac{\partial}{\partial t}  U(\exp(t(0; \wh\xi))) \right|_{t=0} 
\ee
It automatically follows from the fact that $U$ forms a representation of $\wt G$ that these $P$'s and $Q$'s satisfy the appropriate algebra, and from the unitarity that they will be self-adjoint (provided that the labels $\xi$ and $\eta$ are real diffeomorphisms).

\subsection{Wavefunction realization}
\label{subsec:genaspectsquantum}

Let us now discuss the general representation theory for the canonical group $\wt G = \avira \rtimes \vira$. Since this group has the form of a semi-direct product between a locally compact separable group $G = \vira$ and an abelian group $\avira$, it is natural to employ Mackey's theory of induced representations~\cite{mackey1969induced}.
(We offer a brief guide to Mackey's theory in App.~\ref{app:mackey}.) It should be stressed, nonetheless, that Mackey's theory formally only applies to finite dimensional groups, so we must be mindful that the existence, irreducibility and exhaustivity of the representations produced by Mackey's algorithm are not guaranteed (see however \cite{mackey1963infinite}). 
We note that the representation theory of $\BMS$ has been studied recently 
\cite{oblak2017bms,oblak2015characters,barnich2014notes,barnich2015notes,campoleoni2016bms}, from Mackey's perspective, motivated by the important place of this group in the context of asymptotically-flat gravity in three dimensions.

According to Mackey's theory, an irreducible unitary representation of $\wt G = \avira \rtimes \vira$ is characterized by a choice of an orbit $\ca O$ on $\dvira$ (generated by the dual action of $\vira$ on $\dvira$, in this case the coadjoint action),  together with a choice of an irreducible unitary representation of the little group $H_{\ca O}$ associated with $\ca O$ (i.e., $\ca O \sim \vira/H_{\ca O}$). The Hilbert space $\ca H_{\ca O}$ is realized as the space of sections $\Psi$ of an associated bundle $\ca S \hookrightarrow \wt G \times_{\scr U} \ca S \rightarrow \ca O$,\footnote{A fiber bundle with total space $E$, base manifold $M$ and fibers $F$ is denoted by $F \hookrightarrow E \rightarrow B$. Note that the second arrow corresponds to the bundle projection map from $E$ to $B$, and the first (hooked) arrow only indicates that $F$ can be embedded into $E$ as a fiber (although this embedding is not canonical).} where $\scr U : H_{\ca O} \rightarrow \text{Aut}(\ca S)$ is an irreducible unitary representation of the corresponding little group $H_{\ca O}$ on a ``little'' Hilbert space $\ca S$. 
Informally, we can think of this associated bundle as a space defined by gluing a copy of $\ca S$ to each point of $\ca O$, and accordingly think of the quantum states $\Psi$ as wavefunctions ``living'' on the orbit $\ca O$ valued in $\ca S$ --- thus $\ca S$, if non-trivial, describes some sort of ``internal states'' or, in the language of particle physics, ``intrinsic spin'' degrees of freedom.
We explain below how the group acts on this representation.

The group structure of $\avira \rtimes \vira$, expressed in \eqref{cangroupprodrule}, is that $\vira$ acts as the adjoint map on $\avira$. In the notation of \eqref{sdprodnotation},  $\delta : \vira \rightarrow \text{Aut}(\avira)$ is
\be
\delta_{\wh \psi} \wh\eta := \ad_{\wh\psi} \wh\eta
\ee
where $\wh\psi \in \vira$ and $\wh\eta \in \avira$. The corresponding dual action on $\dvira$, denoted by $\wt\delta : \vira \rightarrow \dvira$, is precisely the coadjoint map,
\be
\wt\delta_{\wh\psi} \wt\alpha = \coad_{\wh\psi}\wt\alpha
\ee
where $\wt\alpha \in \dvira$. Given an element $\wt\alpha_0 \in \dvira$, let $\ca O$ denote its orbit, i.e., the set of all points $\wt\delta_{\wh\psi}\wt\alpha_0 = \coad_{\wh\psi}\wt\alpha$ in $\dvira$, for all $\wh\psi \in \vira$. The associated little group is
\be
H_{\ca O} := \{ \wh\chi \in \vira,\, \coad_{\wh\chi}\wt\alpha_0 = \wt\alpha_0 \}
\ee
Consequently, $\ca O$ is homeomorphic to $\vira/H_{\ca O}$. 
Now consider the principal bundle $H_{\ca O} \hookrightarrow \vira \rightarrow \ca O$, in which the projection map $q: \vira \rightarrow \ca O \sim \vira/H_{\ca O}$ is simply the group quotient, $q(\wh\psi) = [\wh\psi]$, where $[\wh\psi] = [\wh\psi \wh\chi]$ for all $\wh\chi \in H_{\ca O}$. More concretely, the quotient map can also be realized as $\wh\psi \mapsto \coad_{\wh\psi}\wt\varepsilon$. Given an irreducible unitary representation $\scr U : H_{\ca O} \rightarrow \text{Aut}(\ca S)$ of $H_{\ca O}$ on some vector space $\ca S$, the associated bundle $\ca S \hookrightarrow \vira \times_{\scr U} \ca S \rightarrow \ca O$ is vector bundle over $\ca O$ with fibers $\ca S$, defined as follows: let the total space be the set of equivalence classes
\be\label{ViraxUSdef}
\vira \times_{\scr U} \ca S := \left\{ [\wh\psi, \varsigma] = [\wh\psi \wh\chi, \scr U(\wh\chi^{-1})\varsigma];\text{ where $\wh\psi \in \vira$, $\wh\chi \in H_{\ca O}$ and $\varsigma \in \ca S$} \right\}
\ee
and the projection map $q^{(\scr U)} : \vira \times_{\scr U} \ca S \rightarrow \ca O$ be 
\be
q^{(\scr U)}([\wh\psi, \varsigma]) := q(\wh\psi) = [\wh\psi]
\ee
This bundle has a natural linear structure defined by
\be\label{ViraxUlinear}
[\wh\psi, \lambda \varsigma + \lambda' \varsigma'] =\lambda [\wh\psi, \varsigma] + \lambda' [\wh\psi, \varsigma']
\ee
Also, the $\vira$ action on $\ca O$ has a natural lift to a $\vira$ action on the associated bundle, denoted by $L : \vira \rightarrow \text{\sl Diff}(\vira \times_{\scr U} \ca S)$ and defined by
\be\label{Lliftdef}
L_{\wh\phi} [\wh\psi, \varsigma] := [\wh\phi\wh\psi, \varsigma]
\ee
where $\wh\phi \in \vira$.
Finally, assume that $\ca O$ admits a measure $\mu$ that is quasi-invariant under $\vira$.\footnote{If $f : M \rightarrow N$ is a measurable map and $\mu$ is a measure on $M$, the push-forward of $\mu$ to $N$ is defined
by $f_*\mu[B] := \mu[f^{-1}(B)]$, 
where $B$ is any Borel subset of $N$ and $f^{-1}$ designates the pre-image under $f$. A measure $\mu$ on a homogeneous space $G/H$ is said to be {\sl quasi-invariant} with respect to $G$ if $\mu$ and its push-forwards $\mu_{g} := \delta_{g *} \mu$, for all $g \in G$, are equivalent to each other (i.e., have the same sets of measure zero).}
Representations defined for equivalent measures (i.e., having the same sets
of measure zero) are unitarily equivalent, and under quite general conditions a homogeneous space admits a unique (up to equivalence) quasi-invariant measure (see footnote \ref{fn:uniquequasiinv}).
In App.~\ref{app:topologyQ} we discuss a possible quasi-invariant measure for the orbit $\ca O = \ca Q$.
The Hilbert space $\ca H$ is defined as the space of sections of the associated bundle,
\be
\ca H := \Gamma(\vira \times_{\scr U} \ca S) := \{ \Psi : \ca O \rightarrow \vira \times_{\scr U} \ca S; \text{ satisfying $q^{(\scr U)}(\Psi(\wt\a)) = \wt\alpha$ for all $\wt\alpha \in \ca O$}\} 
\ee
with the inner product given by
\be
\langle \Psi , \Psi' \rangle := \int_{\ca O} d\mu(\wt\alpha) \lla \Psi(\wt\alpha), \Psi'(\wt\alpha) \rra := \int_{\ca O} d\mu(\wt\alpha) \left( \varsigma(\wt\alpha), \varsigma'(\wt\alpha) \right)
\ee
where $\Psi(\wt\alpha) = [\wh\psi(\wt\alpha), \varsigma(\wt\alpha)]$, $\Psi'(\wt\alpha) = [\wh\psi(\wt\alpha), \varsigma'(\wt\alpha)]$ and $(\,,)$ denotes the inner product in $\ca S$. (Note that $\Psi$ and $\Psi'$ must be expressed in terms of the same $\wh\psi \in q^{-1}(\wt\alpha)$ in this formula.)
The irreducible unitary representation produced by this construction, $U : \avira \rtimes \vira \rightarrow \text{Aut}(\ca H)$, is then
\be\label{MackeyRep}
 U(\wh\eta, \wh\phi) \Psi (\wt\alpha) = e^{-i \wt\alpha(\wh\eta)} \sqrt{\frac{d\mu_{\wh\phi}}{d\mu}(\wt\a)} \, L_{\wh\phi}\! \left(\Psi(\coad_{\wh\phi^{-1}}\wt\a)\right) 
\ee
where $(\wh\eta, \wh\phi) \in \avira \rtimes \vira$ and $d\mu_{\wh\phi}/d\mu$ is the Radon-Nikodym derivative associated with the measure $\mu$. 
The form of this representation is compatible with our intuition in the sense that $\avira$ corresponds classically to momentum translations and is quantum-mechanically represented as a pointwise phase rotation of the wavefunction, and $\vira$ corresponds classically to configuration translations and is quantum-mechanically represented as translations of the wavefunction (i.e., notice that $L_{\wh\phi}$ maps $\Psi(\coad_{\wh\phi^{-1}}\wt\a)$, which belongs to the fiber over $\coad_{\wh\phi^{-1}}\wt\a$, to a point on the fiber over $\coad_{\wh\phi}(\coad_{\wh\phi^{-1}}\wt\a) = \wt\a$.)

Let us see how the conditions from the Casimir matching principle manifests here. 
Considering the group element $(t\wh c; \wh I)$, where $t \in\bb R$ and $\wh I = (I, 0)$ is the identity of $\vira$, formula \eqref{MackeyRep} reads
\be
 U(t\wh c; \wh I) \Psi (\wt\alpha) = e^{-i t \wt\alpha(\wh c)} L_{\wh I}\! \left(\Psi(\wt\a)\right) = e^{-i t \wt\alpha(\wh c)} \Psi(\wt\a)
\ee
since $L_{\wh I}$ acts trivially on the bundle. Comparing with \eqref{CasimircondQ} we conclude that
\be
\wt\a(\wh c) = 1
\ee
that is, the orbit $\ca O$ must be chosen so that its central component is $1$. 
Now consider the central element of $\vira$, $(0;(0,r))$, where $r \in\bb R$.
Notice that since it acts trivially via the coadjoint map on $\dvira$, it always appear in $H_{\ca O}$ for any orbit. That is, $H_{\ca O}$ is always a central extension of the corresponding little group $K_{\ca O}$ of $\diff$, $H_{\ca O} = K_{\ca O} \timesext \bb R$, and moreover $\ca O \sim \vira/H_{\ca O} = \diff/K_{\ca O}$.
According to \eqref{MackeyRep} we have
\ba
 U(0; (0,r)) \Psi (\wt\alpha) &=  L_{(I, r)}\! \left(\Psi(\wt\a)\right) \no
&= [(I,r) \wh\psi(\wt\a), \varsigma(\wt\a)] \no
&= [\wh\psi(\wt\a) (I,r), \varsigma(\wt\a)] \no
&= [\wh\psi(\wt\a), \scr U((I,r)) \varsigma(\wt\a)] \no
&= [\wh\psi(\wt\a), e^{i\lambda r} \varsigma(\wt\a)] \no
&= e^{i\lambda r}[\wh\psi(\wt\a), \varsigma(\wt\a)] \no
&= e^{i\lambda r} \Psi(\wt\a)
\ea
where in the first line we used that $(0;(0,r))$ acts trivially on $\ca O$ (and, in particular, the Radon-Nikodym derivative is $1$); in the second line we wrote $\Psi(\wt\a) = [\wh\psi(\wt\a), \varsigma(\wt\a)]$ (where, technically, $\wt\a \mapsto \wh\psi(\wt\a)$ is a {\sl local} section) and used the definition \eqref{Lliftdef} of the lifted action; in the third line we used that the central element commutes with any other; in the fourth line we used that $(0;(0,r)) \in H_{\ca O}$ so the equivalence relation \eqref{ViraxUSdef} defining the associated bundle can be applied; in the fifth line we used that $(0;(0,r))$, being central, must be unitarily represented as $e^{i\lambda r}$ for some $\lambda \in \bb R$; in the sixth line we used the linear property \eqref{ViraxUlinear} of the associated bundle; and in the last line we simply returned to the $\Psi$ notation. The condition \eqref{CasimircondP} thus implies that $\lambda = 0$, 
\be
\scr U((I, r)) = 1
\ee
i.e., the central factor of $H_{\ca O}$ must be represented trivially on $\ca S$. 
In summary, the Casimir matching principle applied to the central elements $T$ and $R$ implies that the representations of the canonical group should be restricted to those where the orbit $\ca O$ has unit central component in $\dvira$ and the central factor of $H_{\ca O} = K_{\ca O} \timesext \bb R$ is trivially represented on $\ca S$. Formula \eqref{MackeyRep} then reduces to
\be\label{MackeyRep2}
U((\eta, t); (\phi,r)) \Psi (\wt\alpha) =  e^{-i t} e^{-i \alpha(\eta)}  \sqrt{\frac{d\mu_{\phi}}{d\mu}(\wt\a)} \, L_{\phi}\! \left(\Psi(\coad_{\phi^{-1}}\wt\a)\right) 
\ee
where $\wt\a = \alpha(\theta) d\theta^2 + \wt c$ and, as usual, $\alpha(\eta) = \int d\theta\, \alpha(\theta) \eta(\theta)$. Notice that now only $\phi \in \diff$ (and not the central component of $\wh\phi$) appears in the Radon-Nikodym derivative, $L$ and $\coad$.

From \eqref{QdeffromU} and \eqref{PdeffromU} we can compute the action of the quantized versions of the (non-central) canonical charges, $\wh Q_{\eta}$ and $\wh P_{\xi}$, on the wavefunctions. 
For the configuration charges we have
\ba
\wh Q_{\eta} \Psi (\wt\a) &= i \frac{\partial}{\partial \lambda} U(\exp(\lambda(\eta,0); 0))) \Psi (\wt\a) \no
&= i \frac{\partial}{\partial \lambda} e^{-i\lambda \alpha(\eta)} \Psi(\wt\a) \no
&= \a(\eta) \Psi(\wt\a) \label{whQwaveaction}
\ea
where the $\lambda$ derivatives, here and next, are evaluated at $\lambda = 0$. 
For the momentum charges we have
\ba
\wh P_{\xi} \Psi(\wt\a) &= i \frac{\partial}{\partial \lambda}  U(\exp(\lambda(0; \xi))) \Psi(\wt\a) \no
&= i \frac{\partial}{\partial \lambda} \left[\sqrt{\frac{d\mu_{\exp(\lambda\xi)}}{d\mu}(\wt\a)}  L_{\exp(\lambda\xi)}\! \left(\Psi(\coad_{\exp(\lambda\xi)^{-1}}\wt\a)\right) \right] \no
&= - i \scr D_\xi \Psi(\wt\a) \label{whPwaveaction}
\ea
where $\scr D_\xi$ is a derivative operator defined as
\be
\scr D_\xi \Psi(\wt\a) := - \left. \frac{\partial}{\partial \lambda} \left[\sqrt{\frac{d\mu_{\exp(\lambda\xi)}}{d\mu}(\wt\a)}  L_{\exp(\lambda\xi)}\! \left(\Psi(\coad_{\exp(\lambda\xi)^{-1}}\wt\a)\right) \right] \right|_{\lambda = 0}
\ee
As expected from a derivative, this operator is linear under addition, $\scr D_\xi (\Psi + \Psi') = \scr D_\xi\Psi + \scr D_\xi \Psi'$, and satisfies a form of Leibniz rule under multiplication by scalars, $\scr D_\xi (f \Psi) = f \scr D_\xi + (X_\xi f) \Psi$, where $f : \ca O \rightarrow \bb C$ and $X_\xi$ is the vector field on $\ca O$ induced by $\xi \in \adiff$. Moreover, they form an anti-representation of $\diff$, $[\scr D_\xi , \scr D_{\xi'} ] = - \scr D_{[\xi, \xi']}$.

Finally, note that we should in principle consider the universal cover of the group in order to describe all projective representations.
The group $\wt G = \avira \rtimes \vira$ has fundamental group $\bb Z$, due to the $\vira$ factor inheriting it from $\diff$, according to our definitions in Sec.~\ref{subsec:Vira}.
As described in App.~\ref{app:topologyQ}, the universal cover of $\diff$, $\un\diff$, can be characterized 
as the space of diffeomorphisms of the real line, $\un\psi : \bb R \rightarrow \bb R$, satisfying the condition 
\be
\un\psi(\t +2\pi) = \un\psi(\t) + 2\pi
\ee
Thus, elements of the universal cover of $\vira$, $\un\vira$, can be expressed as
\be
\wh{\un\psi} = (\un\psi, x) \,\in\, \un\vira
\ee
The center $\ca Z$ of $\un\diff$ consists of functions $\un\psi(\theta) = \theta + 2\pi n$, $n \in \bb Z$, all of which project to the identity in $\diff$. Let us indicate the quotient $\un\diff \rightarrow \un\diff/{\ca Z} = \diff$ simply by $\un\psi \mapsto \psi$. 
Similarly, the center of $\un\vira$ is $\ca Z \times \bb R$, i.e., consisting of elements $\wh{\un\psi} = (\theta + 2\pi n, r)$. 
We can describe the relationship between these two extensions (i.e., one by the 2-coclycle $\bb R$ and other by the fundamental group $\bb Z$) in terms of a commutative diagram
\begin{center}
\begin{tikzcd}
\wh{\un\psi} \in \un\vira \arrow{d}[swap]{\sfrac{}{\bb Z}} \arrow{r}{\sfrac{}{\bb R}} & \un\psi \in \un\diff \arrow{d}{\sfrac{}{\bb Z}} \\
\wh\psi \in \vira  \arrow{r}[swap]{\sfrac{}{\bb R}} & \psi \in \diff
\end{tikzcd}
\end{center}
where the arrows indicate group projections (or, in the reverse direction, group extensions).
The diagram also helps elucidating the notation, i.e., see that ``hat'' denotes central extensions by 2-cocyles and ``underline'' denotes unwrapping by the fundamental group $\bb Z$ (the two accents are independent).

Evidently, the center $\ca Z \times \bb R \subset \un\vira$ acts trivially through the coadjoint map from $\un\vira$ to $\dvira$, so
\be
\coad_{\wh{\un\psi}}  = \coad_{\wh\psi} = \coad_\psi
\ee
where the coadjoint map on the right-hand side is just the one from $\vira$ to $\dvira$, given in \eqref{coadpsiexpl}.
This means that the coadjoint orbits of $\un\vira$ are exactly the same as the ones of $\vira$, the only difference being that the little group for $\un\vira$ of an orbit $\ca O$, denoted by $\un H_{\ca O}$, is some $\bb Z$-cover of the little group $H_{\ca O}$ for $\vira$.
Therefore, the wavefunctions are still based on the same set of orbits, but they may carry ``internal indices'' in some {\sl projective} unitary irreducible representations of the little group $H_{\ca O}$. 
In summary, the wavefunctions describing a projective unitary irreducible representations are sections of the bundle $\ca S \hookrightarrow \un\vira \times_{\scr U} \ca S \rightarrow \ca O$, where 
$\scr U : {\un H}_{\ca O} \rightarrow \text{Aut}(\ca S)$ is a unitary irreducible representation of the little group ${\un H}_{\ca O}$.

\subsection{The monodromy and winding number Casimirs}
\label{subsec:monocasi}

If there are any Casimir invariants associated with the canonical algebra of observables, we can use the Casimir matching principle (explained at the end of Sec.~\ref{subsec:isham}) to filter out some representations. 
We have seen that the central $T$ and $R$ elements of $\wt G$, which are the simplest Casimir invariants, imply that the orbits must be restricted to those with central component equal to $1$ ($\wt\alpha = \a(\theta) d\t^2 + \wt c$) and the central $\bb R$ factor of the little group must be trivially represented. 
In this section we discuss a family of non-trivial Casimir invariants of $\bms$ associated with the monodromy structure of coadjoint orbits of Virasoro.

Note that a  classical observable that depends only on the $Q_n$  charges, $C(\{Q_n\})$, can be unambiguously quantized to an operator $\wh C := C(\{\wh Q_n\})$, as there are no operator-ordering issues by virtue of the commutativity of the $Q_n$'s. Furthermore, \eqref{whQwaveaction} implies that this operator acts on the wavefunctions by multiplication,
\be
\wh C \Psi (\wt\a) := C[\{\wh Q_n \}] \Psi (\wt\a) = C[\{Q_n (\wt\a) \}] \Psi (\wt\a)
\ee
where $Q_n(\wt\a) = \a(e^{in\theta} \partial_\t) = \int d\theta\, e^{in\theta} \a(\theta) =: 2\pi\a_n$. These numbers $\a_n$ are of course just the Fourier components of $\wt\a$,
\be
\wt\a = \sum_{n\in \bb Z} \a_n e^{-in\theta} d\theta^2 + \wt c
\ee
In particular, note that if we define the following $\ddiff$-valued operator,
\be
\fr  Q := \frac{1}{2\pi}\sum_{n\in \bb Z} \wh Q_n e^{-in\theta} d\theta^2 + \wt c
\ee
we get, in a formal sense,
\be
\fr Q \Psi (\wt\a) = \wt\a \Psi (\wt\a)
\ee
whose meaning is that, given any function $C : \ddiff \rightarrow \bb R$, we can associate to it a quantum operator $\wh C$ defined by
\be
\wh C := C(\fr Q)
\ee
and it follows that
\be
\wh C \Psi (\wt\a) := C(\wt\a) \Psi (\wt\a)
\ee
Note that such an operator is defined in {\sl any} irreducible representation of $\avira \rtimes \vira$ in which $T = 1$. 

If one finds a function $C : \ddiff \rightarrow \bb R$ that is constant on every orbit $\ca O \subset \dvira$, i.e.,
\be
C(\wt\a) = C(\coad_\psi \wt\a)\,,\,\text{ for all $\psi \in \diff$}
\ee
which is therefore a classical Casimir observable (by restricting it to $\ca O = \ca Q$), 
then $\wh C = C(\fr Q)$ will also be a quantum Casimir operator. 
In fact, a family of such functions exists \cite{oblak2017bms} and is given by the trace of the $k$-th power of the {\sl monodromy matrix}, $\bf M$. 
More explicitly, the function $C_k$ is taken as
\be
C_k(\wt\a) :=  \tr[{\bf M}^k] =  \tr \left[ \scr P\!\exp \int_0^{2\pi}\!d\theta\, k
\begin{pmatrix}
0 & 1 \\
-\frac{\a(\theta)}{4} & 0
\end{pmatrix}
\right]
\ee
where $k \in \bb N$, $\wt\a = \a(\theta) d\theta^2 + \wt c$, and $\tr$ and $\scr P\!\exp$ are respectively the trace and the path ordered exponential (on the space of $2\!\times\!2$ matrices).
We shall refer to this Casimir, which in the quantum theory is given by
\be
\wh C_k := C_k(\fr Q) = \tr \left[ \scr P\!\exp \int_0^{2\pi}\!d\theta\, k
\begin{pmatrix}
0 & 1 \\
-\frac{\sum_{n\in \bb Z} \wh Q_n e^{-in\theta}}{8\pi} & 0
\end{pmatrix}
\right]
\ee
as the {\sl Monodromy operator} of $k$-th order.\footnote{To be more precise, $\wh C$ is not a standard Casimir of $\avira \rtimes \vira$, because $T$ is replaced by the value $1$. The ``closest'' one could get to defining a Casimir element of the universal enveloping algebra of $\avira \rtimes \vira$ would be to replace $\frac{1}{8\pi}\sum_{n\in \bb Z} \wh Q_n e^{-in\theta}$ by $\frac{1}{8\pi T} \sum_{n\in \bb Z} \wh Q_n e^{-in\theta}$, which does not make sense since one cannot divide by the abstract symbol $T$. Nevertheless, this operator deserves to be called a ``quantum Casimir'' since  it is the direct quantization of the classical Casimir $C[\{Q_n\}]$, is 
well-defined in any physically relevant irreducible representation (i.e., those where $T = 1$), and is represented as a multiple of the identity in each of these representations.}

Let us recall the how the monodromy matrix is defined. (See, e.g., \cite{oblak2017bms} for details.)  Associated to each element $\wt\a = \a(\t)d\t^2 + a\wt c \in \dvira$ there is a Hill's operator $\scr H_{\wt\a} := 4a \frac{\partial^2}{\partial\theta^2} + \a(\theta)$ acting on the space of real densities of weight $-1/2$ on $\bb R$ (i.e., objects that transform under diffeomorphisms like $F(\theta) = f(\theta) (d\theta)^{-\frac{1}{2}}$, where $f$ is a scalar). The space of solutions of $\scr H_{\wt\a} F = 0$ forms a two-dimensional space, which can be displayed as a vector 
\be
{\bf F} := \begin{pmatrix} F_1 \\ F_2 \end{pmatrix}
\ee
It is assumed that these solutions are normalized with respect to the Wronskian, $W(F_1, F_2) := F_1 F_2' - F_2 F_1' = -1$.
These solutions are not $2\pi$-periodic in general, but due to the $2\pi$-periodicity of $\a(\theta)$ it follows that a $2\pi$-translation corresponds to a change of basis in the space of solutions, i.e., there exists a matrix ${\bf M} \in \text{SL}(2,\bb R)$ such that
\be
{\bf F}(\theta + 2\pi) = {\bf M} {\bf F}(\theta)
\ee
This ${\bf M}$ is called the monodromy matrix associated with $\wt\a$ in the basis ${\bf F}$.
A change of basis ${\bf F} \mapsto {\bf S} {\bf F}$, where ${\bf S} \in \text{SL}(2,\bb R)$ (so that the Wronskian norm is preserved), induces a change ${\bf M} \mapsto {\bf S} {\bf M} {\bf S}^{-1}$. Consequently, there is a basis-independent map from $\dvira$ to the space of conjugacy classes of $\text{SL}(2,\bb R)$, $\wt\a \mapsto [{\bf M}]$. Moreover, it is true that $\scr H_{\wt\a} F = 0 \Leftrightarrow \scr H_{\coad_\psi \wt\a}(\psi_* F) = 0$, for any $\psi \in \diff$. It follows that this map from $\dvira$ to conjugacy classes of $\text{SL}(2,\bb R)$ is constant along each orbit $\ca O$ and, in particular, the traces of powers of ${\bf M}$ are real functions constant on the coadjoint orbits of Virasoro.

For an orbit with constant representative $\wt\a_0 = \a_0 d\theta^2 +\wt c$ this expression evaluates to
\be
C_k(\wt\a_0) = 2\cos \left(k\pi \sqrt{\alpha_0} \right)
\ee
For the orbit $\ca O = \ca Q$, the representative is, as given in \eqref{alphaorbit}, $\wt\varepsilon = d\theta^2 + \wt c$, and therefore it is associated with the Casimir value
\be
C_k(\wt\varepsilon) = 2 (-1)^k
\ee
The Casimir matching principle thus restricts the possible quantum theories to those associated with orbits whose $k$-th power of the monodromy matrix has trace $2 (-1)^k$. This leaves three possibilities for the $\text{SL}(2,\bb R)$ conjugacy classes of ${\bf M}$, whose representatives are
\[
\begin{pmatrix}
-1 & 0 \\
0 & -1
\end{pmatrix}
\,,\,\,
\begin{pmatrix}
-1 & 1 \\
0 & -1
\end{pmatrix}
\,,\,\,
\begin{pmatrix}
-1 & -1 \\
0 & -1
\end{pmatrix}
\]
It appears that there are no other {\sl scalar} matrix-invariants that can be used to distinguish these cases. 

It is possible to extend the Casimir matching principle to include other types of invariant objects, which we call {\sl generalized Casimir operators}.\footnote{A possible formulation of a more generally applicable Casimir matching principle is as follows. Suppose that classically there is a function of the canonical observables, $C(\{H_{\wt\xi}\})$, valued in some space $\scr S$, which is constant on the phase space. Given any function $f:\scr S \rightarrow \bb R$, it is clear that $f\circ C$ is constant on the phase space and therefore it is classical Casimir observable. Say that $C$ admits a natural quantization $\wh C$ in the sense that, given any function $f:\scr S \rightarrow \bb R$, there is a naturally associated quantum operator $\wh C_f := f \circ C(\wh H_{\wt\xi})$ that commutes with all canonical observables, $[\wh C_f, \wh H_{{\wt\xi}}] = 0$, and is compatible with composition by real functions, $\wh C_{\varphi \circ f'} = \varphi(\wh C_{f'})$, where $\varphi : \bb R \rightarrow \bb R$. Each $\wh C_f$ is therefore a quantum Casimir and must be irreducibly represented as a multiple of the identity. If, in a given representation, there exists $\fr c \in \scr S$ such that $\wh C_f = f(\fr c)$, for all $f$, then we say that $\wh C$ has a $\scr S$-eigenvalue $\fr c$ in that representation. It is then possible, in principle, to measure the $\scr S$-value of $\wh C$, and the result of such measurement is the same for all quantum states, so we assume that this value should be matched with the classical $\scr S$-value of $C$.} 
Let us define the {\sl monodromy class operator}, ${\fr M}$, valued in the space of conjugacy classes of $\text{SL}(2,\bb R)$, by
\be
{\fr M} := \left[ \scr P\!\exp \int_0^{2\pi}\!d\theta\,
\begin{pmatrix}
0 & 1 \\
-\frac{\sum_{n\in \bb Z} \wh Q_n e^{-in\theta}}{8\pi} & 0
\end{pmatrix}
\right]
\ee
This generalized operator is well-defined, both mathematically and physically: each entry of the $2\!\times\!2$ matrix is a function of the $\wh Q_n$'s, which are commuting operators and can therefore be simultaneously diagonalized (or measured), and then the $\text{SL}(2,\bb R)$ conjugacy class of these ``eigenmatrices'' can be evaluated. 
In fact, the wavefunctions $\Psi(\wt\a)$ are precisely the eigenvectors of this operator, and we can formally write
\be
\fr M \Psi(\wt\a) = [{\bf M}_{\wt\a}]\Psi(\wt\a)
\ee
The corresponding classical invariant is simply the $\text{SL}(2,\bb R)$ conjugacy class of the monodromy matrix, ${\bf M}$, which in the case of $\ca Q = \diff/\psl$ evaluates to
\be
[{\bf M}_{\ca Q}] = \left[ \begin{pmatrix}
-1 & 0 \\
0 & -1
\end{pmatrix} \right]
\ee
Thus we assume that the quantum theory should realize ${\fr M} = [{\bf M}_{\ca Q}]$, 
which selects the orbits $\ca O$ whose monodromy class is
\be
[{\bf M}_{\ca O}] = \left[ \begin{pmatrix}
-1 & 0 \\
0 & -1
\end{pmatrix} \right]
\ee
The only coadjoint orbits in this monodromy class are those with the following constant representatives \cite{oblak2017bms}
\be\label{wtalphan}
\wt\varepsilon^{(n)} := n^2 d\t^2 + \wt c \,,\,\, \text{where $n \in 2\bb Z + 1$}
\ee
These orbits have topology $\diff/\text{PSL}^{(n)}(2, \bb R)$, where $\text{PSL}^{(n)}(2, \bb R)$ is the subgroup of $\diff$ generated by the subalgebra $\partial_\t$, $\sin(n\theta)\partial_\t$ and $\cos(n\theta)\partial_\t$.

There is another invariant associated with coadjoint orbits of Virasoro, the {\sl winding number} \cite{oblak2017bms}. 
This is a discrete parameter also associated with Hill's equation. Given a (Wronskian-normalized) solution vector ${\bf F}$, define the map 
\ba
&[0, 2\pi) \rightarrow \bb{RP}^1 \no
&\theta \mapsto [{\bf F}(\theta) \sim \lambda {\bf F}(\theta);\, \lambda \in \bb R]
\ea
which should be seen  as going from $S^1$ (the $[0, 2\pi)$ interval of $\bb R$) to $S^1$ (identified with $\bb{RP}^1$).
The number of {\sl complete} wrappings of the domain $[0, 2\pi)$ into the codomain $\bb{RP}^1$ induced by this map is called the winding number, $w$. By definition, $w \in \bb Z$. The winding number is independent of the basis of solutions ${\bf F}$ and invariant under the coadjoint action, $w_{\wt\a} = w_{\coad_\psi\wt\a}$. Therefore, $w$ is constant along each coadjoint orbit. 
Classically, the winding number of $\ca Q$ is $1$.
Quantum mechanically the winding number operator, $\fr w$, can be constructed solely from the $\wh Q_n$'s, and it acts on wavefunctions as
\be
\fr w \Psi(\wt\a) = w_{\wt\a} \Psi(\wt\a)
\ee
From the (generalized) Casimir principle, we conclude that the representations should be restricted to those with orbits with unit winding number,
\be
w_{\ca O} = 1
\ee
From the family of orbits in \eqref{wtalphan}, the only candidate that has winding number $1$ is $\wt\varepsilon^{(1)} = \wt\varepsilon = d\t^2 + \wt c$. 
That is, we conclude that the quantum theory should be based on the orbit $\ca O = \ca Q = \diff/\psl$, which we call the natural orbit.

\subsection{The natural orbit and $\psl$ indices}
\label{subsec:naturalorbit}

The natural choice for the orbit is 
\be
\ca O = \diff/\psl
\ee
so that the wavefunctions ``live'' on the configuration space (i.e., their domain is $\ca Q$). 
This was justified in the previous section from the Casimir matching principle, using the central element $T = 1$, the monodromy operator $\fr M$ and winding number operator $\fr w$. 
The little group is $K = \psl$,
so that the wavefunctions carry an internal index in some (projective) irreducible unitary representations of $\psl$.

The representation theory of $\psl$ is well-known \cite{bargmann1947irreducible,gel1947unitary,harish1952plancherel,kitaev2017notes}, so we briefly recapitulate it here to set up the notation. 
First note that since $\psl$ is non-compact simple Lie group, the only finite-dimensional irreducible unitary representation is the trivial one \cite{hilgert2011structure}. The trivial representation would produce a quantization of the causal diamonds characterized by $\bb C$-valued wavefunctions living on $\ca Q$, which undoubtedly would be the ``simplest guess'' for the Hilbert space for the phase space $T^*\ca Q$. 
The other possibilities unveiled by the general quantization thus correspond to wavefunctions carrying infinitely many ``internal states''. 
The $\apsl$ subalgebra is generated by the elements $\partial_\t$, $e^{i\t} \partial_\t$ and $e^{-i\t} \partial_\t$ of $\adiff$, so in an irreducible unitary representation, $\scr U$, the respective self-adjoint generators,
$\upsilon_0$, $\upsilon_+$ and $\upsilon_-$,\footnote{More precisely, $\scr U(\exp((c_0 + c_+ e^{i\t}+ c_- e^{-i\t})\partial_\t)) = e^{-i(c_0 \upsilon_0 + c_+ \upsilon_+ + c_- \upsilon_-)}$.} will satisfy the algebra
\ba
[\upsilon_{0}, \upsilon_{+}] &= \upsilon_{+}  \no
[\upsilon_{0}, \upsilon_{-}] &= - \upsilon_{-} \no
[\upsilon_{-}, \upsilon_{+}] &= 2\upsilon_{0}
\ea
There is one (independent) Casimir operator
\be
\zeta := \upsilon_{0}^2 - \frac{1}{2} \left( \upsilon_{-}\upsilon_{+} + \upsilon_{+}\upsilon_{-} \right)
\ee
which must be represented as a multiple of the identity,
\be
\zeta = \mu^2 - \frac{1}{4}
\ee
where $\mu^2 \in \bb R$, and $\mu$ serves as a label for the representation. Now, as $\upsilon_0$ is self-adjoint, we can assume that it has at least one eigenvector $|j;\mu\ra$ (possibly in the weak sense, i.e., as a non-normalizable limit of a sequence of vectors) with eigenvalue $j \in \bb R$,
\be
\upsilon_0 |j; \mu\ra = j |j; \mu \ra
\ee
Since $\upsilon_0$ generates the $SO(2)$ subgroup of $\psl$, we must have, in a {\sl true} representation,
\be\label{psltrue}
e^{i2\pi \upsilon_0} = 1
\ee
so $j \in \bb Z$. We will later comment on its projective representations. 
From the algebra we see that $\upsilon_{\pm}$, if non-trivial, act as ladder operators for $\upsilon_0$, i.e., $(\upsilon_{\pm})^k |j; \mu\ra \propto |j \pm k; \mu\ra$. This is compatible with the previous observation that the spectrum of $\upsilon_0$ is a subset of $\bb Z$. Using this ladder structure to build a basis we have
\be\label{pslirrep}
\upsilon_{\pm} |j; \mu\ra = \left(j \pm \left(\mu + \frac{1}{2}\right) \right) |j \pm 1; \mu\ra
\ee
where the vectors $|j;\mu\ra$ are not necessarily normalized.\footnote{Since the basis is discrete, these vectors can be normalized. By computing $\big|\!\big| \upsilon_+ |j;\mu\ra \big|\!\big|^2 = \la j; \mu| \upsilon_- \upsilon_+ |j;\mu\ra$ with formula \eqref{pslirrep}, one finds the relation 
\[
\left( j + \frac{1}{2} - \mu \right) \la j; \mu |j;\mu\ra =  \left( j + \frac{1}{2} + \mu \right) \la j+1; \mu|j+1;\mu\ra
\] \label{fnpslnorm}}
We have the following exhaustive list of possibilities:
\begin{enumerate}
\item \emph{Trivial representation:} If $\mu = -1/2$ (so $\zeta = 0$) and $0 \in \text{\sl Spectrum}(\upsilon_0)$, then all the generators annihilate $|0;\mu\ra$. This is one-dimensional and, in fact, the only finite-dimensional unitary representation.

\item \emph{Discrete series representation:} If $k := \mu + 1/2 \in \bb Z - \{0\}$, then notice that states with $j = \pm k$ will be annihilated by $\upsilon_{\mp}$. Thus, for each $k$, there are two distinct representations: one where $\upsilon_-|k;\mu\ra = 0$, so $\text{\sl Spectrum}(\upsilon_0) = \{k,\, k+1, \,\ldots\}$; and another where $\upsilon_+|-k;\mu\ra = 0$, so $\text{\sl Spectrum}(\upsilon_0) = \{\ldots,\, -k-1,\, -k\}$. 

\item \emph{Limit of discrete series representation:} If $k := \mu + 1/2 = 0$ and $0 \notin \text{\sl Spectrum}(\upsilon_0)$, then there are two distinct representation: one where $\upsilon_-|1;\mu\ra = 0$, so $\text{\sl Spectrum}(\upsilon_0) = \{1,\, 2, \,\ldots\}$; and another where $\upsilon_+|-0;\mu\ra = 0$, so $\text{\sl Spectrum}(\upsilon_0) = \{\ldots,\, -2,\, -1\}$. Notice that, unlike the previous case, the spectrum does not start from $k$.

\item \emph{Principal (spherical) series representation:} If $\mu^2 < 0$, then there is no minimal or maximum weight, i.e., $\text{\sl Spectrum}(\upsilon_0) = \bb Z$. The representations for $\mu$ and $-\mu$ are equivalent.

\item \emph{Complementary series representation:} If $\mu \in (-1/2,0) \cup (0, 1/2)$ this also defines a representation in which $\text{\sl Spectrum}(\upsilon_0) = \bb Z$. The reason that not all $\mu \in \bb R - \frac{1}{2}\bb Z$ is permissible is that, as can be seen from \eqref{pslirrep} and particularly footnote \eqref{fnpslnorm}, some states $|j;\mu\ra$ required to be included in the basis would have negative norm, so the representation would not be unitary. The representations for $\mu$ and $-\mu$ are equivalent.
\end{enumerate}

\noindent 
If we consider projective representations of the canonical group, where the $\vira$ factor is unwrapped to $\un\vira$, the little group of the natural orbit $\ca Q$ would become $\un K := \un\psl$, i.e., the universal cover of $\psl$, where the $SO(2) \sim S^1$ subgroup of $\psl$ is unwrapped to $\bb R$. 
In this way, the wavefunctions would also carry internal labels in {\sl projective} representations of $\psl$. 
The only modification in the analysis above is that condition \eqref{psltrue} should not be imposed \cite{kitaev2017notes}.
The consequence is that the spectrum of $\upsilon_0$ is shifted by a fixed parameter, i.e., $\text{\sl Spectrum}(\upsilon_0) = s + \bb Z$, where $s \in [0,1)$. Thus, together with $\mu$, $s$ is another label classifying the unitary irreducible representations, which have the form
\ba
&\upsilon_0 |j; \mu,s\ra = (j+s) |j; \mu,s\ra \no
&\upsilon_{\pm} |j; \mu, s\ra = \left(j + s \pm \left(\mu + \frac{1}{2}\right) \right) |j \pm 1; \mu, s\ra
\ea
with $j$ in (some subset of) $\bb Z$.
The trivial representation obviously only exists when $s=0$, but the other unitary irreducible representations listed above are described in a completely analogous manner for each $s$, only with the appropriate shifts in the conditions for $\mu$. For example, in the discrete series representation the condition would be $k = \mu +1/2 \mp s \in \bb Z -\{0\}$, thus in one case $\upsilon_-|k;\mu,s\ra = 0$, so $\text{\sl Spectrum}(\upsilon_0) = \{s+k,\, s+k+1, \,\ldots\}$; and in the other $\upsilon_+|-k;\mu,s\ra = 0$, so $\text{\sl Spectrum}(\upsilon_0) = \{\ldots,\, s-k-1,\, s-k\}$.
We anticipate that, from the analysis of the spin in Sec.~\ref{subsec:spinquantization}, together with the assumption that the quantum theory incorporates a CMC time-reversal symmetry, we find that only the $s=0$ and $s=1/2$ cases are allowed. This corresponds to restricting only to the double cover of $\psl$, i.e., $\text{SL}(2, \bb R)$.

Finally, let us discuss one simple manner to relate this representation of the little group with the canonical charges of the diamond, particularly, $P_0$, $P_1$ and $P_{-1}$. Given a wavefunction $\Psi(\wt\a)$,
let us study how the momentum operators act when evaluated at the conformal class of the identity, $[\psi] = [I]$, i.e., at the point $\wt\varepsilon = d\theta^2 + \wt c$ of $\ca Q \subset \dvira$. Write $\Psi(\wt\varepsilon) = [I, \varsigma]$, where $\varsigma \in \ca S$. The $\psl$ subgroup of $\diff$ stabilizes this point, so for any $\chi \in \psl$, formula \eqref{MackeyRep2} gives
\ba
U(0; (\chi,0)) \Psi (\wt\varepsilon) &=   L_{\chi}\! \left(\Psi(\wt\varepsilon)\right) \no
&= L_{\chi} [I, \varsigma] \no
&= [\chi, \varsigma] \no
&= [I, \scr U(\chi) \varsigma]
\ea
Writing $\chi = \exp[(c_0 + c_+ e^{i\t}+ c_- e^{-i\t})\partial_\t]$ and differentiating with respect to the coefficients, we obtain
\ba
&\wh P_0 \Psi(\wt\varepsilon) = [I, \upsilon_0 \varsigma] \no
&\wh P_1 \Psi(\wt\varepsilon) = [I, \upsilon_+ \varsigma] \no
&\wh P_{-1} \Psi(\wt\varepsilon) = [I, \upsilon_- \varsigma]
\ea
or, if we define $\Psi^{j;\mu,s}(\wt\varepsilon) := [I, |j;\mu,s\ra]$, we have
\ba
&\wh P_0 \Psi^{j;\mu,s}(\wt\varepsilon) = (j + s) \Psi^{j;\mu,s}(\wt\varepsilon) \no
&\wh P_1 \Psi^{j;\mu,s}(\wt\varepsilon) = \left(j + s + \left(\mu + \frac{1}{2}\right) \right)\Psi^{j+1;\mu, s}(\wt\varepsilon) \no
&\wh P_{-1} \Psi^{j;\mu,s}(\wt\varepsilon) = \left(j + s - \left(\mu + \frac{1}{2}\right) \right)\Psi^{j-1;\mu, s}(\wt\varepsilon) \label{PSLonPsi}
\ea
and also
\be
\wh\zeta \Psi^{j;\mu,s}(\wt\varepsilon) = \left( \mu^2 - \frac{1}{4} \right) \Psi^{j;\mu,s}(\wt\varepsilon)
\ee
where
\be
\wh\zeta := \wh P_{0}^2 - \frac{1}{2} \left( \wh P_{-1} \wh P_1 + \wh P_1 \wh P_{-1} \right)
\ee
Note that $\wh\zeta$ is not a Casimir for the canonical group $\wt G$, and therefore we cannot use the Casimir matching principle to deduce its value. 

We emphasize that the relations above, connecting the action of the $\apsl$ momenta to the little group representation $\scr U$, are valid on any wavefunction but only when evaluated at the specific point $\wt\a = \wt\varepsilon$ (i.e., $[\psi] = [I]$). 
Nevertheless, for wavefunctions localized at the conformal class of the identity, $[I]$, the formulas above are true operator actions (i.e., valid for all $\wt\a \in \ca O$). More precisely, consider the state $\Psi_{[I]}^{j;\mu,s}$ defined as
\be\label{PsiI}
\Psi_{[I]}^{j;\mu,s}(\wt\a) := \left\{
\begin{array}{ll}
\big[\wh\psi(\wt\a), \delta\big(\wt\a, \wt\varepsilon\big) |j; \mu,s\ra\big] & \text{for $\wt\a \in \ca V$} \\
0 &  \text{for $\wt\a \notin \ca V$} 
\end{array}\right.
\ee
where $\ca V$ is any open neighborhood  of $[I] \in \ca Q$, $\wh\psi: \ca V \rightarrow \vira$ is any local section of $\psl \times \bb R \hookrightarrow \vira \rightarrow \diff/\psl$ satisfying $\wh\psi(\wt\varepsilon) = \wh I$, and $\delta$ is a Dirac delta on $\ca Q$. 
This state is non-normalizable, so it must be understood as ``generalized wavefunction''.\footnote{See ``Rigged Hilbert spaces'' \cite{de2001quantum,gelfand1964vilenkin}.}

\subsection{The spectrum of $\wh Q_0$}
\label{subsec:Q0spec}

An interesting observable to discuss in the quantum theory is the $Q_0$ charge. Classically we have seen that it should correspond to some conformal invariant of the spatial metric (assuming the phase space constraints are satisfied). Moreover, it commutes with the spin $P_0$ and therefore is rotation-invariant in that sense. Quantum mechanically, in a representation based on the natural orbit, the spectrum of $\wh Q_0$ is continuous, unbounded from below and bounded from above by $2\pi$, as we explain in this section. Lastly, we remark that the spectrum of $\wh Q_0$ will be used in the next section to evaluate the spectrum of the spin/twist, $\wh P_0$.

In the context of asymptotically-flat spacetimes in 2+1 gravity, the group of asymptotic symmetries at the null infinity is $\BMS$ \cite{barnich2007classical}. The $Q_0$ charge would play the role of (minus) the total energy. The total energy generates an asymptotic diffeomorphism along the future null infinity, i.e., translations in the $u$ parameter (where $u$ is the retarded null coordinate, $u = t - r$).\footnote{In the so-called BMS gauge, the (vacuum) metric takes the form $ds^2 = 8G \varepsilon(\theta) du^2 - 2dudr + 8G \left( j(\theta) + u \partial_\t \varepsilon(\t) \right)dud\t + r^2 d\t^2$, where $\varepsilon(\t)$ and $j(\t)$ are interpreted as densities of energy and angular momentum, respectively. The total energy is $E = \frac{1}{2\pi} \int_0^{2\pi}\!d\t \, \varepsilon(\t)$. This reveals a quirk of asymptotically-flat gravity in 2+1 dimensions: the asymptotic diffeomorphism  generated by the total energy $E$, $\partial_u$, is not necessarily timelike. In particular, for the Minkowski solution, $\partial_u$ is timelike only if $E < 0$. When $E > 0$ the solutions correspond to the somewhat exotic ``flat space cosmologies''\cite{cornalba2002new,cornalba2003time}. It is debatable whether $E$ should still be called ``energy'' in the cases where it generates a spacelike flow.} 
It was shown by analysis of the wavefunction representation of $\bms$ that the energy has a continuous spectrum, always unbounded from above and, in some representations, bounded from below. This is regarded as proof of energy positivity in 2+1 asymptotically-flat gravity \cite{barnich2014holographic}.

We can straightforwardly import this result \cite{oblak2017bms} to our theory of causal diamonds. Let us succinctly state the argument here. 
The action of $\wh Q_0$ on a wavefunction $\Psi(\wt\a)$ is expressed in \eqref{whQwaveaction} as
\be
\wh Q_0 \Psi(\wt\a) = \a(\partial_\t) \Psi(\wt\a)
\ee
and, in the characterization $\wt\a = \coad_\psi \wt\varepsilon = [\psi]$, it follows from \eqref{betacoad} that
\be
\wh Q_0 \Psi([\psi]) = \int_{0}^{2\pi}\!d\theta\, \frac{1 - 2 S[\psi](\psi^{-1}(\t))}{\psi'(\psi^{-1}(\t))^2} \Psi([\psi])
\ee
Therefore, any properties that are classically satisfied by $Q_0$ on the orbit $\ca O = \ca Q$, will also be satisfied quantum-mechanically. (In fact, this is true for any function of the $Q$ charges since they act as multiplication on the wavefunctions.) 
The property of interest here is a general inequality involving the Schwarzian, known as the average lemma \cite{guieu2007algebre,balog1998coadjoint,schwartz1992projectively,oblak2017bms},  which reads
\be
\int_{0}^{2\pi}\!d\theta\, S[\psi](\t) \le \int_{0}^{2\pi}\!d\theta\, \frac{1 - \psi'(\t)^2}{2}
\ee
which is saturated if and only if $\psi \in \psl$. 
Now let us replace $\psi$ be $\psi^{-1}$ in this inequality, and rewrite it as
\be
\int_{0}^{2\pi}\!d\theta\, \left( {\psi^{-1}}'(\t)^2 + 2 S[\psi^{-1}](\t) \right) \le \int_{0}^{2\pi}\!d\theta = 2\pi
\ee
Using relation \eqref{Schprop} to express $S[\psi^{-1}]$ in terms of $S[\psi]$, and also ${\psi^{-1}}'(\t) = 1/\psi'(\psi^{-1}(\t))$, we get
\be
\int_{0}^{2\pi}\!d\theta\, \frac{1 - 2 S[\psi](\psi^{-1}(\t))}{\psi'(\psi^{-1}(\t))^2} \le 2\pi
\ee
From the formula for $Q_0$ above, it follows that the possible eigenvalues are less or equal to $2\pi$. Moreover, it is clear that any real value below $2\pi$ is attainable by at least one $\psi$.\footnote{For example, consider $\psi(\t) = \t + \frac{\kappa}{n}\sin(n\t)$, where $n \in \bb Z$ and $0 < \kappa < 1$. This describes a valid diffeomorphism of $S^1$ since $\psi'(\t) = 1 + \kappa \cos(n\theta) > 0$. If $\kappa \ll 1$, then $Q_0([\psi^{-1}]) = \int_0^{2\pi}\!d\t \left( 2 S[\psi](\t) + \psi'(\t)^2 \right) \approx - 3\pi (\kappa n)^2 + 2\pi$. Since $(\kappa n)^2$ can be any positive real number,  $Q_0([\psi^{-1}])$ can be arbitrarily negative.}
Therefore, we have
\be
\text{\sl Spectrum}(\wh Q_0) = (-\infty, 2\pi]
\ee
While most eigenspaces of $\wh Q_0$ are highly degenerate, as there are many diffeomorphism classes $[\psi]$ on which $Q_0$ takes the same value,  there is one eigenspace that is special: the one associated to the maximum eigenvalue $2\pi$.
From the average lemma, the inequality is saturated only when $\psi \in \psl$, that is, when $[\psi] = [I]$. This implies that the (non-normalizable) wavefunctions $\Psi^{j;\mu}_{[I]}$, introduced in \ref{PsiI}, are the only (generalized) states in which $\wh Q_0 = 2\pi$,
\be
\wh Q_0 \Psi_{[I]}^{j;\mu} = 2\pi \Psi_{[I]}^{j;\mu}
\ee
Moreover, by direct evaluation, we see that these states are annihilated by all other $\wh Q_n$'s,
\be
\wh Q_n \Psi_{[I]}^{j;\mu} = 0\,,\,\text{for $n \ne 0$}
\ee
Classically, these states correspond a causal diamonds whose spatial geometries are in the same conformal class as the symmetric disc.

\subsection{Spin/Twist quantization}
\label{subsec:spinquantization}

The quantum theory reveals a quite interesting result:
the spin, which we have interpreted as the twist of the boundary loop, is quantized. To see this, note that $\wh P_n$ and $\wh Q_n$ satisfy the following commutation relations with $\wh P_0$,
\begin{align}
&[\wh P_0, \wh P_n] =  n \wh P_n \nonumber\\
&[\wh P_0, \wh Q_n] =  n \wh Q_n \label{ladderP0}
\end{align}
We are interested in representations in which $\wh P_0$ is self-adjoint (since $\wh P_0^\dag = \wh P_{-0}$), so it must have a real spectrum. Let us say that $\wh P_0$ has an eigenvector $|s\ra$ with eigenvalue $s \in \bb R$; even though, rigorously, it is not guaranteed a priori that $\wh P_0$ has any eigenvectors/eigenvalues.
The commutation relations above imply that both $\wh P_n$ and $\wh Q_n$ act as ladder operators for $\wh P_0$, raising its eigenvalues by $n$. That is, $\wh P_n|s\ra$ and $\wh Q_n|s\ra$, if non-zero, are eigenvectors of $\wh P_0$ with eigenvalue $s+n$.
Since the canonical algebra must be represented irreducibly, the whole Hilbert space is spanned by acting with all $\wh P_n$'s and $\wh Q_n$'s (and their products) on the any given state, such as $|s\ra$. Also, note that any string of $P$'s and $Q$'s also acts as a ladder operator for $\wh P_0$, raising the eigenvalues by some integer: for example, $\wh P_1 (\wh P_3)^2 \wh Q_{-4} \wh P_2$ raises spin eigenvalues by $1+2\times 3 -4 +2 = 5$. Therefore we conclude that the spectrum of $\wh P_0$ is some subset of $\{ s + n;\, n\in \bb Z \}$, where $s$ is some (fixed) real number.\footnote{At this stage, $s$ should not be confused with the shift parameter of the little group representation described in Sec.~\ref{subsec:naturalorbit}. But, as the notation suggests, they will be eventually related.} Next we show that the spectrum of $\wh P_0$ is in fact equal to $s + \bb Z$, that is, there are no level gaps.

First, note that since the spectrum of $\wh P_0$ is discrete (i.e., consisting of isolated points), it must be that $\wh P_0$ is diagonalizable by (normalizable) eigenvectors, whose eigenvalues coincide with the spectrum \cite{hall2013quantum,kulkarni2008some,bogachev2007measure}. Let $|\Psi_j\ra$ be a (normalizable) eigenvector of $\wh P_0$ with eigenvalue $s + j$, for some $j \in \bb Z$. 
Now {\sl suppose} that either of the values $s + j + 1$ or $s + j -1$ do not belong to the spectrum of $\wh P_0$, and let us show that this leads to a contradiction. 
We consider the case where $s + j - 1$ is assumed not to be in the spectrum, as the other case is analogous.
We should have
\be\label{QPpmanihPsij}
\wh Q_{-1}|\Psi_j\ra = \wh P_{- 1}|\Psi_j\ra = 0
\ee
From the algebra of $\wh P$ and $\wh Q$, as in \eqref{quantumalgebra}, we have
\be
[\wh Q_{1}, \wh P_{-1}] =  2 \big( 2\pi - \wh Q_0 \big)
\ee
and contracting with $|\Psi_j\ra$ we get
\be
\la \wh Q_{-1} \Psi_j | \wh P_{- 1} \Psi_j \ra - \la \wh P_{ 1} \Psi_j | \wh Q_{1} \Psi_j \ra  =  2 \la \Psi_j | 2\pi - \wh Q_0|\Psi_j \ra
\ee
where we used that $\la \Psi_j | \wh Q_{ 1} = \la \Psi_j | \wh Q_{- 1}^\dag = \la  \wh Q_{- 1} \Psi_j |$. 
In Sec.~\ref{subsec:Q0spec} we showed that
the spectrum of $\wh Q_0$, for the natural orbit $\ca O = \ca Q$, is bounded from above by $2\pi$, only attaining the maximum on the non-normalizable states concentrated at the conformal class of the identity, $\Psi_{[I]}$. Therefore,
\be
\la \Psi_j | 2\pi - \wh Q_0 |\Psi_j\ra > 0
\ee
since $\Psi_j$ is normalizable (and thus not equal to $\Psi_{[I]}$). Most importantly, this expectation value is non-zero. Hence, if \eqref{QPpmanihPsij} is true, 
\be
\la \wh P_{ 1} \Psi_j | \wh Q_{ 1} \Psi_j \ra  =  -2 \la \Psi_j | 2\pi - \wh Q_0 |\Psi_j \ra \ne 0
\ee
implying, in particular, that
\be\label{QmpPsijcont}
\wh Q_{ 1} | \Psi_j \ra \ne 0
\ee
On the other hand, note that
\ba
\la \wh Q_{ 1} \Psi_j | \wh Q_{ 1} \Psi_j \ra &= \la \Psi_j | \wh Q_{- 1} \wh Q_{ 1} | \Psi_j \ra \no
&= \la \Psi_j | \wh Q_{1} \wh Q_{- 1} | \Psi_j \ra \no
&= 0
\ea
where we used the commutativity of the $\wh Q$'s. 
From the non-degeneracy of the Hilbert space metric it follows that
\be
\wh Q_{ 1} | \Psi_j \ra = 0
\ee
which is in contradiction with \eqref{QmpPsijcont}. Therefore we conclude that \eqref{QPpmanihPsij} is false, so the value $j+s - 1$ must belong to the spectrum. The same argument applies to any value $j+s$ in the spectrum, implying that its two neighbors are also in the spectrum. 
We have thus proved that the spectrum of $\wh P_0$, in any non-trivial unitary irreducible representation based on $\ca O = \ca Q$, is
\be
\text{\sl Spectrum}(\wh  P_0) = s + \bb Z
\ee
Without loss of generality, $s$ can be taken to be in the interval $[0,1)$. 

The value of $s$ can be restricted by assuming time-reversal symmetry. That is, if one wishes to implement the CMC time-evolution Hamiltonian (discussed in Part I) into the quantum theory, in such a way that its classical time-reversal symmetry $\tau \mapsto -\tau$ is preserved,\footnote{Under time-reversal the ADM variables change as $h_{ab} \mapsto h_{ab}$, $\sigma^{ab} \mapsto - \sigma^{ab}$ and $\tau \mapsto -\tau$. This is because the induced metric does not depend on the time orientation but the extrinsic curvature $K_{ab} = \boldsymbol\nabla_a u_b$ is defined with $u$ being ``future directed'', which flips sign when time is reversed. 
In terms of the conformal variables this translates into $\psi \mapsto \psi$,  $\bar\sigma^{ab} \mapsto - \bar\sigma^{ab}$ and $\tau \mapsto -\tau$.
The time-evolution Hamiltonian, related to the area of the CMC slice, depends only on the solution $\lambda$ to the Lichnerowicz equation which is not sensitive to this change of signs. 
This transformation is an anti-symplectic symmetry.
Finally note that, classically, $P_n$ depends linearly on $\sigma^{ab}$, so it changes sign, and $Q_n$ depends only on $\psi$ which does not change.}
then there must exist an anti-unitary operator $\fr T$ acting on the canonical algebra as
\begin{align}
&\fr T \wh P_n {\fr T}^\dag =  - \wh P_n \nonumber\\
&\fr T \wh Q_n {\fr T}^\dag =  \wh Q_n \label{timereversal}
\end{align}
Naturally, this implies that ${\fr T}^\dag |s\ra$ is an eigenvector of $\wh P_0$ with eigenvalue $-s$, so the spectrum of $\wh P_0$ must be symmetric under a change of sign. Therefore, $s$ must be integer or half-integer, so the spectrum of $\wh P_0$ must be either $\bb Z$ or $\frac{1}{2} + \bb Z$.

The value of $s$ is naturally related to the representation of the little group in the wavefunction realization. 
In the representation based on the natural orbit, $\ca O = \ca Q$, the wavefunctions carry internal indices associated with (projective) unitary irreducible representations of $\psl$, as described in \ref{subsec:naturalorbit}. In particular, consider the state $\Psi_{[I]}$ defined in \eqref{PsiI}.
This state is a (generalized) eigenvector of $\wh P_0$, for according to \eqref{PSLonPsi} we have
\be
\wh P_0 \Psi_{[I]}^{j;\mu} = (j + s) \Psi_{[I]}^{j;\mu}
\ee
where $j$ some integer. Therefore, we see that the shift parameter characterizing the spectrum of $\wh P_0$ is the same $s$ parameter labeling the (projective) representation of the little group $\psl$. 
In particular, notice that the restriction to $s=0$ or $s=1/2$, imposed by the time-reversal symmetry, implies that we need to consider only true representations of the double cover of $\psl$, that is, $\text{SL}(2, \bb R)$. 
The case $s = 0$ is possibly the most natural, associated with a ``bosonic'' diamond, while $s=1/2$ corresponds to a ``fermionic'' diamond (in the sense that a rotation by $2\pi$ flips the sign of the wavefunction, although it is not clear if one should change the grade of the algebra, i.e., replace commutators by anti-commutators). 
Lastly, note that all values of $j \in \bb Z$ are contained in the spectrum of $\wh P_0$, as proven directly from the canonical algebra, even if the representation of $\psl$ is one of those that only include a subset of integers (this can be interpreted as the fact that $P_0$ is the ``total spin'', which is not entirely accounted for by the ``internal spin'' but also receives a contribution from the ``orbital angular momentum'' of the wavefunction).

Combining the quantization condition above with formula \eqref{P0twist}, relating the spin $P_0$ with the twist $\ca T$ of the boundary loop, yields a quantization condition for the twist
\be\label{twistquantum}
\ca T = \frac{16\pi^2 \ell_P}{\ell} (n + s) \,,\quad n\in \bb Z
\ee
where $s = 0$ or $1/2$.
Therefore we see that the twist can only change in discrete increments defined by the ratio of the Planck length to the boundary length. In particular, this is consistent with our intuition about the ``classical limit'' in the sense that the size of this increment goes to zero as $\ell \gg \ell_P$, so a classical diamond would be in a state with a very large number of ``twist quanta''.

Now let us determine if our choice of boundary metric $\gamma = (\ell/2\pi)^2 d\theta^2$ was merely a convention, or whether it has true physical implications. Note that $P_0 := P_{\partial_\theta}$ is defined with respect to the reference unit disc, so the choice of $\gamma = \gamma(\theta) d\theta^2$ will affect its geometrical interpretation. In particular, the tangent vector to the boundary, $t^a$, with unit norm with respect to the physical metric $\gamma$, would be generally given by
\be
t^a|_\theta = \frac{1}{\sqrt{\gamma(\theta)}} \partial_\theta
\ee
which of course reduces to \eqref{unittangentt} when $\gamma(\theta) = (\ell/2\pi)^2$. Accordingly, formula \eqref{Pxigeometric} is updated to
\be
P_\xi =  - \frac{\ell}{16\pi^2 \ell_P} \int_\partial\!ds\,  K_{ab}  n^a \frac{2\pi\sqrt{\gamma}\xi}{\ell} t^b
\ee
where $\xi = \xi(\theta) \partial_\theta$.
Thus, the charge generating isometric rotations of the boundary, the physical spin $S$, corresponding to a $\xi^a \propto t^a$ with a constant coefficient as given in \eqref{Hspin}, will not in general be equal to $P_0$, but instead
\be
S := - \frac{\ell}{16\pi^2 \ell_P} \int_\partial\!ds\,  K_{ab}  n^a t^b = P_{\xi_\gamma}
\ee
where
\be
\xi_\gamma := \frac{\ell}{2\pi\sqrt{\gamma}} \partial_\theta
\ee
In particular, $S = P_0$ only when $\gamma(\theta) = (\ell/2\pi)^2$. The defining characteristic of $\xi_\gamma$ is that it is a nowhere vanishing vector field on the boundary with total parameter-length $\ell$. In fact, for any two choices of boundary metric, $\gamma$ and $\gamma'$, with the same total length $\ell$, there exists a diffeomorphism $\psi: S^1 \rightarrow S^1$ such that
\be
\gamma' = \psi^*\gamma
\ee
and, consequently, 
\be
\xi_{\gamma'} = \psi_*\xi_\gamma
\ee
In particular, if with one choice of boundary metric the spin gets associated with the operator $\wh P_\xi$, then with another choice it would be associated with the operator $\wh P_{\psi_*\xi}$ instead. The specific choice of the boundary metric, beyond the information contained in the total length $\ell$, would therefore be physically meaningful if the spectrum of $\wh P_{\psi_*\xi}$ were different than that of $\wh P_\xi$. We now show that this is not the case. That is, we will show that there is a unitary map relating $\wh P_{\psi_*\xi}$ with $\wh P_\xi$, so they must have the same spectrum. In fact, as one could naturally guess, the unitary map generating this transformation is associated with the canonical group action by the element $(0; \wh\psi) \in (\avira^*)^* \rtimes \vira$, where $\wh\psi = (\psi, r)$ (in fact the central component of $\wh\psi$ does not matter since it is represented trivially). Take formula \eqref{PrepfromU} defining $\wh P_\xi$ and sandwich it between $U(0; \wh\psi)$ and its adjoint,
\be
U((0; \wh\psi)) U(\exp(t(0;\xi))) U((0; \wh\psi))^\dag = U((0; \wh\psi)) e^{-itP_\xi} U((0; \wh\psi))^\dag
\ee
On the left-hand side we have
\ba
U((0; \wh\psi)) U(\exp(t(0;\xi))) U((0; \wh\psi))^\dag &= U((0;\wh\psi)\exp(t(0;\xi))(0;\wh\psi)^{-1}) \no
&= U(\exp(t\,\ad_{(0;\wh\psi)}(0;\xi))) \no
&= U(\exp(t (0;\psi_*\xi +\Lambda_\psi(\xi) \wh c ))) \no
&= U(\exp(t (0;\psi_*\xi))) U(\exp(t (0;\Lambda_\psi(\xi) \wh c))) \no
&= U(\exp(t (0;\psi_*\xi))) \no
&= e^{-it \wh P_{\psi_*\xi}}
\ea
where on the first line we used the definition of a representation; on the second line that, for any Lie group, $\Ad_g \exp(\xi) = \exp (\ad_g \xi)$; on the third line expression \eqref{advirasymb} for the adjoint action on $\vira$; on the fifth line we factored out the central piece; on the sixth line we used that the central element of $\vira$ is represented trivially; and on the last line we used again \eqref{PrepfromU}.
On the right-hand side we have
\be
U((0; \wh\psi)) e^{-it \wh P_\xi} U((0; \wh\psi))^\dag =  e^{-itU((0; \wh\psi)) \wh P_\xi U((0; \wh\psi))^\dag} 
\ee
so we conclude
\be
U((0; \wh\psi)) \wh P_\xi U((0; \wh\psi))^\dag = \wh P_{\psi_*\xi}
\ee
as we intended to show.

\section{Discussion}
\label{sec:discussion}

In this work we have explored the canonical quantization of causal diamonds in {(2+1)-dimensional} gravity, with a non-positive cosmological constant, via the reduced phase space approach, in conjunction with Isham's group-theoretic quantization method. In this discussion section we recapitulate on what has been achieved, from a general perspective, and comment on possible future directions to explore.

As explained in Part I \cite{e2023quantization}, a motivation for this project was to understand how causal diamonds, which serve as the prototype notion of subsystems in classical gravity, could be described in quantum mechanics. The concept of a subsystem lies at the core of the current philosophical framework of physics, in which one imagines the universe as consisting of a net of subsystems, such that in some regimes certain ``pieces of the universe'' can be satisfactorily described independently from the ``rest of the universe'', and if two such subsystems are allowed to interact then their union becomes the new subsystem under consideration. Naturally this philosophy is intimately related to the principle of locality and causality, where causally-disconnected regions of spacetime can be treated as independent subsystems. 
Nevertheless, quantum gravity points to the necessity of a fundamental revision of this concept due to the absence of compactly-supported gauge-invariant observables. There are two plausible directions to address this point: one would be to quantize gravity in the whole universe and then try to figure out what is the net of subsystems at the quantum level (e.g., by trying to find some ``order'' in the highly intricate algebraic structure of observables); and the other  would be to quantize the classical gravitational notion of subsystems (i.e., causal diamonds) and use the resulting quantum objects as building blocks to construct the theory of quantum gravity in the whole universe. It is unclear which approach is more promising, and whether they would be equivalent in any sense, but here we chose to investigate the latter. 
What we have achieved is only a first step in this program, as we have described a fully non-perturbative kinematical quantization of causal diamonds by treating them as a self-contained system. A next step would be to consider how (or if) quantum causal diamonds can be sewn together to assemble the entire spacetime.

Aligned with the goal of implementing a rigorous, non-perturbative quantization of a gravitational system, a second motivation we had was to continue the exploration of Moncrief's program of quantizing gravity. In this program, one employs a convenient gauge-fixing of time where the spacetime is foliated by surfaces of constant-mean-curvature. This provides a surprisingly general prescription to solve the constraints of gravity in a variety of cases (i.e., those admitting such a foliation), removing the associated gauge ambiguities, and ending with a reduced phase space equal to the cotangent bundle of an appropriate space of conformal geometries on the Cauchy slice. 
Most applications of this idea have been to systems with a closed Cauchy slice, and here we applied it to causal diamonds whose Cauchy slice has a finite boundary. We made some simplifying assumptions, so that the problem could be treated more concretely, such as considering pure gravity (i.e., General Relativity with no matter fields), lower spacetimes dimension (i.e., 2+1) and a trivial topology for the Cauchy slice (i.e., a disc). It could be interesting to extend the formalism to more general causal diamonds, in higher dimensions, with matter and non-trivial topologies. 
We have also considered that the (induced) metric on the boundary is fixed. The only intrinsic parameter characterizing this condition is the total length $\ell$ of the corner, in the sense that two choices of boundary metrics with the same total length are related by a symplectomorphism at the classical level and, according to our quantization, a unitary transformation at the quantum level. This condition was introduced so that the CMC gauge would be attainable (i.e., so that an arbitrary Cauchy slice could always be deformed into a CMC surface), and for that reason it is not clear whether this boundary condition could be altered or removed. 
Lastly, we have decided to carry out the reduction of the phase space from the perspective of the ADM formulation,  described in terms of spatial metrics and extrinsic curvatures, in part because the boundary condition on the metric was expressed very naturally. It could also be interesting to investigate the phase space reduction from the perspective of the Chern-Simons formulation, which is described in terms of a pair of $\text{SL}(2, \bb R)$ connections \cite{witten19882+,carlip2003quantum,Carlip:1998uc,carlip2005conformal,chern1974characteristic}.

To quantize the reduced phase space, $\wt P = T^*(\diff/\psl)$, which is manifestly a homogeneous space for $\diff$, we employed Isham's group-theoretic approach to canonical quantization. 
From the structure of this phase space, we found that a natural canonical group on which to base the quantization is $\wt G = \avira \rtimes \vira$. When realized as symplectic symmetries of the phase space, the group interestingly reduces to $\BMS$  (i.e., the central charge associated with $\vira$ is trivially realized, matching with the algebra obtained in \cite{barnich2007classical}). 
The quantum theory is thus constructed to carry an irreducible unitary representation of the associated algebra, $\bms$, which according to Mackey's theory of induced representations is characterized by a choice of a coadjoint orbit of Virasoro together with a choice of a (projective) irreducible unitary representation of the corresponding little group \cite{oblak2017bms}. In the most natural case, according to the Casimir matching principle, the quantum states are identified with wavefunctions ``living'' on $\diff/\psl$ with an ``internal index'' in some projective irreducible unitary representations of $\psl$ (or, assuming CMC time reversal symmetry, a true irreducible unitary representation of $\text{SL}(2,\bb R)$). 
This canonical group appeared naturally in our reduction procedure, but it is worth noticing that there is a non-trivial symplectomorphism between $T^*\ca Q$, with the natural symplectic form associated with its cotangent bundle structure, and $\ca Q \times \ca Q$, with the natural symplectic structure that each factor $\ca Q$ inherits as a coadjoint orbit of Virasoro \cite{Scarinci:2011np}. 
Thus, while those two phase spaces are isomorphic, the quantization for the latter would most naturally be based on the group $\vira \times \vira$, thus producing a different set of representations. In fact, $\ca Q \times \ca Q$ is the most natural realization for the reduced phase space of asymptotically $\ads$ spacetimes, and $\vira \times \vira$ is the symmetry structure of a $\text{CFT}_2$, which is a manifestation of the holographic principle \cite{maloney2010quantum,witten2007three,carlip2005conformal}.

We emphasize that we have only constructed a non-perturbative {\sl kinematical} quantization of the system, i.e., obtained a family of possible quantum theories that are naturally associated with the classical phase space, carrying a representation of the canonical observables. 
A complete quantization needs in addition a description of the quantum dynamics, i.e., finding a suitable manner to represent the classical time-evolution Hamiltonian as a self-adjoint operator on the Hilbert space. 
While we have a classical description of the Hamiltonian generating evolution in CMC time, whose value has the very simple geometric interpretation of measuring the area of the CMC slice \cite{york1972role}, it is expressed as a highly complicated formula in terms of the reduced phase space variables. In particular, that formula involves the solution to the Lichnerowicz equation, which cannot be written in closed-form. 
We do not see a straightforward way to effectuate such a quantization, at least non-perturbatively, for even if an explicit formula for the Hamiltonian as a function of the canonical charges were to be found, there would likely be serious operator-ordering ambiguities which would need to be resolved.
A more realistic goal would be to address the dynamical portion of the quantization at the perturbative level. As we have seen in Part I, there are regimes where the Hamiltonian can be approximated,  even becoming ``free'' (i.e., quadratic in momentum-like variables) for states sufficiently close to the symmetric diamond, in the limit of large boundary length compared to the $AdS$ length.  
Notice, however, that the symmetric diamond is not a ``classical vacuum state'' for the time-evolution Hamiltonian since there are states with lower ``energy'' (i.e., area), even in a neighborhood of that state. In fact, while the classical Hamiltonian is manifestly bounded from below, it has no minimum, much like the Hamiltonian for Liouville field theory whose potential has an exponential form $V(\phi) = e^{2b\phi}$ (indeed Liouville theory has been linked to three dimensional gravity before~\cite{seiberg1990notes,balog1998coadjoint,krasnov2001three,nakayama2004liouville,faddeev2014zero,erbin2015notes,li2020liouville}).

Another set of questions concerns what a ``quantum causal diamond'' really means. At the classical level we arrived at the interpretation that the states (i.e., points in the reduced phase space) correspond to shapes of causal diamonds embedded into $\ads$ (or $\mink$ if the cosmological constant vanishes). That is, a causal diamond with a round maximal slice and another one with an oval maximal slice are distinct classical states, with physically distinguishable properties (e.g., if one shoots geodesics across the maximal slice, starting normal to the boundary at one point and ending at the other side of the boundary, in the first case they would all have the same length while in the second case some would be longer than others).
In fact, we have shown that there is a map from each point of the reduced phase space to a causal diamond in $\ads$ or $\mink$ (or, more simply, to a loop of length $\ell$ that is the boundary of a spacelike disc). However, for this interpretation to hold it is essential that both the $Q_n$ {\sl and} $P_n$ coordinates are specified simultaneously, which of course is not feasible in the quantum theory since these observables do not commute. 
This apparent breakdown of the fabric of spacetime is expected in a generic theory of quantum gravity, and there is no reason to believe that a quantum spacetime should have properties similar to those of a manifold (except of course in some semi-classical limit). 
In our specific realization, in addition to the non-commutativity of the geometrical variables, note also that: as the CMC time is chosen before quantization, it retains a classical character in the quantum theory, thus standing on a different footing than ``space'' in the quantum picture; second, even the concept of a spatial geometry is deformed since, while the configuration variables (i.e., the $\wh Q_n$'s) do commute, they are not in direct correspondence with the spatial geometry of the CMC slices, but rather with the {\sl conformal} geometry of the slices. 
Accordingly, it would be valuable to better understand what kind of ``spacetime'' emerges, from our quantization, as the ``interior'' of a quantum causal diamond, 
possibly shedding some light on what is the meaning of a ``subsystem of a quantum spacetime'', or even what is a ``quantum spacetime'' itself. 

To make sense of the quantum geometry of causal diamonds, a reasonable first step is to understand the physical meaning of the canonical observables. In this paper we mostly focused on the meaning of the momentum charges (i.e., the $P_n$'s), which were found to be related to Fourier modes of the component $K_{ab} t^a n^b$ of the extrinsic curvature of the CMC slices at the corner, where $t^a$ and $n^b$ are unit vectors on the slice tangent and normal, respectively, to the boundary.
Of distinguished significance is $P_0$ which can be interpreted as the twist of the corner loop, therefore being classically a property of the diamond shape itself (i.e., independent of choices of spatial slices). 
In the quantum theory we showed that the twist is quantized, in any non-trivial representation, in integer (or half-integer) multiples of $16\pi^2 \ell_P/\ell$. This is a non-trivial result since the twist is a continuous parameter classically (in fact, we can see that when the length of the loop is much larger than the Planck length, the spacing between twist levels goes to zero, recovering the expected classical behavior). 
One might speculate that there is some sense in which this result can extrapolated to a general statement about the twist of loops in three-dimensional gravity (and perhaps even in the presence of matter fields, since the ``twist quantum'' is independent of the cosmological constant). 

The physical interpretation of the configuration charges (i.e., the $Q_n$'s) has proved much more elusive. Although we have given explicit formulas for these charges in terms of the $\psl$-class of the boundary diffeomorphism, which in turn is explicitly related to a conformal class of spatial geometries on the CMC slices, we could not provide an interpretation in terms of simple geometrical properties of the shape of the causal diamond. 
A path currently under consideration attempts to understand the structure of corner symmetries directly at the ADM level, by analyzing the differentiability conditions for the ADM Hamiltonians (i.e., the boundary terms that need to be included so that the corresponding Hamiltonians generate regular symplectic flows on the phase space, and consequently have well-defined Poisson brackets with other charges). We find that the $P_n$'s are precisely the charges associated with diffeomorphisms tangent to the corner, but there are another two sets of charges associated with deformations of the corner in the normal directions. These normal deformation charges, when evaluated on the constraint surface, are related to the expansion parameter of the null rays of the future and past horizons at the corner. 
In this paper we have shown that two families of charges ($P$'s and $Q$'s) should suffice to parametrize the reduced phase space, so it may be speculated that the $Q_n$'s are related to some function of the expansion parameters of the null generators of the horizons. 
Another hint in this direction is that $\BMS$ is, as we have mentioned, notably associated with asymptotic symmetries on the null infinity of asymptotically-flat spacetimes in three dimensions. It is thus plausible that there exists an alternative point of view in which the causal diamonds are described by charges directly associated with the horizons in a neighborhood of the corner,  making the analogy with asymptotically-flat spacetimes more explicit.

It is natural to comment on a different set of variables: the charges associated with boundary diffeomorphisms. Those are primary in the formalism of edge modes\cite{donnelly2016local,donnelly2021gravitational,donnelly2023matrix,speranza2018local,geiller2017edge,geiller2018lorentz,freidel2020edgei,freidel2020edgeii,freidel2021edgeiii}. Recently, the boundary diffeomorphism charges have been computed, from a covariant phase space perspective, for causal diamonds in $(2+1)$-dimensions \cite{pulakkat2024charge}.
These charges have nice and simple geometric interpretations.
It should be pointed out, however, that there is an important difference between asymptotic $\ads$ (or other asymptotic spacetimes) and causal diamonds. 
In the former case, the group of asymptotic diffeomorphisms acts legitimately on the reduced phase space. In fact, the asymptotic diffeomorphisms are defined precisely as the set of diffeomorphisms that preserve the class of asymptotic spacetimes under consideration, and thus they always map a state into another state. 
In the case of causal diamonds this does not happen, at least when normal translations of the corner are allowed. In fact, as these boundary diffeomorphisms act by deforming the diamond corner, there is nothing preventing them from ``pushing it too far'', that is, deforming a ``nice corner'' into a loop that is not acausal or that develops crossings and knots, and thus can no longer be the corner of another diamond. In other words, the Hamiltonian flow of generic boundary diffeomorphism charges will not be complete, reaching the ``end of phase space'' in finite parameter-length.
While these charges cover the phase space with regular coordinates, they are not appropriate ``canonical charges'' to be used as the basis of quantization, at least from Isham's perspective.
In particular, it would seem that unitary representations of the corner group are not (entirely) relevant for the ``quantum theory'' of this system, similarly to attempting to quantize a particle on the half-line using coordinates $x$ and $p$, as we explained in Sec.~\ref{subsec:canons}.
Thus, we may have a dilemma in the case of causal diamonds: while boundary diffeomorphism charges are physically natural and have a nice geometric interpretation, then do not form a viable basis for quantization; on the other hand, the $\bms$ charges that we have constructed were derived from a group of symmetries that acts properly on the phase space, and thus can be used for quantization, but unfavorably half of them (namely, the $Q$ charges) seem to lack a simple geometric interpretation.

Finally, we note that it would be interesting to investigate the thermodynamic and entanglement properties of the quantum causal diamonds. A Holy Grail of quantum gravity is still to explain the Bekenstein-Hawking entropy formula from a microscopic point of view, and this entropy is widely believed to correspond to a measure of the entanglement across the black hole horizon. In fact, it is expected that a similar formula for the entanglement entropy should also hold for generic entangling surfaces. Naturally this can be analyzed in the context of causal diamonds, e.g., \cite{jacobson2016entanglement,jacobson2019spacetime,jacobson2019gravitational,jacobson2023entropy,jacobson2023partition,banks2021path,banks2023comments,bousso1999covariant,bousso1999holography,fischler1998holography,martinetti2003diamond}. The research on the thermodynamics of causal diamonds has been mainly perturbative in nature, so having a concrete quantum theory of causal diamonds could be valuable in understanding their thermodynamic and entanglement structures from a non-perturbative perspective.
In the microcanonical ensemble, the entropy is given by the logarithm of the number of states compatible with specified values of some ``macroscopic'' charges, typically the energy and angular momentum. Taking the ``energy'' to mean the CMC time-evolution Hamiltonian, the entropy $\ca S (\ca E, \ca J )$ should be defined as the logarithm of the number of eigenstates of $H$ and $P_0$ with eigenvalues (around), respectively, $\ca E$ and $\ca J$.\footnote{One could also consider the entropy associated with fixing $Q_0$ (and $P_0$), as $Q_0$ can be regarded as some kind of ``quasi-local energy''. However, the spectrum of $Q_0$ is continuous and thus it seems that a corresponding entropy cannot be properly defined.} 
Curiously, in the entanglement equilibrium argument \cite{jacobson2016entanglement,jacobson2019gravitational}, variations of the maximal slice with fixed volume play a key role --- this condition is quite analogous, in our case, to fixing the value of $H$ (at $\tau=0$), as it has the interpretation of being the ``volume'' (in two dimensions) of the maximal slice. 
This is, however, a topic for future research.

\newpage
\section*{Acknowledgments} 

I am especially grateful to Ted Jacobson for motivating this project, for collaboration on parts of it, and for useful discussions and valuable insights.
I also thank Stefano Antonini, 
Luis Apolo, Abhay Ashtekar, Batoul Banihashemi,
Steve Carlip, Gong Cheng, Laurent Freidel,
Marc Henneaux, Jim Isenberg, 
Alex Maloney, 
Blagoje Oblak,
Pranav Pulakkat, Gabor Sarosi, Antony Speranza, Raman Sundrum, 
Yixu Wang and Edward Witten for helpful discussions.

This research was supported in part by the National Science Foundation under Grants PHY-1708139 and 
PHY-2012139 at the University of Maryland; and
in part by the National Science Foundation under Grants No. NSF PHY-1748958 and PHY-2309135 at the Kavli Institute for Theoretical Physics. I am grateful for the hospitality of Perimeter Institute for Theoretical Physics where part of this work was carried out. (Research at Perimeter Institute is supported by the Government of Canada through the Department of Innovation, Science and Economic Development and by the Province of Ontario through the Ministry of Research, Innovation and Science.)

\newpage
\appendix

\section{Glossary, symbols and conventions}
\label{app:GSandC}

This appendix features a quick reference guide to recurrent terms and symbols found in this paper. It is organized roughly in the order of appearance in the text, but also in such a way that later entries only refers to terms and symbols already introduced in earlier entries. Each entry is indicated by an \underline{\emph{underlined italic name}} followed by the explanation; in cases where relevant symbols are introduced in the explanation, those symbols are displayed on the left margin for easier reference.
Despite our efforts to keep the notation uniform throughout, there are variations between Part I and II. For example, $\Psi$ is used in Part I to denote diffeormorphisms of the disc, while in Part II it is used to denote quantum states (particularly in their wavefunction realization). We hope that the context will prevent confusion in those instances. Some general conventions are also referenced in this section.
\\

\noindent{\bf\emph{Units:}}~~ We adopt units in which the speed of light and Planck's constant are $1$,
\[
c = \hbar = 1
\]

\renewcommand\stackalignment{r}% RIGHT ALIGNED STACKS
\renewcommand\stacktype{L}% MAKE STACKS OBEY FIXED BASELINESKIP
\strutlongstacks{T}% TO GET PROPER SPACING FOR SINGLE-LINE ITEMS

\begin{itemize}[
	label=,
%	labelwidth=3em,
     labelsep=1.7em,
     itemsep=7pt,
     leftmargin=2em,
%     rightmargin=2cm,
     itemindent=-2em
]

\item[] \underline{\emph{Causal diamonds}}~~ The domain of dependence of an acausal spacelike disc is referred to as a causal diamond.

\item[$\ell$, $\lads$, $\ell_P$] \underline{\emph{Length scales}}~~ The length of the diamond corner is $\ell$, the Anti-de Sitter radius is $\lads = 1/\sqrt{-\Lambda}$ (if $\Lambda < 0$), and the Planck length is $\ell_P = \hbar G$.

\item[$T^a_{\,\,b}$] \underline{\emph{Tensor indices}}~~ The abstract index notation is used for spacetime tensors, where subscript Latin letters indicate covariant tensor slots and upperscript Latin letters indicate contravariant slots (typically we reserve for this purpose letters from the first half of the alphabet like $a$, $b$, $c$ etc). In the present paper (Part II), most tensors are associated with the phase space, configuration space or groups and algebras, and these will typically be denoted without any indices (the specific notation for some these objects is featured below). 

\item[$f_*$, $f^*$] \underline{\emph{Push-forward and pull-back}}~~ Given a smooth map between two manifolds, $f : \ca M \rightarrow \ca N$, the push-forward operator mapping vectors (or contravariant tensors) on $\ca M$ to $\ca N$ is denoted by $f_*$ and the pull-back operator mapping 1-forms (or covariant tensors) from $\ca N$ to $\ca M$ is denoted by $f^*$. If $f$ is a diffeomorphism (smooth invertible map) then $f^* = f^{-1}_*$.

\item[$\pounds$, $d$, $\ii$] \underline{\emph{Lie, exterior and interior derivatives}}~~ The Lie derivative along a vector field $X$ is denoted by $\pounds_X$. The exterior derivative of a form is denoted by $\delta$; and the interior derivative (a.k.a., interior product), with respect to a vector $X$, by $\ii_X$.

\item[$\wt{\ca P}$, $\ca Q$] \underline{\emph{Phase and configuration space}}~~ The (reduced) phase space is denoted by $\wt{\ca P}$. If it has a cotangent bundle structure, $\wt{\ca P} = T^*\ca Q$, the base space $\ca Q$ is referred to as the configuration space. For the causal diamond $\ca Q = \diff/\psl$.

\item[\smash{\Longunderstack{$\omega$, $\theta$\\$H$, $X$\\$\{\,,\}$}}] \underline{\emph{Symplectic structure and Poisson brackets}}~~ The symplectic 2-form on the phase space is denoted by $\omega$. The symplectic potential, denoted by $\theta$, is a 1-form satisfying $\omega = d\theta$. The phase space function $H$ is associated with the Hamiltonian vector field $X$ on phase space via $dH = - \ii_X \omega$. The Poisson brackets of two functions $f$ and $g$ is defined by $\{f, g\} := - \omega(X_f, X_g)$.

\item[\smash{\Longunderstack{$\wt G$, $G$\\$\wt{\fr g}$, $\fr g$\\$\Gamma_{\wt g}$, $\delta_g$, $\wt\delta_g$}}] \underline{\emph{Groups and algebras of symmetry}}~~ The canonical group, acting transitively as symplectomorphisms of the phase space, will typically be denoted by $\wt G$. A group acting transitively on the configuration space, used as the basis for constructing the canonical group, will typically be called $G$. The Lie algebras of $\wt G$ and $G$ will be denoted by $\wt{\fr g}$ and $\fr g$, respectively. The action of $\wt G$ on the phase space $\wt{\ca P}$ is denoted by $\Gamma_{\wt g}$; the action of $G$ on the configuration space $\ca Q$ is typically denoted by $\delta_g$, and its lift to the phase space (cotangent bundle) is $\wt\delta_g$. (Alternatively, sometimes we write simply $gx$ for the action of $g$ on $x$.)

\item[$\rtimes$, $\sdplus$] \underline{\emph{Semidirect product and sum}}~~ The semidirect product of a group $G$ and a group $N$, with respect to a homomorphism $\varphi : G \rightarrow \text{Aut}(N)$, is denoted by $N \rtimes_\varphi G$ (or, omitting $\varphi$, simply $N \rtimes G$). It has the topology of $N \times G$, its element are denoted by $(n;g)$ and the product rule is $(n, g)(n', g') = (n \varphi_g(n); gg')$. The Lie algebra of this group corresponds to a semidirect sum of the respective Lie algebras, which is denoted by $\fr n \sdplus \fr g$. (To remember the notation, note that the arrow in $\rtimes$ points from $G$ to $N$, since $G$ acts on $N$; for the algebra, note that $\sdplus$ looks like a ``rounded'' arrow from $\fr g$ to $\fr n$.)

\item[$\wh G$, $\wh{\fr g}$] \underline{\emph{Central extensions}}~~ The central extension of a group $G$ by  a real 2-cocycle is denoted by $\wh G$. Its elements are $\wh g = (g,r)$, where $g \in G$ and $r \in \bb R$. The product rule has the form $(g,r)(g',r') = (gg', r + r' + W(g,g'))$, where $W: G \times G \rightarrow \bb R$. (Note that $(e,r)$ belongs to the center of $\wh G$.) The central extension of a Lie algebra $\fr g$ by a real 2-cocycle is denoted by $\wh{\fr g}$. Its elements are denoted by $\wh \xi = \xi + r \wh c$, where $\xi \in \fr g$, $r \in \bb R$ and $\wh c$ is the central element of $\wh{\fr g}$.

\item[\smash{\stackunder{$\diff$}{$\psi$, $\phi$}}] \underline{\emph{Circle diffeomorphisms}}~~ The group of (orientation-preserving) diffeomorphisms of the circle is denoted by $\diff$. Its elements are typically denoted by $\psi : S^1 \rightarrow S^1$ or $\phi$.

\item[\smash{\Longunderstack{$\adiff$\\$\xi = \xi(\theta)\partial_\t$\\$\eta = \eta(\t)\partial_\t$}}] \underline{\emph{Algebra of circle diffeormorphisms}}~~ The Lie algebra of $\diff$ is denoted by $\adiff$. Its elements are identified with vector fields on $S^1$ and typically denoted by $\xi = \xi(\theta) \partial_\theta$ or $\eta = \eta(\t)\partial_\t$.
The product rule is $[\xi, \eta] := [\xi, \eta]_\fr{diff} := [\eta, \xi]_{S^1} := \left(\eta(\theta) \partial_\theta \xi(\theta) - \xi(\theta) \partial_\theta \eta(\theta) \right) \partial_\theta$.

\item[\smash{\stackunder{$\ddiff$}{$\a = \a(\t) d\t^2$}}] \underline{\emph{Dual algebra of circle diffeormorphisms}}~~ The dual Lie algebra of $\diff$ is denoted by $\ddiff$. Its elements can be identified with quadratic forms $S^1$ and typically denoted by $\alpha = \alpha(\theta) d\theta^2$ (or $\beta$).
The pairing between $\ddiff$ and $\adiff$ is $\alpha(\xi) := \int \alpha(\theta) d\theta^2 (\xi(\theta) \partial_\theta) = \int d\theta \,\alpha(\theta) \xi(\theta)$.

\item[\smash{\stackunder{$\psl$, $\apsl$}{$\chi$, $\upsilon\quad\!$}}\!\!\!\!\!\!] \underline{\emph{Projective special linear group and algebra}}~~ The group of $2\!\times\!2$ real matrices $S$ with unit determinant, where $S$ is identified with $-S$, defines $\psl$, whose elements are typically denoted by $\chi$. 
Its Lie algebra is denoted by $\apsl$, and here identified with the subalgebra of $\adiff$ generated by the elements $\partial_\theta$, $\sin\theta \partial_\theta$ and $\cos\theta \partial_\theta$. The elements of $\apsl$ are sometimes denoted by $\upsilon$. By exponentiating this subalgebra, $\psl$ is realized as a subgroup of $\diff$. 

\item[\smash{\Longunderstack{$\diff/\psl$\\$[\psi]\in\ca Q\quad\!$}}\!\!\!\!\!\!] \underline{\emph{Configuration space}}~~ The configuration space for the causal diamond is the quotient space $\ca Q = \diff/\psl$, and its points are denoted by $[\psi] = [\psi\chi]$. 

\item[\smash{\Longunderstack{$\vira$\\$\wh\psi = (\psi, r)$\\$\wh I = (I, 0)$}}] \underline{\emph{Virasoro group}}~~ The Virasoro group, $\vira = \wh{\diff}$, is a central extension of $\diff$. Its elements are typically denoted by $\wh\psi = (\psi, r)$, where $\psi \in \diff$ and $r \in \bb R$. The identity element of $\vira$ is denoted by $\wh I = (I, 0)$, where $I$ is the identity diffeomorphism of $S^1$.

\item[\smash{\Longunderstack{$\avira$\\$\wh\xi = \xi+x\wh c$\\$\wh\eta = \eta + y \wh c$}}] \underline{\emph{Virasoro algebra}}~~ The Virasoro algebra, $\avira = \wh{\adiff}$, is a central extension of $\adiff$. Its elements are typically denoted by $\wh\xi = \xi + x \wh c$, where $\xi = \xi(\theta) \partial_\theta \in \adiff$, $x \in \bb R$ and $\wh c$ is the central element.

\item[\smash{\stackunder{$\dvira$}{$\wt\alpha = \alpha+a\wt c$}}] \underline{\emph{The dual of Virasoro algebra}}~~ The dual of the Virasoro algebra is denoted by $\dvira$. Its elements are typically denoted by $\wt\a = \a + a \wt c$, where $\a = \a(\theta) d\theta^2 \in \ddiff$, $a \in \bb R$ and $\wt c$ is the dual of $\wh c$ (i.e., $\wt c(\xi + x \wh c) = x$).

\item[$\wt\varepsilon = d\theta^2 + \wt c$] \underline{\emph{The representative of $\ca Q$}}~~ In the realization of $\ca Q = \diff/\psl$ as a coadjoint orbit of $\vira$, $\wt\varepsilon := d\theta^2 + \wt c \in \dvira$ is point on the orbit (corresponding to $[I] \in \ca Q$).

\item[\smash{\Longunderstack{$\un\psi \in \un\diff$\\ \raisebox{-0.2em}{$\wh{\un\psi} \in \un\vira$}}}] \underline{\emph{Universal covers}}~~ The universal cover of a group $G$ is denoted by an underline, $\un G$. Accordingly, the universal cover of $\diff$ is denoted by $\un\diff$, and its elements by $\un\psi$. The universal cover of $\vira$ is denoted by $\un\vira$, and its elements by $\wh{\un\psi} = (\un\psi, x)$.

\item[${S[\psi](\theta)}$] \underline{\emph{Schwarzian derivative}}~~ The Schwarzian derivative maps circle diffeomorphisms $\psi$ into real functions on the circle, $\psi \mapsto S[\psi](\theta)$.

\item[$\exp$] \underline{\emph{Lie Exponential}}~~ The Lie group exponential is typically denoted by $\exp : \fr g \rightarrow G$. The group it refers to should be understood from the context, but typically will be the Virasoro group.

\item[\smash{\Longunderstack{$\text{Ad}_g$\\ $\ad_g$, $\ad_\xi$}}] \underline{\emph{Adjoint maps}}~~ The adjoint action of a group element $g \in G$ on the group $G$ is denoted by $\text{Ad}_g : G \rightarrow G$ and defined by $\text{Ad}_g(g') := gg'g^{-1}$, where $g' \in G$. The adjoint action of a group element $g \in G$ on its Lie algebra $\fr g$ is denoted by $\ad_g : \fr g \rightarrow \fr g$ and defined by $\ad_g := (\text{Ad}_g)_*$, seen as map from $T_eG$ to itself. The adjoint action of an algebra element $\xi \in \fr g$ on the Lie algebra $\fr g$ is denoted by $\ad_\xi : \fr g\rightarrow \fr g$ and defined by $\ad_\xi  := \left. \frac{d}{dt} \ad_{\exp(t\xi)}\right|_{t=0}$. (It is also true that $\ad_\xi\eta = [\xi, \eta]$.)

\item[$\coad_g$, $\coad_\xi$] \underline{\emph{Coadjoint maps}}~~ The coadjoint action of a group element $g \in G$ on its dual Lie algebra $\fr g^*$ is denoted by $\coad_g : \fr g^* \rightarrow \fr g^*$ and defined by $\coad_g \a := \text{Ad}^*_{g^{-1}}\a$, where $\a \in \fr g^*$. The coadjoint action of an algebra element $\xi \in \fr g$ on the dual Lie algebra $\fr g^*$ is denoted by $\coad_\xi : \fr g^*\rightarrow \fr g^*$ and defined by  $\coad_\xi  := \left. \frac{d}{dt} \coad_{\exp(t\xi)}\right|_{t=0}$.

\item[\smash{\Longunderstack{$\wt G = \avira \rtimes \vira$\\$(\wh\eta; \wh\phi)$\\$(\eta+y \wh c; (\phi, r))$}}] \underline{\emph{Canonical group}}~~ The canonical group for the causal diamonds is taken to be $\wt G = \avira \rtimes \vira$. The $\vira$ factor act as ``configuration translations'' while the abelian factor, $\avira$, act as ``momentum translations''.
Elements of $\avira \rtimes \vira$ are denoted as $(\wh\eta; \wh\phi)$. The semi-colon separates the two components of $\wt G$, so the notation is more transparent when $\wh\phi$ and $\wh\eta$ are written explicitly, i.e., $(\wh\eta; \wh\phi) = (\eta+y \wh c; (\phi, r))$.

\item[\smash{\Longunderstack{$\wt{\fr g} = \avira^c \sdplus \avira$\\$(\wh\eta; \wh\xi)$\\$(\eta+y \wh c; \xi + x \wh c)$}}] \underline{\emph{Canonical algebra}}~~ The canonical algebra is the Lie algebra of $\wt G$, $\wt{\fr g} = \avira^c \sdplus \avira$, where $\avira^c$ is the Lie algebra of abelian group $\avira$ (i.e., $\avira^c$ is isomorphic to $\avira$ as vector space but has a commutative algebraic structure). Elements of $\avira^c \sdplus \avira$ are denoted by $(\wh\eta; \wh\phi) = (\eta+y \wh c; \xi + x \wh c)$.

\item[\smash{\Longunderstack{$L_n$, $R$\\$K_n$, $T$}}] \underline{\emph{Fourier basis for $\wt{\fr g}$}}~~ A basis for $\wt{\fr g}$ is defined as: $L_n := (0; e^{in\t}\partial_\t)$, $R := (0; \wh c)$, $K_n := (e^{in\t}\partial_\t;0)$ and $T := (\wh c; 0)$. Note that $L_n$ and $R$ are generators of the $\vira$ factor of $\wt G$ (``configuration translations''), and $K_n$ and $R$ are the generators of the $\avira$ factor of $\wt G$ (``momentum translations''). Also, $R$ and $T$ belong to the center of $\wt{\fr g}$.

\item[\smash{\Longunderstack{$P_{\wh\xi}$, $Q_{\wh\eta}$\\$P_n$, $Q_n$}}] \underline{\emph{Canonical charges}}~~ The canonical charges associated with the $\vira$ factor of $\wt G$ are called {\sl momentum charges}, $P_{\wh\xi} := H_{(0; \wh\xi)}$, and the canonical charges associated with the $\avira$ factor of $\wt G$ are called {\sl configuration charges} $Q_{\wh\eta}:= H_{(\wh\eta; 0)}$. In the Fourier basis we have $P_n := P_{L_n}$ and $Q_n := Q_{K_n}$, while the central charges are realized as $P_R = 0$ and $Q_T = 1$. 

\item[\smash{\Longunderstack{$\BMS$\\$\bms$}}] \underline{\emph{The Bondi-Metzner-Sachs group and algebra}}~~ The $\bms$ algebra is a reduction of $\wt{\fr g} = \avira^c \sdplus \avira$ in which $R = (0; \wh c)$ is removed, and the $\BMS$ group is its exponentiation. Note that $\avira \rtimes \vira$ is a central extension of $\BMS$.

\item[$\pi$, $q$] \underline{\emph{Projection maps}}~~ The projection map in the cotangent bundle is typically denoted by $\pi : T^*\ca Q \rightarrow \ca Q$. The quotient by $\psl$ from $\diff$ to $\diff/\psl$ is denoted by $q$. By an abuse of notation, since $\ca Q = \diff/\psl$ is identified with a coadjoint orbit of Virasoro, we also use $q$ to denote the map from $\diff$ (or $\vira$ since the central element acts trivially) to $\ca Q \subset \dvira$ defined by $q(\psi) := \coad_{\psi} \wt\varepsilon$.

\item[\smash{\Longunderstack{$\ca O$, $H_{\ca O}$}}] \underline{\emph{Orbits and little groups}}~~ When a group $G$ acts on a manifold $\ca M$, the orbit $\ca O$ of $x \in \ca M$ is the set of points $gx$ for all $g \in G$. The subgroup $H$ of $G$ that fixes a point $x$ (i.e., $gx = x$) is called the little group of $x$. The little group of points on the same orbit are the same (up to conjugation) so we denote it by $H_{\ca O}$ (and it is true that $\ca O$ is homeomorphic to $G/H_{\ca O}$).
We are typically interested in coadjoint orbits of Virasoro, so $\ca O \subset \dvira$. 

\item[$F \hookrightarrow E \rightarrow B$] \underline{\emph{Fiber bundles}}~~ A fiber bundle with total space $E$, base manifold $M$ and fibers $F$ is denoted by $F \hookrightarrow E \rightarrow B$. The second arrow corresponds to the bundle projection map from $E$ to $B$, and the first (hooked) arrow indicates that $F$ can be embedded into $E$ as a fiber (although not uniquely).

\item[\smash{\Longunderstack{$\ca S \hookrightarrow G \times_{\scr U} \ca S \rightarrow \ca O$\\$[\wh\psi, \varsigma]\quad\!$\\$L_{\wh\psi}\quad\!$}}\!\!\!\!\!] \underline{\emph{Associated vector bundle}}~~ The vector bundle associated with the principal bundle $H \hookrightarrow G \rightarrow \ca O$, where $\ca O = G/H$, with respect to the linear representation $\scr U : H \rightarrow \text{\sl Aut}(\ca S)$, is denoted by $\ca S \hookrightarrow G \times_{\scr U} \ca S \rightarrow \ca O$. Its elements are classes of equivalence $[g, \varsigma] = [gh, \scr U(h^{-1})\varsigma]$, where $g \in G$, $h \in H$ and $\varsigma \in \ca S$. The action of $G$ on $\ca O$ lifts to an action on the bundle defined by $L_{g'}[g,\varsigma] := [g'g; \varsigma]$. Typically we will consider the bundle $\ca S \hookrightarrow \vira \times_{\scr U} \ca S \rightarrow \ca O$, whose elements will be denoted by $[\wh\psi, \varsigma]$, and the lifted action is denoted by $L_{\wh\psi}$ (or the one based on the universal cover of $\vira$, $\ca S \hookrightarrow \un\vira \times_{\scr U} \ca S \rightarrow \ca O$).

\item[\smash{\stackunder{$\ca H$}{$\Psi(\wt\a)$, $\Psi([\psi])$}}] \underline{\emph{Hilbert space and wavefunctions}}~~ The Hilbert space $\ca H$ carries a (projective) irreducible unitary representation of the canonical group. In the wavefunction realization, $\ca H$ is identified with the space of sections on $\ca S \hookrightarrow G \times_{\scr U} \ca S \rightarrow \ca O$, whose elements are denoted by $\Psi(\wt\a)$ where $\wt\a \in \ca O$. For the natural representation, based on $\ca O = \ca Q$, we can also use the notation $\Psi([\psi])$ where $[\psi] \in \diff/\psl$. 

\item[\smash{\stackunder{$\ca S$}{$\varsigma$}}] \underline{\emph{``Little'' Hilbert space}}~~ In the wavefunction realization, the ``little'' Hilbert space $\ca S$ carries an irreducible unitary representation $\scr U$ of the little group $H$ associated to a coadjoint orbit of Virasoro. Its elements will be denoted by $\varsigma$, and they correspond to ``internal states'' of the wavefunction (i.e., like an intrinsic spin in quantum field theory). In the natural representation $\ca S$ carries a (projective) unitary irreducible representation of $\psl$.

\item[\smash{\Longunderstack{$\wh P_n$, $\wh Q_n$}}] \underline{\emph{Quantum operators}}~~ The quantum version of a classical observable, represented on the Hilbert space, is typically denoted by a \raisebox{-0.2em}{$\,\,\,\wh{}\,\,\,$} accent above the classical symbol. For example, the quantum operators associated with the canonical momentum and configuration charges, $P_n$ and $Q_n$, are denoted by $\wh P_n$ and $\wh Q_n$, respectively.

\item[\smash{\Longunderstack{$\ca T$}}] \underline{\emph{Twist}}~~ The twist of the corner of the diamond, as embedded in spacetime, is denoted by $\ca T$.

\item[$\text{[}a\sim b\text{]}$] \underline{\emph{Classes of equivalence}}~~ The space of classes of equivalence of all objects $a$ and $b$, belonging to some space $S$, identified under the relation ``$\sim$'' are typically denoted as $[a\sim b; a,b\in S]$. Sometimes the space $S$ is clear from the context and omitted in the notation, $[a \sim b]$. Often the equivalence relation comes from a group $G$ acting on $S$; then $[a \sim ga; a\in S, g \in G]$ is also called the space of $G$-orbits on $S$.

\item[$l_g$, $r_g$] \underline{\emph{Group translations}}~~ The left group translation $l_g : G \rightarrow G$, by a group element $g \in G$, is defined as $l_g(g') := gg'$. The right group translation $r_g: G \rightarrow G$, by $g$, is defined as $r_g(g') := g'g$.

\item[$\Xi$] \underline{\emph{Maurer-Cartan form}}~~ The Maurer-Cartan form is denoted by $\Xi$. It is a Lie algebra-valued 1-form on the Lie Group defined by $\Xi(X) = l_{g^{-1}*}X$, where $X \in T_gG$. (In App.~\ref{app:projrep} we use an alternative definition based on the right group translation,  $\Xi(X) = r_{g^{-1}*}X$.)

\item[$\ac\sigma \in \whddiff$] \underline{\emph{``Sigma-circle''}}~~ The subspace of $\ddiff$ that annihilates $\apsl$ is denoted by $\whddiff$ and its elements are typically denoted by $\ac\sigma$. (See Part I \cite{e2023quantization}.)

\item[$\wh{\ca S}$, $J$] \underline{\emph{Partially-reduced phase space}}~~ In some sections we refer to the partially-reduced phase space $\wh{\ca S} = \diff \times \whddiff$. The projection map to the (fully) reduced phase space is denoted by $J : \wh{\ca S} \rightarrow \wt{\ca P}$. (See Part I \cite{e2023quantization}.)

\end{itemize}

\section{Construction of the Virasoro group}
\label{app:Vira}

In this appendix we offer more details on the construction of the Virasoro group (from the exponentiation of the Virasoro algebra) and its adjoint and coadjoint representations, complementary to the exposition in Sec.~\ref{subsec:Vira}.

Recall that, in our notation, elements of the Virasoro algebra are written as
\be
\wh \xi = \xi(\theta)\partial_\t + x \wh c \,\in\, \avira
\ee
where $\wh c$ is an element in the center.
The product rule is
\be\label{viraproda}
[\xi\partial_\t + x \wh c, \eta \partial_\t + y \wh c] = \left( \eta \xi' - \xi \eta' \right) \partial_\t + \wh c \int\!d\t \left( \eta \xi''' - \xi \eta''' \right)
\ee
The global topology of the Virasoro group, defined as the exponentiation of $\avira$, is not completely fixed. We shall take it to have topology $\diff \times \bb R$, so that it can be characterized by pairs 
\be
\wh\psi = (\psi, r) \,\in\, \vira
\ee
where $\psi \in \diff$ and $r \in \bb R$.
Note that an alternative option would be to define the exponentiated group as being simply connected. This corresponds to ``unwrapping'' the $\diff$ factor (i.e., taking its universal cover), given that $\diff$ has fundamental group $\bb Z$ due to its $SO(2)$ subgroup of rotations (see App.~\ref{app:topologyQ}). 
Its universal cover will instead be denoted by $\un\vira$.

The Lie algebra $\avira$ can be exponentiated into the group $\vira$ with the help of Baker-Campbell-Hausdorff formula, which reads
\be\label{BCHform}
\exp(\wh\xi) \exp(\wh\eta) = \exp\!\left( \wh\xi + \wh\eta + \frac{1}{2} [\wh\xi , \wh\eta] + \frac{1}{12} [\wh\xi, [\wh\xi , \wh\eta]] - \frac{1}{12} [\wh\eta, [\wh\xi , \wh\eta]] + \cdots \right)
\ee
where $\wh\xi,\,\wh\eta \in \avira$ and the ``$\cdots$'' consists of higher-order algebra products. This formula allows us to reconstruct the group, at least in a neighborhood of the identity.\footnote{Rigorously speaking, this is only true for finite-dimensional algebras, since in this case it can be shown that the exponential map yields a diffeomorphism between a neighborhood of $0$ in the algebra and a neighborhood of $e$ in the group. In fact, the exponential map from $\adiff$ into $\diff$ is not surjective, even in a neighborhood of the identity. This happens because there are diffeomorphims $\psi$, arbitrarily close to the identity, that do not admit a ``square root'', i.e., there is no $\phi$ such that $\psi = \phi^2$. If $\psi$ were equal to the exponential of some algebra element, $\psi = \exp(\xi)$, then $\phi = \exp(\xi/2)$ would be the square root of $\psi$.\label{fn:expdiff}} From the Baker-Campbell-Hausdorff formula, it is clear that when restricting to $\wh\xi = \xi\partial_\t$ the exponential map yields $\diff$, and when restricting to $\wh \xi = x \wh c$ the exponential map yields $(\bb R, +)$. So let us define
\be\label{viraexpdef}
(\psi_\xi, x) := \exp(\xi\partial_\t+x \wh c)
\ee
in which
\be
\psi_\xi := \exp_\fr{diff} (\xi\partial_\t)
\ee
where $\exp_\fr{diff}$ denotes the exponential from $\adiff$ into $\diff$. Using that $\wh c$ is a central element, the Baker-Campbell-Hausdorff formula also gives
\be
\exp(\xi\partial_\t) \exp(x \wh c) = \exp(\xi\partial_\t+x \wh c)
\ee
which in the notation above reads
\be
(\psi_\xi, x) = (\psi_\xi, 0)(I, x)
\ee
With a natural extrapolation we can pose
\be\label{viradeco}
(\psi, r) = (\psi, 0)(I, r)
\ee
for all $\psi \in \diff$ and $r \in \bb R$, which implies that every element of $\vira$ can be decomposed as a product of an element of $\diff \subset \vira$ and an element of the center $\bb R \subset \vira$. 

Next we derive the general form for the product rule in $\vira$. First, write \eqref{viraprod} in the compact form $[\xi\partial_\t + x \wh c, \eta \partial_\t + y \wh c] = [\xi \partial_\t, \eta \partial_\t]_\fr{diff} + w(\xi, \eta) \wh c$, where $w(\xi, \eta)$ is the bilinear functional of $\xi$ and $\eta$ given by
\be
w(\xi, \eta) := \int\!d\t \left( \eta \xi''' - \xi \eta''' \right)
\ee
From Baker-Campbell-Hausdorff formula we get,
\begin{align}
&\exp(\xi\partial_\t+x \wh c) \exp(\eta\partial_\t+y \wh c) = \nonumber\\
&= \exp\!\bigg[\left( \xi  + \eta  + \frac{1}{2} [\xi , \eta ]_\fr{diff} + \cdots \right)\partial_\t +  \left( x + y + \frac{1}{2} w(\xi, \eta) + \frac{1}{12} w(\xi, [\xi, \eta]_\fr{diff}) + \cdots \right) \wh c \, \bigg] \nonumber\\
&= \exp\! \left[ \left( \xi  + \eta  + \frac{1}{2} [\xi , \eta ]_\fr{diff} + \cdots \right)\partial_\t  \right] \exp\!\bigg[ \left( x + y + W(\xi, \eta) \right) \wh c \,\bigg]
\end{align}
where $W(\xi, \eta) := \frac{1}{2} w(\xi, \eta) + \frac{1}{12} w(\xi, [\xi, \eta]_\fr{diff}) + \cdots$ is a (real) function of $\xi,\, \eta \in \adiff$. The first factor can be identified as the product rule in $\diff$, while the second factor gives a central element,
\be
\exp(\xi\partial_\t+x \wh c) \exp(\eta\partial_\t+y \wh c) = (\psi_\xi \psi_\eta, 0) (I, x + y + W(\xi, \eta)) = (\psi_\xi \psi_\eta,  x + y + W(\xi, \eta))
\ee
Since $\xi$ determines $\psi_\xi$, we can by a slight abuse of notation write $W(\xi, \eta)$ as $W(\psi_\xi, \psi_\eta)$. The general product rule in $\vira$ then reads
\be
(\psi, a)(\phi, b) = (\psi\phi, a + b + W(\psi, \phi))
\ee
Note that this derivation only defines $W$ when acting on diffeomorphisms in the image of the exponential map (see footnote \ref{fn:expdiff}), but we can smoothly extend the domain of $W$ so as to include the exceptional points. This function is called the {\sl Bott 2-cocycle}, and it can be explicitly expressed as $W(\psi, \phi) = \int_{S^1} \log ( \psi \circ \phi)' d \log \phi'$, where the prime denotes derivative with respect to the angle \cite{kosyak2018regular}.

\subsection{Adjoint representation}
\label{subapp:ViraAd}

Let us now consider the adjoint representation of the Virasoro group on its Lie algebra, and the corresponding adjoint representation of the Lie algebra on itself. As always, the adjoint representation of the Lie algebra on itself is given by the algebra product, 
\be\label{adxieta}
\ad_{\xi\partial_\t + x \wh c} ( \eta \partial_\t + y \wh c ) = \left( \eta \xi' - \xi \eta' \right) \partial_\t + \wh c \int\!d\t \left( \eta \xi''' - \xi \eta''' \right)
\ee
As the adjoint action of the algebra on itself also corresponds to the ``derivative'' of adjoint action of the group on the algebra, we can think of $\ad_{\wh\xi}\wh\eta$ as the infinitesimal change of $\wh\eta$ as one deform the identity in the group along a curve tangent to $\wh\xi$ (see the ``$\Ad_g$, $\ad_g$ and $\ad_\xi$'' entries in App.~\ref{app:GSandC}). Therefore we can ``integrate'' the action above to obtain the adjoint representation of the group. First note that $\wh c$ is represented trivially, i.e., $\ad_{\wh c} = 0$. Upon integration, it is clear that $(I, r)$ will also be represented trivially, i.e., $\ad_{(I, r)} = 1$, where $1$ denotes the identity operator on $\avira$. Using the decomposition property in \eqref{viradeco}, and the fact that the adjoint action forms a representation of the group, we have
\be
\ad_{(\psi, r)} = \ad_{(\psi, 0)(I, r)} = \ad_{(\psi, 0)} \ad_{(I, r)} = \ad_{(\psi, 0)}
\ee
that is, the adjoint action does not depend on the central component of $\wh\psi = (\psi, r)$. Accordingly, we can simplify the notation and just write $\ad_\psi$ instead of $\ad_{\wh\psi}$, while keeping in mind that this refers to the adjoint representation of $\vira$, not $\diff$. Note also that, since $\ad_{\wh\xi} \wh c = 0$, the adjoint map should act trivially on $\wh c$, i.e., $\ad_\psi \wh c = \wh c$. Thus we have,
\be\label{adpsieta}
\ad_\psi \wh\eta = \ad_\psi(\eta \partial_\t) + y \wh c
\ee
and we can focus on $\ad_\psi (\eta \partial_\t)$. Note that $\ad_{\xi} (\eta \partial_\t)$ changes both the $\partial_\theta$ and the central components of $\eta$. Since the way it changes the $\partial_\theta$ component is exactly the same as the action of the adjoint map in $\diff$, we conclude that the ``integrated'' action must have the form
\be\label{advirasymb}
\ad_\psi(\eta \partial_\t) = \psi_*(\eta \partial_\t) + \Lambda_\psi(\eta) \wh c
\ee
where $\Lambda$ is some $\psi$-dependent linear functional of $\eta$. We can get a clue about what is $\Lambda_\psi$ by computing the product of two adjoint maps,
\be
\ad_\psi \ad_\phi \eta = \ad_\psi ( \phi_*\eta + \Lambda_\phi(\eta) \wh c ) = \psi_*\phi_* \eta + \left( \Lambda_\psi(\phi_*\eta) + \Lambda_\phi(\eta) \right) \wh c
\ee
where, for notational simplicity, $\eta$ is being written in place of $\eta(\t)\partial_\t$. But since the adjoint map forms a representation of the group, 
\be
\ad_\psi \ad_\phi = \ad_{(\psi, 0)} \ad_{(\phi, 0)} = \ad_{(\psi, 0)(\phi, 0)} = \ad_{(\psi \phi, W(\psi, \phi))} = \ad_{\psi\phi}
\ee
which gives
\be
\ad_\psi \ad_\phi \eta = (\psi\phi)_* \eta + \Lambda_{\psi\phi}(\eta) \wh c
\ee
Hence we conclude that $\Lambda$ must satisfy the relation
\be\label{LambdarelSch}
\Lambda_{\psi\phi} = \Lambda_\psi \circ \phi_* + \Lambda_\phi
\ee
A solution to this relation can be obtained from the {\sl  Schwarzian derivative} \cite{ovsienko2004projective}, a map from $\diff$ into $C^\infty(S^1, \bb R)$ defined as
\be
S[\psi](\t) := \frac{\psi'''(\t)}{\psi'(\t)} - \frac{3}{2} \left( \frac{\psi''(\t)}{\psi'(\t)} \right)^2
\ee
in which $\psi$ is being realized as a monotonous map from $[0, 2\pi]$ into $\bb R$ satisfying $\psi(2\pi) = \psi(0) + 2\pi$. It satisfies the relation
\be\label{Schprop}
S[\psi \circ \phi] = \left( S[\psi] \circ \phi \right) (\phi')^2 + S[\phi]
\ee
that is, $S[\psi \phi](\t) = S[\psi] (\phi(\t)) \phi'(\t)^2 + S[\phi](\t)$. Comparing with  \eqref{LambdarelSch}, we infer
\be
\Lambda_\psi(\eta) = \kappa \int\!d\t\, S[\psi](\t) \, \eta(\theta)
\ee
where $\kappa$ is a coefficient to be adjusted. If $\bar\t = \phi(\t)$, we have
\be
(\bar\eta \partial_\t) \big|_{\bar\theta} := \phi_*(\eta \partial_\t) = \eta(\t) \phi'(\t) \partial_\t
\ee
that is, $\bar\eta(\bar\t) =  \phi'(\t)\eta(\t)$. Then,
\begin{align}
\Lambda_{\psi\phi}(\eta) &= \kappa \int\!d\t\, S[\psi\phi](\t) \, \eta(\theta) \nonumber\\
&= \kappa \int\!d\t\, \left( S[\psi] (\phi(\t)) \phi'(\t)^2 + S[\phi](\t) \right) \eta(\theta) \nonumber\\
&= \kappa \int\!d\t\, \phi'(\t)  S[\psi](\phi(\t)) \phi'(\t) \eta(\t) + \Lambda_\phi(\eta) \nonumber\\
&= \kappa \int\!d\bar\t\,  S[\psi] (\bar\t)) \bar\eta(\bar\t) + \Lambda_\phi(\eta) \nonumber\\
&= \Lambda_\psi(\phi_*\eta) + \Lambda_\phi(\eta)
\end{align}
where in the second line we used property \eqref{Schprop} and in the fourth line we changed the integration variable from $\t$ to $\bar\t$. Finally, in order to determine $\kappa$, we compute the derivative of \eqref{adpsieta} and demand that it matches with \eqref{adxieta}. If $t \mapsto \psi_t$ is a curve in $\diff$ tangent to $\xi \in \adiff$, then we have
\be
\ad_\xi(\eta) = \frac{d}{dt} \ad_{\psi_t} (\eta) = [\xi, \eta]_\fr{diff} + \wh c \kappa \int\!d\t\, \frac{d}{dt} S[\psi_t](\t) \, \eta(\theta)
\ee
where the derivative is evaluated at $t=0$. If $\xi = \xi(\theta)\partial_\t$, then a possible choice for the curve is $\psi_t(\theta) = \theta + t\xi(\theta)$, so that $S[\psi_t](\theta) = t\xi'''(\t) + \ca O (t^2)$. Thus,
\be
\ad_\xi(\eta) =  \left( \eta \xi' - \xi \eta' \right) \partial_\t + \wh c \kappa \int\!d\t\, \xi''' \eta
\ee
Note that, by integrating by parts, we have $\int\!d\t\, \xi''' \eta = - \int\!d\t\, \xi \eta'''$, and so we see that the choice $\kappa = 2$ reproduces \eqref{adxieta}. That is, 
\be\label{adviraA}
\ad_{\wh\psi} \wh\eta = \psi_*(\eta\partial_\t) + \wh c \left( y + 2 \int\!d\t\, S[\psi](\t) \, \eta(\theta) \right)
\ee
where $\wh\psi = (\psi, r)$ and $\wh\eta = \eta\partial_\t + y \wh c$.

\subsection{Coadjoint representation}
\label{subapp:ViraCoad}

The coadjoint representation of $\vira$ on its dual Lie algebra, $\dvira$, and the corresponding coadjoint representation of $\avira$ on $\avira^*$ are derived here. Similar to our manner of characterizing elements of $\ddiff$, we characterize elements of $\avira^*$ as $\wt\a = \a + \a_0 \wt c$ where $\a := \a(\t) d\t^2$ is a quadratic form on $S^1$ and $\wt c$ is dual to $\wh c$ in the sense that $\wt c(\eta + y \wh c) = y$. The pairing of $\wt\a \in \avira^*$ with $\wh\eta \in \avira$ is defined as
\be
\wt\a (\wh\eta) := \int_{S^1} \alpha(\eta) + \alpha_0 y = \int\!d\t\, \alpha(\t) \eta(\t) + \alpha_0 y
\ee	
From the definition of the coadjoint action (see ``$\coad_g$'' entry in App.~\ref{app:GSandC}) we have
\be
\coad_{(\psi, r)} \wt\alpha \left(\wh\eta\right) = \wt\alpha \left(\ad_{(\psi, r)^{-1}} \wh \eta \right) =  \wt\alpha \left(\ad_{(\psi^{-1}, -r - W(\psi, \psi^{-1}))} \wh \eta \right)
\ee
which from \eqref{adviraA} gives
\begin{align}
\coad_{(\psi, r)} \wt\alpha \left(\wh\eta\right) &= \wt\alpha \left( \psi^{-1}_*\eta + \wh c \left( y + 2 \int\!d\t\, S[\psi^{-1}](\t) \, \eta(\theta) \right) \right) \nonumber\\
&= \int \alpha ( \psi^{-1}_*\eta ) + 2\alpha_0 \int\!d\t\, S[\psi^{-1}](\t) \, \eta(\theta)  + \alpha_0 y \nonumber\\
&= \big( \psi_*(\alpha d\t^2 )+ 2\a_0 S[\psi^{-1}] d\t^2 + \alpha_0 \wt c \big) (\wh\eta )
\end{align}
where, as usual, $\psi_* = \psi^{-1*}$. Therefore,
\be\label{coadpsiexplA}
\coad_{(\psi, r)} \wt\alpha =  \psi_*(\alpha d\t^2)  + 2\a_0 S[\psi^{-1}] d\t^2 + \alpha_0 \wt c 
\ee
Note that $r$ also acts trivially through the coadjoint map, so we can use the shortened notation $\coad_\psi$. 

The coadjoint representation of the group leads to a coadjoint representation of the algebra, obtained by differentiation. The simplest way to derive this representation is by using directly the adjoint representation of the algebra, given in \eqref{adxieta}, since the coadjoint action is (minus) the dual of the adjoint action. We have,
\begin{align}
\coad_{\wh\xi} \wt\alpha (\wh \eta) = - \wt\alpha(\ad_{\wh\xi}\wh \eta) &= - \wt\alpha \left( \left( \eta \xi' - \xi \eta' \right) \partial_\t + \wh c \int\!d\t \left( \eta \xi''' - \xi \eta''' \right) \right) \nonumber\\
&= - \int\!d\t\, \alpha  \left( \eta \xi' - \xi \eta' \right) - \alpha_0  \int\!d\t \left( \eta \xi''' - \xi \eta''' \right) \nonumber\\
&= - \int\!d\t\, \eta \left( 2 \a \xi' + \xi \a' + 2 \a_0 \xi''' \right)
\end{align}
from which we read
\be
\coad_{\wh\xi} \wt\alpha  = - \left( 2 \a \xi' + \xi \a' + 2 \a_0 \xi''' \right) d\t^2
\ee
Naturally, this can also be obtained by explicitly differentiating \eqref{coadpsiexplA}.

\section{Topology of $\ca Q = \diff/\psl$}
\label{app:topologyQ}

We have argued that a more sophisticated method of quantization, such Isham's group-theoretic method, was necessary to handle the phase space $T^*(\diff/\psl)$ due to its non-linear structure. In this appendix we explain that this space does have a ``trivial'' topology (more precisely, contractible\footnote{We thank E. Witten for bringing this to our knowledge.}), and it even admits a global chart of ``Cartesian'' coordinates (i.e., coordinates ranging from $-\infty$ to $+\infty$ that cover the entire space), but it appears that there is no preferred choice of coordinates due to the absence of an underlying linear structure. 

Let us begin by quoting an important result in homotopy theory \cite{frankel2011geometry}. Given a fiber bundle $F \hookrightarrow E \rightarrow M$, where the base space $M$ is connected, there is an associated {\sl exact} sequence of homotopy groups 
\be\label{longexactseq}
\ldots \rightarrow \pi_n(F) \rightarrow \pi_n(E) \rightarrow \pi_n(M) \rightarrow \pi_{n-1}(F) \rightarrow \ldots \rightarrow \pi_0(F) \rightarrow 0
\ee
where $0$ denotes the trivial group.
The maps featuring in this sequence are the following: $\pi_n(F) \rightarrow \pi_n(E)$ is the homomorphism induced from the fiber-into-bundle inclusion map $F \hookrightarrow E$ (while $F$ is not canonically embedded as a fiber of $E$, any such embedding defines the same homomorphism between the homotopy groups since the base space is connected); $\pi_n(E) \rightarrow \pi_n(M)$ is the homomorphism induced from the bundle projection map $E \rightarrow M$; and $\pi_n(M) \rightarrow \pi_{n-1}(F)$ is the boundary homomorphism (sometimes denoted by $\partial$) obtained by lifting (marked) $n$-spheres $s$ on $M$ to $n$-balls $b$ on $E$, whose boundaries $\partial b$ are $(n-1)$-spheres on the fiber $F$ over the mark of $s$. This sequence is exact, meaning that the image of one map coincides with the kernel of the next, and all the maps are homomorphisms (with the possible exception of the last two since $\pi_0(F)$ may not be a group --- however, as we will be interested in principal bundles, $\pi_0(F)$ has a group structure inherited from $F$).

Now we recall that $\diff$ has fundamental group $\bb Z$. To see this, note that the group $\text{\sl Diff}^+\!(\bb R; 2\pi)$ of (orientation-preserving) diffeomorphisms $f: \bb R \rightarrow \bb R$ of the real line ($f'(x) > 0$), satisfying the condition
\be
f(x +2\pi) = f(x) + 2\pi
\ee
is a (group) covering of $\diff$. The projection map $\rho: \text{\sl Diff}^+\!(\bb R; 2\pi) \rightarrow \diff$ is defined from the identification $S^1 = \bb R/(2\pi\bb Z)$, where $\psi_f := \rho(f)$ acts on $\theta \in [0, 2\pi)$ as $\psi_f(\theta) = f(\theta) \mod 2\pi$. 
The kernel of $\rho$, $\text{ker}(\rho)$, which is a normal subgroup of $\text{\sl Diff}^+\!(\bb R; 2\pi)$, consists of all translations by multiples of $2\pi$, i.e., $f_n(x) = x+ 2\pi n$ where $n \in \bb Z$, and thus it is isomorphic to the abelian group of integers $\bb Z$.
Moreover, as to be expected, this (discrete) subgroup coincides with the center of $\text{\sl Diff}^+\!(\bb R; 2\pi)$. Accordingly, we can think of $\diff$ as the coset group
\be
\diff = \text{\sl Diff}^+\!(\bb R; 2\pi)/\text{ker}(\rho) = \text{\sl Diff}^+\!(\bb R; 2\pi)/\bb Z
\ee
so that $\text{\sl Diff}^+\!(\bb R; 2\pi)$ is a (principal) fiber bundle over the base $\diff$ with fibers (structure group) $\bb Z$, which we write $\bb Z \hookrightarrow \text{\sl Diff}^+\!(\bb R; 2\pi) \rightarrow \diff$. 
Since $\text{\sl Diff}^+\!(\bb R; 2\pi)$ is simply-connected, it is actually the universal cover of $\diff$,
\be
\un\diff  = \text{\sl Diff}^+\!(\bb R; 2\pi)
\ee
Inserting that $\pi_1(\text{\sl Diff}^+\!(\bb R; 2\pi)) = 0$ and $\pi_0(\bb Z) = \bb Z$ into \eqref{longexactseq}, we obtain the exact (sub)sequence
\be
0 \rightarrow \pi_1(\diff) \rightarrow \bb Z \rightarrow 0
\ee
That $\bb Z$ maps to $0$ implies that $\pi_1(\diff) \rightarrow \bb Z$ is surjective, and that $0$ maps to $\pi_1(\diff)$ implies that $\pi_1(\diff) \rightarrow \bb Z$ is injective. Thus the fundamental group of $\diff$ is
\be
\pi_1(\diff) = \bb Z
\ee
Evidently the non-contractible loops are associated with the $SO(2) \sim S^1$ subgroup of $\diff$, and the homotopy classes are characterized by the winding number. 

Finally, let us consider $\ca Q = \diff/\psl$.\footnote{This space is a submanifold of the so-called {\sl universal Teichm{\"u}ller space}.} One can anticipate that that $\ca Q$ is simply-connected as the $SO(2)$ inside $\diff$ is ``cancelled out'' by the $SO(2)$ in $\psl \sim \bb R^2 \rtimes SO(2)$.
In fact, a rotation by $2\pi n$ in $\psl$ is mapped to a rotation by $2\pi n$ in $\diff$ when $\psl$ is embedded as a subgroup of $\diff$; therefore $\pi_1(\psl) \rightarrow \pi_1(\diff)$ is an isomorphism.
From \eqref{longexactseq}, using that $\pi_0(\psl) = 0$, we get
\be
\bb Z \overset{\text{id}}{\rightarrow} \bb Z \rightarrow \pi_1(\ca Q) \rightarrow 0
\ee
From the map $\pi_1(\ca Q) \rightarrow 0$ we conclude that $\bb Z \rightarrow \pi_1(\ca Q)$ must be a surjection, and from the fact that $\bb Z \rightarrow \bb Z$ is an identity (id)  it follows that $\bb Z \rightarrow \pi_1(\ca Q)$ must be trivial. Therefore $\pi_1(\ca Q) = 0$.
It is straightforward to show that all homotopy groups of $\ca Q$ are trivial, so $\ca Q$ is topologically contractible. 

While not guaranteed that a contractible space is homeomorphic to a vector space, in this case one can construct global charts of ``Cartesian coordinates'' for $\ca Q$. A possible construction follows. 
First notice that given any two sets of three points on $S^1$, there is always a $\psl$ transformation that maps one set into the other. We can therefore use this property to ``gauge-fix'' the $\psl$ action
by restricting to diffeomorphisms that keep any three particular points fixed, e.g., $\theta = 0,\,2\pi/3,\,4\pi/3$. 
This is not a complete "gauge-fixing", since there is a $\bb Z_3$ subgroup of $SO(2) \subset \psl$ that maps those three points to themselves. Therefore we can think of $\diff/\psl$ as a trio of diffeomorphisms on a closed interval, up to cyclic permutation, 
\be\label{diffpsldiff3z3}
\diff/\psl = \text{\sl Diff}^+\!([0, \pi])^3/\bb Z_3
\ee
Also, $\text{\sl Diff}^+\!([0, \pi])$ can be characterized by functions $f : [0,\pi] \rightarrow [0,\pi]$ satisfying $f(0) = 0$, $f(\pi) = \pi$ and $f'(x) > 0$. A manner to automatically enforce the last condition is to write $f'(x)$ as an exponential of a periodic function, which can be expanded in Fourier modes,
\be
f'(x) = \kappa \, e^{\sum_n \left[a_n \sin(nx) + b_n \cos(nx) \right]}
\ee
where $n \ge 1$ and $\kappa := e^{b_0}$. Then $f(x) = \int_0^x dy f'(y)$ and the only constraint now is $f(\pi) = \int_0^\pi dy f'(y) = \pi$, which fixes $\kappa$ in terms of $a$'s and $b$'s,
\be
\kappa = \frac{\pi}{\int_0^\pi dx\, e^{\sum_n \left[a_n \sin(nx) + b_n \cos(nx) \right]}}
\ee
Thus, up to the discrete quotient by $\bb Z_3$, we can think of  $\ca Q$ as being coordinatized by these six families of real parameters (i.e., for each of the three diffeomorphisms, families of $a_n$ and $b_n$). 

These coordinates can in principle be extended to the phase space, $\wt{\ca P} = T^*\ca Q$, by introducing the canonically conjugated pairs (i.e., the components of the 1-forms with respect to the coordinate basis, $p = p_i dq^i$). The symplectic form would locally take the Darboux format --- the coordinates are not global because of the $\bb Z_3$ quotient, which would need to be addressed if one chooses to proceed with Dirac's quantization. 
It is important to stress, however, that these coordinates are arbitrary, and since it appears that there is no choice that is physically preferred, 
this approach cannot be used to justify a natural quantization. 

Let us comment on a quasi-invariant measure on $\diff/\psl$.\footnote{If $G$ is a separable locally compact group and $H$ is a closed subgroup, then the homogeneous space $G/H$ admits a {\sl unique} (up to equivalence) quasi-invariant measure with respect to $G$ \cite{mackey1952induced,raczka1986theory}. (We are not aware if there are subtleties associated with infinite-dimensional groups.)\label{fn:uniquequasiinv}} First recall the classic {\sl Wiener measure} on $C([0,1])$, the space of continuous real functions on the interval $[0,1]$, defined as follows \cite{wiener1923differential,kuzmin2007circle,kosyak2018regular}. Let $0 \le x_1 < x_2 < \cdots < x_n \le 1$ be any ordered set of $n$ points in $[0,1]$, and $\{A_1, A_2,\, \ldots A_n\}$ be a collection of Borel sets on $\bb R$. A basis for the Borel $\sigma$-algebra on $C_0([0,1])$ consists of the sets $B(x_1,\ldots x_n; A_1, \ldots A_n)$ of functions $y \in C([0,1])$ satisfying $y(x_i) \in A_i$. The Wiener measure is defined on each basis element as
\ba
\mu_W & \big[ B(x_1,\ldots x_n; A_1, \ldots A_n) \big] := \no
&\quad \int_{A_1}\!dy_1\, p(x_1;y_1) \int_{A_2}\!dy_2\, p(x_2- x_1;y_2 - y_1) \cdots \int_{A_n}\!dy_n\, p(x_n - x_{n-1};y_n - y_{n-1}) 
\ea
where
\be
p(x;y) := \frac{e^{y^2/2x}}{\sqrt{2\pi x}} 
\ee
As proposed by Shavgulidze \cite{shavgulidze1989distributions}, the Wiener measure can be pushed (for definition see footnote \ref{pushedmeasuredef}) to a measure $\mu_S$ on $\text{\sl Diff}^+([0,1])$ via the correspondence: 
\ba
C([0,1]) &\rightarrow \text{\sl Diff}^+([0,1]) \no
y &\mapsto f(x) = \frac{\int_0^x\!dt\, e^{y(t)}}{\int_0^1\!dt\, e^{y(t)}}
\ea
This measure is quasi-invariant with respect to the left-action by diffeomorphisms. 
We can define the product measure $\mu_S \otimes \mu_S \otimes \mu_S$ on $\text{\sl Diff}^+\!([0, \pi])^3$ and push it under the $Z_3$ quotient to a measure $\mu$ on the configuration space, as given in \eqref{diffpsldiff3z3}, which is quasi-invariant with respect to the left-action by Virasoro (based on a similar result from \cite{kosyak1994irreducible,kosyak2018regular}).

Lastly, we remark that there exists a surprising, non-trivial symplectomorphism between $T^*\ca Q$, with the symplectic form associated with its cotangent bundle structure, and $\ca Q \times \ca Q$, with the symplectic structure that each factor $\ca Q$ inherits as a coadjoint orbit of Virasoro. This map is called the {\sl Mess map}, first discovered in \cite{mess2007lorentz}, and proven to be a symplectomorphism in \cite{Scarinci:2011np}. (This map actually refers to the universal Teichm{\"u}ller space, $\ca T(1)$, but modulo potential mathematical subtleties it descends to its submanifold $\diff/\psl$.)  
It is interesting that $\ca Q \times \ca Q$ is the most natural realization for the (reduced) phase space of vacuum asymptotically-$\ads$ gravity~\cite{witten2007three,Scarinci:2011np,krasnov2007minimal,bonsante2010maximal}. However, the natural group to be quantized in that case is $\vira\times\vira$, which is of course the conformal group in two-dimensions (displaying a clear holographic aspect). Thus, while the underlying phase spaces may be the same, the structure of the symmetry groups $\vira \times \vira$ and $\avira \rtimes \vira$ are distinct.

\section{Mackey's theory}
\label{app:mackey}

In this appendix we outline the idea behind Mackey's theory of induced representations \cite{mackey1965induced,mackey1978unitary,mackey1952induced,mackey1953induced}. In particular, we discuss how the concept of systems of imprimitivity can be used to prove irreducibility and exaustivity of the induced representations of semi-direct products of groups (at least in finite dimensions --- but see \cite{mackey1963infinite}).

\subsection{Induced representations}

First let us recall how to construct a representation of a group $G$ by inducing it from a representation of a subgroup $H$. Let $ \scr U : H \rightarrow \text{Aut}(\ca S)$ denote a representation of $H$ on a Hilbert space $\ca S$. From the bundle $H \rightarrow G \rightarrow G/H$, we can construct the associated bundle $\ca S \rightarrow G \times_{\scr U} \ca S \rightarrow G/H$ by ``gluing'' a copy of $\ca S$ at each point of $G/H$. More precisely, $G \times_{\scr U} \ca S$ is defined as the set of equivalence classes $[g, v] = [gh, \scr U_{h^{-1}} v]$, where $g \in G$, $h \in H$ and $v \in \ca S$, with projection map $\pi([g, v]) = gH \in G/H$. The induced representation from $H$ to $G$, $U : G \rightarrow \text{Aut}(\ca H)$, acts on the space of cross sections, $\ca H$, of this associated bundle as
\be\label{ind}
(U_g \Psi)(x) = \sqrt{\frac{d\mu_g}{d\mu}(x)}\, L_g (\Psi(g^{-1} x))
\ee
where $x \in G/H$, $\Psi : G/H \rightarrow G \times_{\scr U} \ca S$ is a cross section ($\pi(\Psi(x)) = x$, for all $x$) and $L_g$ is the $G$-action induced on the associated bundle defined by $L_g [g', v] := [gg', v]$. The Jacobian-like factor $\frac{d\mu_g}{d\mu}$, associated with a given quasi-invariant (Borel) measure $\mu$ on $G/H$, is the Radon-Nikodym derivative, with respect to $\mu$, of the pushed measure $\mu_g$ through the group action, defined by $\mu_g[B] = \mu[g^{-1}B]$ for all Borel subsets $B$ of $G/H$.\footnote{More generally, given a measurable map $\rho: X \rightarrow Y$ and a measure $\mu$ on $X$, its push-forward of $\mu$ to $Y$ is defined as $\rho_*\mu[B] = \mu[\rho^{-1}(B)]$, where $B$ is any Borel subset of $Y$ and $\rho^{-1}$ denotes the pre-image under $\rho$.\label{pushedmeasuredef}} The inner product on $\ca H$ is defined by
\be
\langle \Psi , \Psi' \rangle = \int_{G/H} d\mu(x) \lla \Psi(x), \Psi'(x) \rra
\ee
where the inner product inside the integral $\lla \,, \rra$ comes from the inner product on $\ca S$ $(\,,)$ and is defined by $\lla \Psi(x), \Psi'(x) \rra = \lla [g, v(x)], [g, v'(x)] \rra := (v(x), v'(x))$, where  $g$ is any element of $G$ that projects to $x$ under the quotient. Note that this is well-defined provided that $\scr U$ is unitary. Also note that $U$ is unitary with respect to this inner product, as can be seen from
\begin{align}
\la U_g \Psi, U_g \Psi' \ra &= \int d\mu(x) \lla U_g \Psi(x) , U_g \Psi'(x) \rra \nonumber\\
&= \int d\mu(x) \left\la\!\!\!\left\la \sqrt{\frac{d\mu_g}{d\mu}(x)}\, L_g (\Psi(g^{-1} x)), \sqrt{\frac{d\mu_g}{d\mu}(x)}\, L_g (\Psi'(g^{-1} x)) \right\ra\!\!\!\right\ra \nonumber\\
&= \int d\mu(x) \frac{d\mu_g}{d\mu}(x) \lla  L_g (\Psi(g^{-1} x)),  L_g (\Psi'(g^{-1} x)) \rra  \nonumber\\
&= \int d\mu_g \lla  L_g (\Psi(g^{-1} x)),  L_g (\Psi'(g^{-1} x)) \rra \nonumber\\
&= \int d\mu_g \lla  \Psi(g^{-1} x),  \Psi'(g^{-1} x) \rra  \nonumber\\
&= \int d\mu(x) \lla  \Psi(x),  \Psi'( x) \rra \nonumber\\
&= \la \Psi, \Psi' \ra
\end{align}
In the third line we just changed measures, $d\mu \frac{d\mu_g}{d\mu} = d\mu_g$, and in the sixth line we used the definition of the pushed measure, which implies $\int_B d\mu_g(x) f(x) = \int_{g^{-1}B} d\mu(x) f(gx)$ for any function $f$.

\subsection{Systems of imprimitivity}

Now we introduce the concept of systems of imprimitivity. But first let us define a {\sl projection-valued measure} in analogy with the usual definition of a measure. Consider a manifold $\ca M$ and a Hilbert space $\ca H$. A projection-valued measure $P$ is a map associating each Borel set $B$ of $\ca M$ to an operator $P_B$ on $\ca H$, satisfying:
\\

\noindent $(i)$ $P_{\ca M} = 1$ (i.e., identity operator on $\ca H$)

\noindent $(ii)$ $P_{B \cap B'} = P_B P_{B'}$ for any Borel sets $B$ and $B'$

\noindent $(iii)$ $P_{B_1 \cup B_2 \cup \cdots} = P_{B_1} + P_{B_2} + \cdots$ for any disjoint collection of Borel sets $B_1$, $B_2$, ...
\\

\noindent Let $G$ be a group that has an action on $\ca M$ and $U : G \rightarrow \text{Aut}(\ca H)$ be a representation of $G$ on a Hilbert space $\ca H$. A {\sl system of imprimitivity} for $U$ based on $\ca M$ is a projection-valued measure $P$ that transforms by conjugation under the group action, that is,
\be
P_{gB} = U_g P_B U_{g^{-1}}
\ee
for all Borel sets $B$ and $g \in G$. We shall refer to a system of system of imprimitivity by the pair $(U, P)$, where the Hilbert space $\ca H$ (associated with $U$) and the manifold $\ca M$ (associated with $P$) are implicit. 

Note that any representation of $G$ induced from $H$ is associated with a ``canonical'' system of imprimitivity based on $G/H$, with a natural projection-valued measure given by
\be\label{csi}
P_B\Psi(x) := H_B(x) \Psi(x)
\ee
where $H_B : \ca M \rightarrow \bb R$ is the Heaviside function
\be
H_B(x) = \left\{
\begin{array}{l l}
1 \,,  &  x \in B \\
0 \,, &  x \notin B
\end{array}
\right.
\ee
That this constitutes a system of imprimitivity can be shown as follows:
\begin{align}
( U_g P_B U_{g^{-1}} \Psi )(x) &= \sqrt{\frac{d\mu_g}{d\mu}(x)}\, L_g \left[ (P_B U_{g^{-1}} \Psi)(g^{-1}x) \right] \nonumber\\
&= \sqrt{\frac{d\mu_g}{d\mu}(x)}\, L_g \left[ H_B(g^{-1}x) (U_{g^{-1}} \Psi)(g^{-1}x) \right] \nonumber\\
&= H_B(g^{-1}x) \sqrt{\frac{d\mu_g}{d\mu}(x)}\, L_g \left[ \sqrt{\frac{d\mu_{g^{-1}}}{d\mu}(g^{-1}x)}\, L_{g^{-1}} (\Psi(x)) \right] \nonumber\\
&= H_B(g^{-1}x) L_g L_{g^{-1}} (\Psi(x)) \nonumber\\
&= H_B(g^{-1}x) \Psi(x) = H_{gB} \Psi(x) \nonumber\\
&= (P_{gB}\Psi)(x)
\end{align}
In the third line the Radon-Nikodym derivatives cancels because they combine into $\frac{d\mu_{g g^{-1}}}{d\mu} = \frac{d\mu}{d\mu} = 1$, following from the general formula $\frac{d\mu_{g g'}}{d\mu}(x) = \frac{d\mu_{g'}}{d\mu}(g^{-1}x) \frac{d\mu_{g}}{d\mu}(x)$. From the fourth to the fifth line we used $L_g L_{g'} = L_{gg'}$. 

Given a group $G$, two systems of imprimitivity, $(U, P)$ and $(U', P')$, based on the same manifold $\ca M$, are said to be equivalent if the representations, $U$ and $U'$, and the projection-valued measures, $P$ and $P'$, are related via conjugation by an isometry. More precisely, there exits an isometry $T : \ca H \rightarrow \ca H'$ such that
\begin{align}
&U' = T U T^{-1} \\
&P' = T P T^{-1}
\end{align}
It is interesting to comment here on the measure introduced in the definition of the induced representation (\ref{ind}). We said that $\mu$ is {\sl a} given quasi-invariant measure on $G/H$. In principle, each choice of $\mu$ could lead to a different representation. But this choice does not correspond to true arbitrariness because any two measures $\mu$ and $\mu'$ that are equivalent (i.e., have the same sets of measure zero) define equivalent representations. Moreover, they lead to equivalent systems of imprimitivity, as defined in (\ref{csi}). In fact, if $G$ is locally compact, the choice is unique in the sense that all quasi-invariant measures on $G/H$ are equivalent to each other \cite{bourbakielements}. This justifies the term ``canonical'' for the system of imprimitivity defined by (\ref{csi}). To see this equivalence, we note that any two equivalent measures are related by a Radon-Nikodym derivative, $d\mu(x) = \frac{d\mu}{d\mu'}(x) d\mu'(x)$, and it follows that
\be
\frac{d\mu'_g}{d\mu'}(x) = \frac{\frac{d\mu}{d\mu'}(x)}{\frac{d\mu}{d\mu'}(g^{-1}x)} \frac{d\mu_g}{d\mu}(x) 
\ee
The two representations $U_g$ and $U'_g$ defined as in (\ref{ind}), using $\mu$ and $\mu'$, act on the same Hilbert space $\ca H$. So the intertwiner $T$ must be a unitary operator on $\ca H$. Let us try to define it as a multiplication by a positive function, $(T\Psi)(x) := \lambda(x) \Psi(x)$. We have,
\begin{align}
(TU_gT^{-1} \Psi)(x) &= \lambda(x) (U_gT^{-1} \Psi)(x) \nonumber\\
&= \lambda(x) \sqrt{\frac{d\mu_g}{d\mu}(x)}\, L_g (T^{-1} \Psi(g^{-1} x)) \nonumber\\
&= \frac{\lambda(x)}{\lambda(g^{-1}x)} \sqrt{\frac{d\mu_g}{d\mu}(x)}\, L_g (\Psi(g^{-1} x))
\end{align}
so, with the choice $\lambda(x) = \sqrt{\frac{d\mu}{d\mu'}(x)}$, this gives $(U'_g\Psi)(x)$. It is also clear that $P'_B = T P_B T^{-1}$. Thus $(U, P) \sim (U', P')$.

Mackey's fundamental theorem is stated as follows:

\vskip 0.5em
\noindent\underline{\emph{The imprimitivity theorem}}~~  Let $G$ be a locally compact, separable group and $H$ a closed subgroup of $G$. Let $U$ be a unitary representation of $G$ on a Hilbert space $\ca H$ and $P$ be a system of imprimitivity for $U$ on $\ca M = G/H$. Then there exists a unitary representation $\scr U$ of $H$ on a Hilbert space $\ca S$ such that the canonical system of imprimitivity for the representation induced on $G$ is equivalent to $(U, P)$. 
\vskip 0.5em

It follows that if every unitary representation of a certain group $G$ can be associated with a {\sl transitive} system of imprimitivity (i.e., based on a homogeneous space for $G$), then they all come from unitary representations of the corresponding ``little groups'' $H$ by induction.

\subsection{Semi-direct products}

Here we consider groups of the form $A \rtimes G$, where $A$ is a vector space\footnote{Mackey's theory actually applies for abelian groups. When $A$ is not a vector space, we just need to replace below the dual space $A^*$ by the space of unitary characters $\text{Char}(A)$.} and $G$ is a locally compact, separable group. The product rule is given by
\be\label{sdprodnotation}
(\alpha, g) (\alpha', g') = (\alpha + \delta_g \alpha', gg')
\ee
where $\alpha \in A$, $g \in G$ and $\delta : G \rightarrow \text{Aut}(A)$ is a left $G$-action on $A$ (which need not be linear). 

Note that a generic element of the group can be decomposed as
\be
(\alpha, g) = (\alpha, e)(0, g)
\ee
where $e$ is the identity element of $G$. Therefore, the operators representing $A \rtimes G$ on a Hilbert space $\ca H$ will factorize accordingly $U(\alpha, g) = U(\alpha, e) U(0, g)$. We can define $V(\a) := U(\a, e)$ and $D(g) := U(0,g)$, so that
\be\label{VD}
U(\alpha, g) = V(\alpha)D(g)
\ee
Hence, in order to classify the representations of $A \rtimes G$, we can effectively study the representations of $A$ and $G$ separately, as we shall explain next.

Let us begin with $A$. We can define the self-adjoint generator $N(\a)$ by
\be
V(\a) =: e^{-i N(\a)}
\ee
Since $A$ is abelian, we have
\be
V(\a)V(\a') = V(\a + \a') \quad \Rightarrow \quad e^{-i N(\a)} e^{-i N(\a')} = e^{-i N(\a + \a')}
\ee
and, since $N(\a)$ commutes with $N(\a')$, the product of exponentials is the exponential of the sum, so $N(\a) + N(\a') = N(\a + \a')$. Moreover, as $N((t+t')\a) = N(t\a) + N(t'\a)$ for any real numbers $t$ and $t'$, $N(t\a)$ must be linear in $t$, so $N(t\a) = t N(\a)$.\footnote{More rigorously, define the operator-valued function of a real variable, $F(t) := N(t\a)$. As the representation of a Lie group is required to be smooth (i.e., a smooth homomorphism from the group into the space of linear operators), this function must be differentiable. Taking the $s$-derivative of $F(t+s) = F(t) + F(s)$, and evaluating at $s=0$, gives $F'(t) = F'(0)$, where $F'$ is the derivative of $F$ with respect to its argument. Integrating from $0$ to $t$ gives $F(t) - F(0) = \int_0^t dt' F'(0) = t F'(0)$. But $F(0) = N(0) = 0$ since $U(0)$ is the identity, so $N(t\a) = t F'(0)$. Finally, evaluating this expression at $t=1$ gives $F'(0) = N(\a)$, proving the result.} Thus,
\be
N(t\alpha + t'\alpha') = t N(\a) + t' N(\a')
\ee
meaning that $N$ is a linear map from $A$ to a space of self-adjoint operators on $\ca H$. We can therefore think of $N$ as an operator-valued element of the dual space $A^*$. Accordingly, a simultaneous basis of eigenvectors of $N(\a)$ can be labeled by elements $w \in A^*$ as
\be
N(\a) |w\ra = w(\alpha) |w\ra
\ee
Note that the eigenvalues of $N(\a)$ need not be non-degenerate, so each $|w\ra$ may be a vector in a Hilbert (sub)space $\ca S_w \subset \ca H$; and not all $w \in A^*$ need to be included in a given representation. What can we say about the relation between $\ca S_w$'s for different $w$? Note that, for every element $(\a' , g')$ of the group,
\be
(\a', g')(\a, e)(\a', g')^{-1} = (\delta_{g'}\a, e)
\ee
so $A$ is a normal subgroup of $A \rtimes G$. At the level of the representation, we have $U(\a', g') N(\a) U(\a', g')^{-1} = N(\delta_{g'}\a)$, so
\be
N(\a) U(\a', g') |w\ra = U(\a', g') N(\delta_{g'^{-1}}\a)|w\ra = w(\delta_{g'^{-1}}\alpha) U(\a', g') |w\ra = \widetilde\delta_{g'}w(\a) U(\a', g') |w\ra
\ee 
where $\widetilde\delta_{g'} w (\a) := \delta_{g'^{-1}}^*w (\a) = w(\delta_{g'^{-1}}\a)$ is the dual action of $G$ in $A^*$. This means that $U(\a', g')$ maps $\ca S_w$ isometrically into $\ca S_{\widetilde\delta_{g'} w}$. In other words, $\ca S_w$ is isomorphic to $\ca S_{w'}$ as long as $w$ and $w'$ belong to the same $G$-orbit $\ca O$ in $A^*$. However, the isomorphism is not canonical for there is no unique group element that maps $w$ into $w'$.

Considering the diagonalization above, the Hilbert space $\ca H$ is given by a direct ``sum'' of $\ca S_w$ over $A^*$,
\be\label{snag}
\ca H = \int_{A^*}\!d\mu(w)\, \ca S_w
\ee
for some (Borel) measure $\mu$ on $A^*$. This is formalized by the Stone-Naimark-Ambrose-Godement theorem \cite{raczka1986theory,folland2016course}. 
A generic state $|\Psi \ra \in \ca H$ can be expanded as
\be
|\Psi \ra = \int_{A^*} d\mu(w) |\Psi(w)\ra
\ee
where $|\Psi(w)\ra \in S_w \subset \ca H$. We can define a projector onto each $\ca S_w$ as
\be
P_w |\Psi \ra = |\Psi(w)\ra
\ee
and, for each Borel set $B \subset A^*$, define 
\be\label{PBdef}
P_B = \int_{B} d\mu(w) P_w
\ee
The action on the each $\ca S_w$ can also be written, using the Heaviside function, as $P_B|\Psi(w)\ra = H_B(w) |\Psi(w)\ra$. Note that,
\begin{align}
U(\a, g) P_B U(\a, g)^{-1} |\Psi\ra &= \int_{A^*} d\mu(w) U(\a, g) P_B U(\a, g)^{-1} |\Psi(w)\ra \nonumber\\
&= \int_{A^*} d\mu(w) U(\a, g) H_B(\widetilde\delta_{g^{-1}}w) U(\a, g)^{-1} |\Psi(w)\ra \nonumber\\
&= \int_{A^*} d\mu(w) H_B(\widetilde\delta_{g^{-1}}w) |\Psi(w)\ra \nonumber\\
&= \int_{A^*} d\mu(w)  H_{\widetilde\delta_g B}(w) |\Psi(w)\ra = \int_{A^*} d\mu(w)  P_{\widetilde\delta_g B} |\Psi(w)\ra \nonumber\\
&= P_{\widetilde\delta_g B} |\Psi\ra \label{invpvm}
\end{align}
In the second line we used that $U(\a, g)^{-1} |\Psi(w)\ra \in \ca S_{\widetilde\delta_{g^{-1}}w}$. Hence, this is a projection-valued measure $P$ on $A^*$ that transforms by conjugation under $G$. This implies that $(D, P)$ is a system of imprimitivity for $G$, based on $A^*$, where $D$ is the restriction of $U$ to $G$ introduced in (\ref{VD}). 

The system defined above is likely not transitive (unless $G$ acts transitively on $A^*$), so we cannot apply the imprimitivity theorem yet. If the action is not transitive, this means that there must exist subsets $Z$ of $A^*$ which are invariant under the action of $G$, that is, $Z = \widetilde\delta_g Z$ for all $g \in G$. Assuming those are Borel sets, (\ref{invpvm}) implies that $P_Z$ commutes with all operators $U(\a, g)$ of the representation. If we are interested in irreducible representations of $A \rtimes G$, then $P_Z$ must be either $0$ or $1$ by Schur's lemma\footnote{Schur's lemma implies that $P_Z = \lambda$, but $P_Z = P_{Z \cap Z} = P_Z P_Z$, so $\lambda(\lambda - 1) = 0$.}. Of course, any two invariant sets $Z$ and $Z'$ must be disjoint and thus we must have $P_{Z \cup Z'} = P_Z + P_{Z'}$. Consequently, there must exist at most one invariant subset $Z \subset A^*$ that is equal to $1$, all others must be $0$. This means that the action of $G$ on $A^*$ is ergodic\footnote{A group action on a space is said to be {\sl ergodic} if every invariant Borel subset is either of measure zero or the complement of a set of measure zero.} with respect to the measure $\mu$. Each such measure can lead to a different irreducible representation of $A \rtimes G$, so to exhaust all the possibilities we must classify all the inequivalent, quasi-invariant\footnote{Note that (\ref{invpvm}) implies that $P_B = 0 \Leftrightarrow P_{\widetilde\delta_g B} = 0$. From the definition of $P_B$,  (\ref{PBdef}), this means that $\mu[B] = 0 \Leftrightarrow \mu[\widetilde\delta_g B] = 0$.} measures on $A^*$ with respect to which $G$ acts ergodically. We shall consider this next.

Note that the orbits $\ca O$ in $A^*$ are invariant subspaces. We have two possibilities for the measure: 

$(i)$ The measure is ``concentrated'' on a single orbit $\ca O$, i.e., $P_{\ca O} = 1$ for some orbit $\ca O$ (and $0$ for all other orbits);

$(ii)$ All orbits have measure zero. (This case is called {\sl strictly ergodic}.)

\noindent
The case $(i)$ is the ``simple'' one. Since the measure on $A^*$ is concentrated on a single orbit $\ca O$, it can be naturally restricted to $\ca O$, $\bar\mu := \mu|_{\ca O}$. Hence, the Hilbert space in (\ref{snag}) can be reduced to
\be
\ca H_{\ca O} = \int_{\ca O} d\bar\mu(w)\, \ca S_w
\ee
Note that this is the Hilbert space that we would have constructed by starting with a single vector $|\Psi(w)\ra \in \ca S_w$ and acting with all operators $U(\a, g)$ on it. If we take $P$ to be the projection-valued measure obtained by integrating over Borel sets of $\ca O$ with respect to $\bar\mu$, then $(D, P)$ will be a transitive system of imprimitivity for $G$ based on $\ca O \sim G/H$, where $H$ is the ``little group'' associated with $\ca O$. Now we can apply the imprimitivity theorem to conclude that $D$ must be induced from a unitary representation of $H$. This fully determines the irreducible representation $U(\a, g) = V(\a)D(g)$ on $\ca H_{\ca O}$. More concretely, realizing $\ca H_{\ca O}$ as sections of the bundle $\ca S \rightarrow G \times_{\scr U} \ca S \rightarrow \ca O$, where $\scr U : H \rightarrow \text{Aut}(\ca S)$ is the corresponding irreducible unitary representation of $H$ on a Hilbert space $\ca S$ (isomorphic to any $\ca S_w$, $w \in \ca O$), then
\be
\left( U(\a, g) \Psi \right)(w) = \left( V(\a) D(g) \Psi \right)(w) = e^{-i w(\a)} \sqrt{\frac{d\mu_g}{d\mu}(w)} \, L_g\! \left(\Psi(\widetilde\delta_{g^{-1}} w)\right) 
\ee
where $w \in \ca O \subset A^*$ and $\mu$ is a quasi-invariant measure on $\ca O$.

To guarantee that there are no other representations, we need to ensure that case $(ii)$ does not occur. One simple property which is sufficient is the following. A semi-direct product $A \rtimes G$ is said to be {\sl regular} if the space of $G$-orbits in $A^*$, $A^*/G$, is measurable and there exists a measurable map $\zeta : A^*/G \rightarrow A^*$ associating to each orbit $\ca O$ a dual vector $w \in \ca O$. This map allows one to ``pull-back'' the measure $\mu$ in $A^*$ to the orbits $\ca O$ in such a way that $\mu$ can be recovered by integration \cite{bourbakielements,mackey1965induced,raczka1986theory}. More precisely, for any Borel set $B$ in $A^*$,
\be
\mu[B] = \int_{A^*/G} \! d\sigma(r) \, \bar\mu[B \cap \ca O_r]
\ee
where $r \in A^*/G$, $\sigma$ is the measure on $A^*/G$, $\ca O_r \subset A^*$ is the orbit passing through $\zeta(r)$ and $\bar\mu$ is a measure on $\ca O_r$. In that case,
\be
P_{A^*} = \int_{A^*} d\mu(w) P_w = \int_{A^*/G} \! d\sigma(r) \int_{\ca O_r} d\bar\mu(x) P_w = \int_{A^*/G} \! d\sigma(r) P_{\ca O_r}
\ee
where $x \in \ca O_r$ and $P_{\ca O_r}$ was defined by integrating $P_w$ over $\ca O_r$ with $\bar\mu$. But if $P_{\ca O_r} = 0$ for all orbits (i.e., all $r$), then $P_{A^*} = 0$, which is a contradiction. Hence, the case $(ii)$ does not occur in regular semi-direct products. 

Another definition of regularity is the following. A semi-direct product $A \rtimes G$ is said to be {\sl regular} if there is a countable family of Borel subsets $Z_i$ of $A^*$, each a union of $G$-orbits, such that every orbit $\ca O$ is the intersection of a subfamily $Z^{\ca O}_s$ containing $\ca O$. The proof that this condition prevents case $(ii)$ is similar to the above. If $P_{\ca O} = 0$ for every orbit, then
\be
P[\cap_s Z^{\ca O}_s] = \prod_s P[Z^{\ca O}_s] = P[\ca O] = 0
\ee
Since $P_{Z_i} = 0$ or $1$, by Schur's lemma, then the formula above implies that at least one member of $\{Z^{\ca O}_s\}$, say $s = \bar s$, satisfies $P[Z^{\ca O}_{\bar s}] = 0$. Now consider the subfamily of $Z_i$ composed of such members $Z^{\ca O}_{\bar s}$. Since they cover $A^*$, we would have $P_{A^*} = 0$, which is a contradiction.

Let us consider examples of a regular and a irregular semi-direct product. First, consider $\bb R^2 \rtimes SO(2)$ where $SO(2)$ acts as usual rotations. The dual action on $\bb R^{2*} \sim \bb R^2$ is also a rotation. The orbits decompose into circles (plus the origin), so $\bb R^2/SO(2) \sim \bb R^+ \cup \{0\}$. A possible choice of $\zeta$ is to associate the radius $r$ of the orbit $C_r = \{x \in \bb R^2, |x| = r\}$ with the point it crosses the $x$-axis, i.e., $\zeta(r) = (r, 0)$. This is clearly a measurable map (given appropriate measures on $\bb R^+ \cup \{0\}$ and $\bb R^2$). For instance, if $\mu$ is the usual (Euclidean) measure on $\bb R^2$, the measure of a Borel set $B$ can be decomposed as follows
\be
\mu[B] = \int_0^\infty \! dr \, \bar\mu[B \cap C_r]
\ee
where $\bar\mu$ is the measure associated with $r d\theta$, with $0 \le \theta < 2\pi$. Thus $\bb R^2 \rtimes SO(2)$ is regular. It is also regular with respect to the second definition because we can take $Z_i$ to correspond to the family of all discs $D_r$, $r \in \bb Q$, together with their complements $\overline D_r = \bb R^2 - D_r$. 

Second, let us consider $\bb R^2 \rtimes \bb Z$ where the cyclic group $\bb Z$ acts on $\bb R^2$ as rotations by irrational multiples of $\pi$, i.e., $(r, \theta) \mapsto (r, \theta + \pi\gamma)$ with $\gamma$ irrational. The circles $C_r$ are invariant under this action, but every orbit inside $C_r$ is countable and thence have measure zero in the usual measure of the circle. The action is strictly ergodic for this measure. In particular, there exists irreducible unitary representations of $\bb R^2 \rtimes \bb Z$ carried by wavefunctions on $C_r$, such as 
\be
U(\alpha, n)\Psi(x) = e^{-i\alpha \cdot x} \Psi(n^{-1}x)
\ee
where $x \in C_r \subset \bb R^2$ and $nx$, with $n \in \bb Z$, denotes $x$ rotated by $n\pi\gamma$. Since $C_r$ is not an orbit of $\bb Z$, this is an ``extra'' representation of the group. This semi-direct product is not regular since no measurable map $\zeta$ exists, given that the space of orbits is $(S^1/\gamma\pi\bb Z) \times \bb R^+ \cup \{0\}$ is non-measurable.

\section{Projective representations}
\label{app:projrep}

In this appendix we define projective representations of a Lie group, and  explain their relationship with true representations of central extensions (by 2-cocycles) of the group. The intention is to offer a simple review of the subject, adapted to our notation. 
For details see \cite{bargmann1954unitary} or, for a more informal discussion, \cite{StackEx}.
Rigorously, the results here are only valid for finite-dimensional groups, even though we apply them to the quantization of causal diamonds, where the canonical group, $\avira \rtimes \vira$, is infinite-dimensional.
To simplify the language, we shall use ``unirrep'' as a short for ``unitary irreducible representation''.

\subsection{Definition}

If a physical symmetry acts on a Hilbert space $\mathcal H$ (i.e., so that it preserves all expectation values), then according to Wigner's theorem \cite{wigner1931gruppentheorie} the action must be unitary (linear) or anti-unitary (anti-linear). Let us consider the unitary case here, as the anti-unitary case is analogous. Since physical states are actually rays in $\mathcal H$, the unitary operators $U(g)$ and $e^{i\phi} U(g)$ implement the same physical transformation. 
Thus, the physically relevant space of transformations is the quotient $P\ca U(\mathcal H):=\ca U(\mathcal H)/U(1)$ of unitary operators on $\mathcal H$ modulo a phase. 
In this manner, a group $G$ of physical symmetries is realized in quantum mechanics as a homomorphism from $G$ into $P\ca U(\mathcal H)$, that is, $[U(g)][U(g')] = [U(gg')]$. 
Such a homomorphism is called a {\sl projective (unitary) representation of $G$ on $\mathcal H$}. 
\begin{center}
\begin{tikzcd}
 & \ca U(\ca H) \arrow{d}{\sfrac{}{U(1)}} \\
G \arrow[dashed]{ru}{\text{\it Proj Rep}} \arrow{r}[swap]{\text{\it Homo}} & P\ca U(\ca H)
\end{tikzcd}
\end{center}

\noindent Given some (arbitrary, local) association $g \mapsto U(g)$, we have 
\be
U(g) U(g') = e^{i\phi(g, g')} U(gg')
\ee
for some real function $\phi : G \times G \rightarrow \mathbb R$.
From associativity, $\phi$ must satisfy 
\be\label{phicocycle}
\phi(g, g') + \phi(gg', g'') = \phi(g', g'') + \phi(g, g'g'')
\ee
which is called the {\sl cocyle condition}. 
Applying this condition for $g' =e$ we see that $\phi(g,e) = \phi(e, g'')$, implying that $\phi(g,e) = \phi(e,g) = \phi(e,e)$. Without loss of generality, let us take $\phi(g, e) = \phi(e, g) = 0$ so that $U(e) = 1$. 
Note that if we choose another association, $g \mapsto U'_g = U_g e^{i \alpha(g)}$, where $\alpha: G \rightarrow \bb R$ is some function, we get a redefinition of $\phi$ given by $\phi'(g, g') = \phi(g, g') + \alpha(g) + \alpha(g') - \alpha(gg')$. This new $\phi$ automatically satisfies the cocycle condition, provided that the old $\phi$ does, and clearly these $\phi$'s define equivalent projective representations. So we have a {\sl cohomology}, where a generic $\phi$ satisfying the cocycle condition is called a 2-cocycle, a 2-cocycle with the particular form $\phi(g, g') = \alpha(gg') - \alpha(g) - \alpha(g')$ is called a 2-coboundary, and the set of equivalence classes in which two 2-cocycles that differ by a 2-coboundary are identified is called a 2-cohomology.

\subsection{Central extension by a 2-cocycle}
\label{appsub:2co}

Given a function $\phi: G \times G \rightarrow \bb R$ satisfying the cocycle condition \eqref{phicocycle}, and for convenience $\phi(e,e) = 1$, we define the group extension 
\be
G_\phi := G \times_\phi \bb R
\ee
by the product rule 
\be
(g, r)(g', r') = (gg', r + r' + \phi(g, g'))
\ee
The cocycle condition is necessary to ensure associativity of $G_\phi$. Note that $(e, r)$ is in the center. A generic element of the extended group can be factorized as $(g, r) = (e, r)(g, 0)$, so a (true) representation of $G_\phi$ will satisfy $U(g, r) = U(e, r)U(g, 0)$, and we can define $V(r) := U(e, r)$ and $D(g) := U(g, 0)$. 
Due to Schur's lemma, in a (complex) irreducible representation the central elements are represented as multiples of the identity. Therefore, $V(r)$ forms a unitary irreducible representation of $\mathbb R$, and consequently $V(r) = e^{i\alpha r}$ for some $\alpha \in \bb R$. Thus, 
\be
U(g, r) = e^{i\alpha r} D(g)
\ee
Clearly, $D(g) D(g') = e^{i\alpha\phi(g, g')} D_(gg')$, so $D$ is a projective representation of $G$ associated with the phase $\alpha \phi$. But if $\phi$ satisfies the cocycle condition, then so does $\alpha\phi$. We thus conclude that every (true) unitary irreducible representation of $G_\phi$ corresponds to a projective unitary irreducible representation of $G$. 
Also, notice that if $\phi'$ and $\phi$ are cohomologous (i.e., differ by a coboundary), then $G_{\phi'}$ and $G_\phi$ are homomorphic, so it follows that equivalent (true) unirreps of $G_\phi$ define equivalent projective unirreps of $G$.

The Lie algebra of $G_\phi$, denoted by 
\be
\fr g_\varphi := \fr g \oplus_\varphi \bb R
\ee
has the product structure
\be
[(\xi, a), (\xi', a')] = ([\xi, \xi], \varphi(\xi, \xi'))
\ee
where $\varphi : \mathfrak g \times \mathfrak g \rightarrow \mathbb R$ is derivative of $\phi$ with respect to both of its arguments (i.e., the push-forward on both the first and second entries of $\phi$). 
The function $\varphi$ is anti-symmetric, bilinear and satisfies the cocycle condition 
\be
\varphi(\xi, [\xi', \xi''])  + 
\varphi(\xi', [\xi'', \xi]) + 
\varphi(\xi'', [\xi, \xi']) = 0
\ee
Note that this cocycle condition can either be seen as following from the cocycle condition \eqref{phicocycle} for $\phi$, or as following directly from the Jacobi identity. There is also an analogous cohomology for the algebra, where the coboundaries (i.e., the trivial elements) are those $\varphi$'s with form $\varphi(\xi, \xi') = f([\xi, \xi'])$, where $f$ is any real linear function on $\mathfrak g$.  Naturally, a central extension of $G$ is only possible if its algebra $\fr g$ admits a central extension by 2-cocycles. If the algebra provides an obstruction to this construction, by not admitting (non-trivial) central extensions, then the group cannot be extended in this manner.

\subsection{Central extension by a discrete group}
\label{appsub:disg}

It may still be possible to further centrally extend $G_\phi$ by a {\sl discrete} abelian group, that is, to ``unwrap'' it to some covering groups. 
True representations of the covering groups also define projective representations of the group. 
The point is that the exponentiation of any unirrep of $\fr g_\varphi$ defines a unirrep of the universal cover of $G_\phi$ and, at the same time, a projective unirrep of $G$, as we will see in this section.
In this subsection we shall omit the labels $\phi$ and $\varphi$, since we are concerned only with discrete extension (that is, assume that $G$ has already been extended by a 2-cocycle, if possible).
We remark that the approach here is inspired by  \cite{weinberg2005quantum}, particularly App.~2.B, but rephrasing it in a more geometrical language.

Let us recall how ``exponentiate'' a representation of the algebra $\fr g$, $A : \fr g \rightarrow \text{SelfAdj}(\ca H)$ into a representation of the (universal cover) of the group. Consider a curve $\gamma : [0, 1] \rightarrow G$ starting at $e$ and ending at $g$, and let $\dot\gamma(t)$ be the vector tangent to it at the parameter $t$. Since $A$ acts  on the algebra, we introduce the (right-invariant) Cartan-Maurer form, $\Xi$, to map $\dot\gamma(t)$ to $\fr g$.\footnote{The {\sl right-invariant} Cartan-Maurer form uses the right-translation, $r_g(g') := g'g$, to map vectors to the identity, i.e., if $v \in T_gG$ then $\Xi(v) := r_{g^{-1}*}v$. Note that this is not the standard definition, which uses left-translations instead. Our choice is better adapted to the conventions of this section, in particular because of the introduction of the complex unit in the exponential, so that $A$ is self-adjoint instead of anti-self-adjoint.} Consider the following equation
\be
\frac{d}{dt} U(\gamma(t)) = - i A(\Xi(\dot\gamma(t))) U(\gamma(t))
\ee
with initial condition $U(\gamma(0)) = U(e) = 1$, and define $U(\gamma) := U(\gamma(1))$. The solution is the path-ordered exponential
\be
U(\gamma) = \ca P\! \exp \left( -i \int_0^1 dt A(\Xi(\dot\gamma(t))) \right)
\ee
This can be interpreted as a parallel transportation in the (trivial) principal bundle 
\be
\ca U(\ca H) \hookrightarrow \ca U(\ca H) \times G \rightarrow G
\ee
with fibers $\ca U(\ca H)$, base manifold $G$, projection map $(U, g) \mapsto g$, and the group structure is $\ca U(\ca H)$ acting from the right as $(U, g)U' = (UU', g)$. 
Accordingly, $-A\Xi$ is interpreted as a (representative of a) connection on this bundle, since it takes a tangent vector at each point on the base and returns a value in the algebra of the structure group. In this way, the formula above is expressing that $U(\gamma)$ is the parallel transport of $1$ along $\gamma$ (i.e., $\gamma(t) \mapsto (U(\gamma(t)), \gamma(t))$ is a horizontal lift).

%image.png?

Now we prove that this connection is flat. Denote the connection representative by $\omega = - A\Xi$. (This representative is associated with the trivial section, $g \mapsto (1, g)$.) The curvature is given by 
\be
F(X, Y) = d\omega(X, Y) + \omega(X) \star \omega(Y)
\ee
where $\star$ denotes the product in the underlying algebra of the bundle, $\text{SelfAdj}(\mathcal H)$, defined by $\star = -i [\,,]$. 
Since this is a tensor, its value at some point $g$ depends only on the local values of the fields $X$ and $Y$, so we might choose them to be right-invariant, i.e., let $X = r_{g*}\xi$ and $Y = r_{g*}\eta$. Thus, $\omega(X) = -A(\xi)$ and $\omega(Y) = -A(\eta)$ are constant functions on $G$. The first term gives $d\omega(X, Y) = X(\omega(Y)) - Y(\omega(X)) - \omega([X, Y]) = A\Xi([X, Y]) = - A([\xi, \eta])$, where we used $[r_{g*} \xi, r_{g*} \eta] = - r_{g*}[\xi, \eta]$. The second term gives $\omega(X) \star \omega(Y) = A(\xi) \star A(\eta) = A([\xi, \eta])$, where we used that $A$ is a homomorphism from $\fr g$ to $\text{SelfAdj}(\ca H)$. Thus, the two terms cancel and we get $F = 0$. This implies that $U(\gamma)$ depends only on the homotopy class of the curve $\gamma$ joining $e$ and $\gamma(1)$.

Here we prove that, given two curves $\gamma$ and $\gamma'$ starting from $e$, we have $U(\gamma) U(\gamma') = U(\gamma\gamma')$, where the product of curves is defined as: go along $\gamma'(t)$ with double speed for half the time, and then continue along $\gamma(t)\gamma'(1)$ with double speed until $\gamma(1)\gamma'(1)$. More precisely, 
\be
\gamma\gamma'(t) := \left\{
\begin{array}{ll}
\gamma'(2t) & \text{for $0 \le t < \frac{1}{2}$} \\ \gamma(2t-1)\gamma'(1) & \text{for $\frac{1}{2} < t \le 1$}
\end{array} \right.
\ee
Denote the vector tangent to $\gamma\gamma'(t)$ by $v(t)$. For the first half of the integration, we have $v(t) = 2 \dot\gamma'(2t)$. By changing variables, $s = 2t$, that we get $U(t = \frac{1}{2}) = U(\gamma'(1)) = U(\gamma')$. For the second half, note that $v(t) = r_{g'*} (2 \dot\gamma(2t -1))$, so that $\Xi(v(t)) = 2\Xi( \dot\gamma(2t -1))$. Multiplying the equation on both sides (from the right) by $U(\gamma')$, we get
\be
\frac{d}{dt} \left( U(\gamma\gamma'(t)) U(\gamma')^{-1} \right) = - i A\left(2\Xi(\dot\gamma(2t - 1))\right) 
\left( U(\gamma\gamma'(t)) U(\gamma')^{-1} \right)
\ee
so changing the variables to $s = 2t - 1$, the solution (at $t=1$) is just $U(\gamma\gamma') U(\gamma')^{-1} = U(\gamma)$. That is,
\be
U(\gamma)U(\gamma') = U(\gamma\gamma')
\ee
revealing that $U$ is a homomorphism from the space of homotopy classes of curves on $G$ (starting at $e$) to unitary operators on $\ca H$. 
But, if $G$ is connected, the space of homotopy classes of curves on $G$ (starting at $e$) is precisely the universal cover, $\wt G$, of $G$.
Therefore $U$ defines a unitary representation of $\wt G$ on $\ca H$.
In fact, as any unitary representation of $\wt G$ defines a self-adjoint representation of $\fr g$, we conclude that there is a one-to-one correspondence between unitary representations of $\wt G$ and self-adjoint representations of $\fr g$.

Finally, let us see how irreps of $\widetilde G$ descend to projective irreps of $G$. 
Denote a curve on $G$ going from $e$ to $g$ by $\gamma_g$.
We can always write $\gamma_g \gamma_{g'} := \eta(g, g') \gamma_{gg'}$, where $\eta$ is a curve going from $e$ to $e$ (i.e., an element of the fundamental group $\pi_e(G)$). Note that $\eta$ is an element on the fiber over $e$ in $\widetilde G$, which forms a normal subgroup of $\widetilde G$ and, being discrete, must be in the center of $\widetilde G$. It follows, by considering the triple product $\gamma_{g}\gamma_{g'}\gamma_{g''}$, that $\eta$ also satisfies a cocycle condition, $\eta(g, g')\eta(gg', g'') = \eta(g', g'') \eta(g, g'g'')$. 
Now consider $U(\gamma_g) U(\gamma_g') = U(\gamma_g \gamma_{g'}) = U(\eta(g, g'))U(\gamma_{gg'})$.  For unirreps, $U(\eta(g, g')) = e^{i \phi(g, g')}$. The cocycle condition for $\eta$ implies that $\phi$ satisfies the cocycle condition \eqref{phicocycle}. Thus, defining $D(g) := U(\gamma_g)$, we have $D(g) D(g') = e^{i\phi(g, g')} D(gg')$.

\subsection{A diagram and an example}

We have seen in Sec.~\ref{appsub:disg} that self-adjoint irreps of $\fr g_\varphi$ are in one-to-one correspondence with unirreps of $\wt G_\phi$, which in turn define projective unirreps of $G_\phi$. 
While have seen in Sec.~\ref{appsub:2co} that (projective) unirreps of $G_\phi$ also define projective unirreps of $G$, the construction does not necessarily go in the other direction, i.e., a projective unirrep of $G$ does not necessarily define a unirrep of $G_\phi$. This is because the association $g \mapsto U(g)$ is local, so it may not be possible to define a map $\phi : G \times G \rightarrow \bb R$ on the entire domain. Nevertheless, such a topological aspect is inconsequential at the level of the algebras, so a projective unirrep of $G$ does define a self-adjoint irrep of $\fr g_\varphi$, for some 2-cocycle $\varphi: \fr g \times \fr g \rightarrow \bb R$. This closes the loop, as depicted in the following diagram:
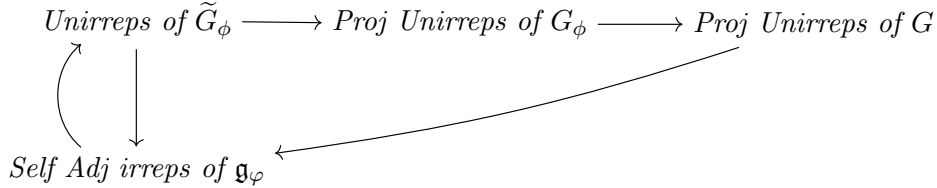
\begin{center}
\begin{tikzpicture}[commutative diagrams/every diagram]
\node (Gp) at (4.3,-0.05) {\it Proj Unirreps of $G_\phi$};
\node (wtGp) at (0,0) {\it Unirreps of $\wt G_\phi$};
\node (gv) at (0,-2) {\it Self Adj irreps of $\fr g_\varphi$};
\node (G) at (9,-0.05) {\it Proj Unirreps of $G$};
\node (wtGp2) at (-0.6,-0.2) {};
\node (gv2) at (-0.6,-1.8) {};
\path[commutative diagrams/.cd, every arrow, every label]
(Gp) edge node {} (G)
(wtGp) edge node {} (Gp)
(gv2) edge[bend left=50] node {} (wtGp2)
(G) edge[bend left=5] node {} (gv)
(wtGp) edge node {} (gv);
\end{tikzpicture}
\end{center}
In conclusion, projective unirreps of a group $G$ are in one-to-one correspondence with unirreps of the universal cover of the central extensions (by 2-cocycles $\phi$) of the group, $\wt G_\phi$, and also in one-to-one correspondence with self-adjoint irreps of the central extensions (by 2-cocycles $\varphi$) of its Lie algebra, $\fr g_\varphi$.

As an example, let us consider a rotation-invariant physical system in three Euclidean dimensions. The group of symmetry is $SO(3)$. It happens that its algebra, $\mathfrak{so}(3)$, does not admit (non-trivial) central extensions by 2-cocycles (in fact, this is true for any semisimple algebra). So the projective unirreps of $SO(3)$ must be in correspondence with true unirreps of its universal cover, $SU(2)$. 
Suppose one continuously rotate the system around some axis until it goes around by $2\pi$. This curve starts and ends at $e \in SO(3)$, but it is not contractible.  Viewing this process in $SU(2)$ the curve is open, as it starts at $e$ and ends at $e'$ (i.e., goes from the north to the south pole in $SU(2) \sim S^3$). Since $e'$ is in the center, $U(e') = e^{i\alpha}$, and since $(e')^2 = e$, then $\alpha$ is $0$ or $\pi$. That is, either the state returns to its original value (boson), $\Psi \mapsto \Psi$, or it flips sign (fermion), $\Psi \mapsto -\Psi$.

\bibliographystyle{ieeetr}
\bibliography{CausalDiamonds}

\end{document}